\documentclass[acmsmall]{acmart}

\AtBeginDocument{%
  }

\usepackage{enumitem}
\usepackage{multirow}
\usepackage{makecell}
\usepackage{booktabs}
\usepackage{pifont}
\usepackage{geometry}
\usepackage{caption}
\usepackage{float}
\usepackage{listings}
\usepackage{array}
\usepackage{subfigure}
\usepackage{subcaption}
\usepackage{caption}
\usepackage{threeparttable}
\usepackage{xspace}

\usepackage{xcolor}
\usepackage[most]{tcolorbox}
\usepackage{hyperref}
\usepackage{wasysym} 
\usepackage[normalem]{ulem}
\usepackage{graphicx}
\usepackage{algorithm}
\usepackage{algorithmicx}
\usepackage[noend]{algpseudocode}

\usepackage{booktabs}
\usepackage{makecell}
\usepackage[dvipsnames]{xcolor}

\newcolumntype{P}[1]{>{\centering\arraybackslash}p{#1}}
\tcbuselibrary{listings,skins,breakable}

\usepackage{capt-of}

\newcommand{\sysname}{\textsf{AgenticScholar}}
\newcommand{\sysnameC}{\sysname$_{noR}$\xspace}
\newcommand{\sysnameR}{\sysname$_{noC}$\xspace}
\newcommand{\matht}{CHTaxo\xspace}

\newcommand{\belowavg}[1]{\textcolor{BrickRed}{#1}}    
\newcommand{\avg}[1]{\textcolor{Blue}{#1}}             
\newcommand{\aboveavg}[1]{\textcolor{ForestGreen}{#1}}

\newcommand{\iconscale}{0.85}

\newcommand{\iconbox}[1]{\raisebox{-0.05ex}{\scalebox{\iconscale}{#1}}}

\newcommand{\full}{\iconbox{\CIRCLE}}
\newcommand{\half}{\iconbox{\LEFTcircle}}
\newcommand{\none}{$\times$}

\newcommand{\our}{\textsf{AgenticScholar}}

\newcommand{\elicit}{Elicit\xspace}
\newcommand{\gdr}{Gemini Deep Research\xspace}
\newcommand{\qwen}{Tongyi Deep Research\xspace}
\newcommand{\open}{Open Deep Researcher\xspace}
\newcommand{\smol}{SmolAgent\xspace}

\newtcolorbox{promptbox}{
    colback=gray!5,
    colframe=black!50,
    boxrule=0.5pt,
    arc=2pt,
    outer arc=1pt,
    top=4pt,
    bottom=4pt,
    left=6pt,
    right=6pt,
    enhanced,
    breakable,
    listing only,
    listing options={
        basicstyle=\footnotesize\ttfamily,
        breakindent=0pt,
        breaklines=true,
        columns=fullflexible,
        numbers=none,
    }
}

\lstdefinestyle{promptcompact}{
  basicstyle=\ttfamily\scriptsize, % tighter than \footnotesize
  breaklines=true,
  breakatwhitespace=true,
  columns=fullflexible,
  keepspaces=true,
  showstringspaces=false,
  aboveskip=2pt, belowskip=2pt,
  literate={_}{\_}1 {–}{-}1 {—}{--}2 {“}{``}1 {”}{''}1 {’}{'}1 {…}{...}3
}

\newtcblisting{evbox}{
  colback=gray!3, colframe=gray!55,
  boxrule=0.35pt, arc=2pt,
  left=3pt,right=3pt,top=3pt,bottom=3pt,
  listing only,
  listing options={
    style=promptcompact,
    basicstyle=\ttfamily\tiny
  }
}

\begin{document}

\title{\sysname{}: Agentic Data Management with Pipeline Orchestration for Scholarly Corpora}

\author{Hai Lan}
\email{h.lan@uq.edu.au}
\orcid{0009-0007-4433-9232}
\affiliation{%
  \institution{School of Electrical Engineering and Computer Science, The University of Queensland}
  \city{Brisbane}
 \state{Queensland}
  \country{Australia}
}

\author{Tingting Wang}
\email{tingting.wang@uq.edu.au}
\orcid{0000-0002-4912-7171}
\affiliation{%
  \institution{School of Electrical Engineering and Computer Science, The University of Queensland}
  \city{Brisbane}
 \state{Queensland}
  \country{Australia}
}

\author{Zhifeng Bao}
% \authornote{Zhifeng Bao is the corresponding author.}
\email{zhifeng.bao@uq.edu.au}
\orcid{0000-0003-2477-381X}
\affiliation{%
  \institution{School of Electrical Engineering and Computer Science, The University of Queensland}
  \city{Brisbane}
 \state{Queensland}
  \country{Australia}
}

\author{Guoliang Li}
\email{liguoliang@tsinghua.edu.cn}
\orcid{0000-0002-1398-0621}
\affiliation{%
  \institution{Department of Computer Science and Technology, Tsinghua University}   
  \city{Beijing}
  %\state{}
  \country{China}
}
\author{Daomin Ji}
\email{d.ji@uq.edu.au}
\orcid{0009-0000-0037-3614}
\affiliation{%
  \institution{School of Electrical Engineering and Computer Science, The University of Queensland}
  \city{Brisbane}
 \state{Queensland}
  \country{Australia}
}

\author{Ge Lee}
\email{ge.lee@uq.edu.au}
\orcid{0009-0002-8790-1495}
\affiliation{%
  \institution{School of Electrical Engineering and Computer Science, The University of Queensland}
  \city{Brisbane}
 \state{Queensland}
  \country{Australia}
}

\author{Feng Luo}
\email{f.luo@uq.edu.au}
\orcid{0009-0005-0448-3462}
\affiliation{%
  \institution{School of Electrical Engineering and Computer Science, The University of Queensland}
  \city{Brisbane}
 \state{Queensland}
  \country{Australia}
}

\author{Zi Huang}
\email{huang@itee.uq.edu.au}
\orcid{0000-0002-9738-4949}
\affiliation{%
  \institution{School of Electrical Engineering and Computer Science, The University of Queensland}
  \city{Brisbane}
 \state{Queensland}
  \country{Australia}
}

\author{Hailang Qiu}
\email{helloqiu@whu.edu.cn}
\orcid{0009-0008-4290-8056}
\affiliation{%
  \institution{School of Computer Science, Wuhan University}
  \city{Wuhan}
  \state{Hubei}
  \country{China}}

\author{Gang Hua}
\email{huagang01@whut.edu.cn}
\orcid{0009-0001-9140-0303}
\affiliation{%
  \institution{School of Computer Science and Artificial Intelligence, Wuhan University of Technology}   
  \city{Wuhan}
  \state{Hubei}
  \country{China}
}

% \setcopyright{cc}
% \setcctype{by}
% \acmJournal{PACMMOD}
% \acmYear{2026} \acmVolume{4} \acmNumber{3 (SIGMOD)} \acmArticle{131} \acmMonth{6} \acmPrice{}\acmDOI{10.1145/3788254}

\setcopyright{cc}
\setcctype{by}
% \setcopyright{none}
% \renewcommand\footnotetextcopyrightpermission[1]{}
\acmJournal{PACMMOD}
\acmYear{2026} \acmVolume{4} \acmNumber{2 (SIGMOD)} \acmArticle{131}
\acmMonth{6} \acmDOI{10.1145/3802008}

\renewcommand{\shortauthors}{Hai Lan et al.}
% Tingting Wang, Zhifeng Bao, Guoliang Li, Daomin Ji, Ge Lee, Feng Luo, Zi Huang, Hailang Qiu and Gang Hua}

\begin{abstract}
Managing the rapidly growing scholarly corpus poses significant challenges in representation, reasoning, and efficient analysis. An ideal system should unify structured knowledge management, agentic planning, and interpretable execution to support diverse scholarly queries -- from retrieval to knowledge discovery and generation -- at scale. Unfortunately, existing RAG and document analytics systems fail to achieve all query types simultaneously. To this end, we propose {\sysname}, an agentic scholarly data management system that integrates a structure-aware knowledge representation layer, an LLM-centric hybrid query planning layer, and a unified execution layer with composable operators. {\sysname} autonomously translates natural language queries into executable DAG plans, enabling end-to-end reasoning over multi-modal scholarly data. Extensive experiments demonstrate that {\sysname} significantly outperforms existing systems in effectiveness, efficiency, and interpretability, offering a practical foundation for future research on agentic scholarly data management.

\end{abstract}

%    <ccs2012>
% <concept_id>10002951.10002952</concept_id>
% <concept_desc>Information systems~Data management systems</concept_desc>
% <concept_id>10010147.10010178.10010199</concept_id>
% <concept_desc>Computing methodologies~Planning and scheduling</concept_desc>
% \ccsdesc[500]{Information systems~Data management systems}
% \ccsdesc[500]{Computing methodologies~Planning and scheduling}


\maketitle

\section{Introduction}\label{sec:intro}

The exponential growth of scholarly corpora represents one of today’s most complex data management challenges~\cite{KatzLG24}. This vast corpus is far more than a collection of text files;
it spans \textbf{\textit{multiple modalities}}, including but not limited to tables, figures, code snippets, and bibliographic metadata, forming an intricately interconnected and semi-structured ecosystem~\cite{LoWNKW20}.
Managing and reasoning over this ecosystem is key to democratizing knowledge and promoting evidence-based research, yet it pushes the boundaries of traditional data management systems.

\begin{figure*}[t]
\setlength{\abovecaptionskip}{0cm}
\setlength{\belowcaptionskip}{-0.5cm}
\centering
\includegraphics[width=1.0\textwidth]{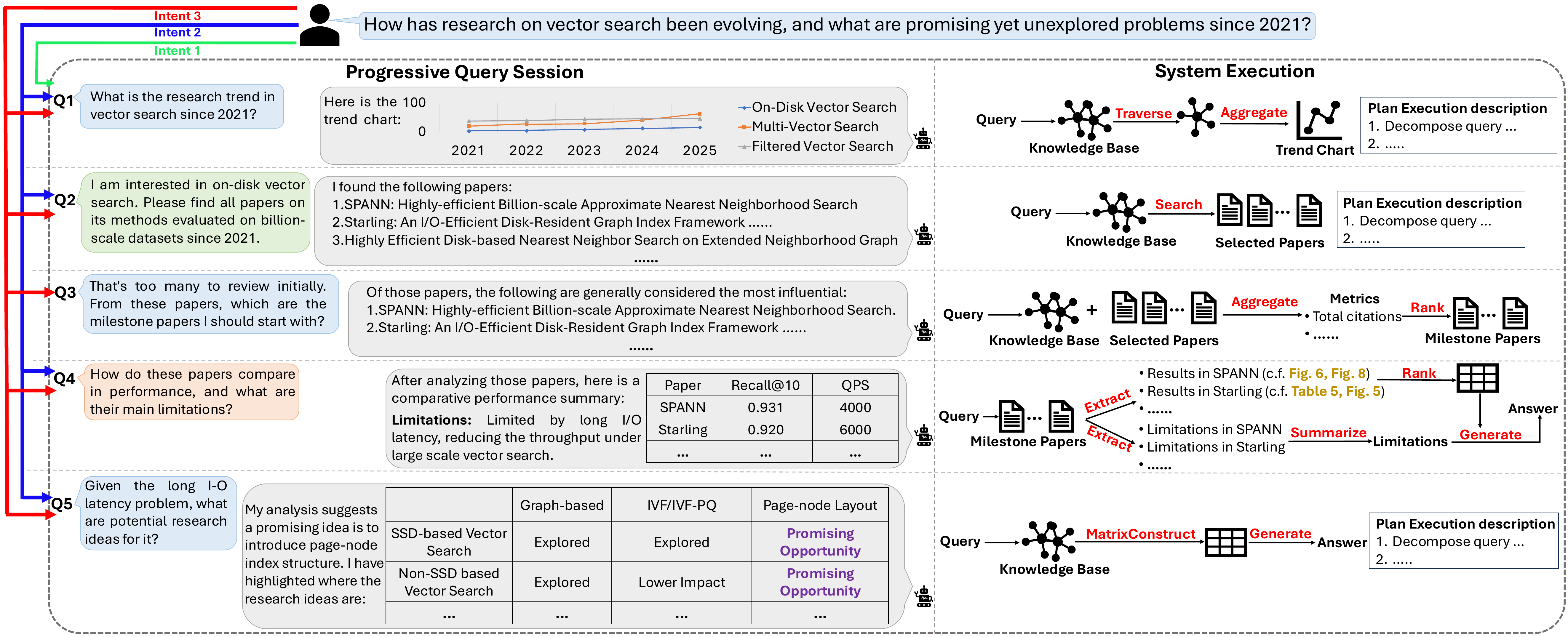}
\caption{The workflow of an agentic scholarly data
management system, illustrating how it responds to diverse query intents.}

\label{fig:example}
\vspace{-0.2em}
\end{figure*}

\newtcolorbox{examplebox}{
   enhanced, breakable,
   boxrule=0.3pt,
   colframe=gray!45,       
   colback=gray!6,          
   colbacktitle=gray!70,   
   coltitle=white,          
   fonttitle=\bfseries,    
   title=An Example Illustrating Scholarly Query Characteristics,           
   arc=2mm,                 left=3pt,right=3pt,top=1pt,  bottom=1pt,
   before skip=2pt, after skip=2pt
}
\begin{examplebox}
In Figure~\ref{fig:example}, a researcher begins with an open question -- ``How has research on vector search been evolving, and what are promising yet unexplored problems since 2021?'' -- and the system must progressively clarify intents and compose a series of analytical steps to generate an evidence-grounded answer. We observe that typical scholarly queries often:

\noindent\emph{1. Start open-ended and evolve diverse intents.}
The researcher may begin with a general trend analysis on ``vector search'' and then diverge into different  paths: stopping at the trend (\textbf{Intent 1}); expanding to performance comparison and research idea exploration across the full literature (\textbf{Intent 2}); or focusing on milestone papers for a targeted synthesis (\textbf{Intent 3}).

\noindent\emph{2. Require multi-step semantic reasoning.} Addressing Q4 involves multiple analytical steps: first \texttt{extract} experimental results and limitations from individual papers, then \texttt{rank} the results and \texttt{summarize} shared limitations across papers, and finally \texttt{generate} a coherent response.

\noindent\emph{3. Depend on multi-modal evidence and structured synthesis}. Resolving Q4 entails extracting quantitative results from tables (e.g., Table 5 of Starling) and figures (e.g., Figures 6 and 8 of SPANN, Figure 5 of Starling), aligning them across papers into comparative matrices, and grounding the synthesis in this integrated, multi-modal evidence.

\noindent\emph{4. Require context-aware knowledge generation.} Addressing Q5 requires generating novel insights that extend beyond summarization. The system must interpret the problem context, analyze the methodological landscape, and infer which problem–method combinations remain underexplored to produce an informed and contextually grounded conclusion.
\end{examplebox}

These query characteristics present a fundamental challenge for existing systems~\cite{abs-2508-05002,PatelJPGAGZ25,abs-2405-04674,ShankarCSPW25}, which will be discussed shortly in Section~\ref{sec:related_system}. While they may address isolated queries in Figure~\ref{fig:example}, they cannot autonomously orchestrate a complete analytical pipeline -- from trend analysis (Q1) through multiple reasoning steps -- to the ultimate stage of research idea exploration (Q5). This gap between capability and necessity motivates our introduction of \sysname{}, \textbf{an agentic scholarly data management system} that autonomously interprets user intents, plans complex analytical pipelines, and executes them end-to-end within the rich semantic context of a scholarly ecosystem. 
We distill four key challenges for an agentic scholarly data management system: % must address:

\smallskip\noindent\textbf{Challenge 1: Semantic Representation and Contextualization of Scholarly Data.} Scholarly documents contain multiple modalities that collectively convey a paper's semantics. The challenge is not recognizing this diversity, but representing and indexing  
it in a way that preserves \emph{both internal structure and semantic connections across components within a paper and those across related papers}. 
Queries like Q5 in Figure~\ref{fig:example} demand reasoning at the topic level, integrating contextual evidence to perform comparative or synthetic analysis. Supporting them requires representations that maintain \emph{contextual organization} beyond individual papers. Efficiently constructing and maintaining such structures, while enabling fine-grained retrieval and reasoning, remains an open data management problem. Existing systems~\cite{abs-2508-05002,WangF25} typically flatten documents into unstructured text or dense embeddings, discarding structure and context essential for complex scholarly analysis.

\smallskip\noindent\textbf{Challenge 2: Planning Complex Analytical Pipeline.} 
As shown in Figure~\ref{fig:example}, scholarly queries exhibit \emph{diverse intents} that go far beyond the rigid patterns of SQL or keyword search. A central challenge is to translate high-level natural language into an accurate and efficient execution plan. The system must infer user intent, decompose it into \emph{a coherent sequence of analytical steps}, and determine an efficient execution strategy amid the uncertainty of operating over multi-modal data. Traditional query optimizers -- built for fixed relational schemas -- lack the flexibility needed for such dynamic and context-aware semantic reasoning.

\begin{figure}[t]
\setlength{\abovecaptionskip}{0cm}
\setlength{\belowcaptionskip}{-0.7cm}
\centering
\includegraphics[width=0.8\textwidth]{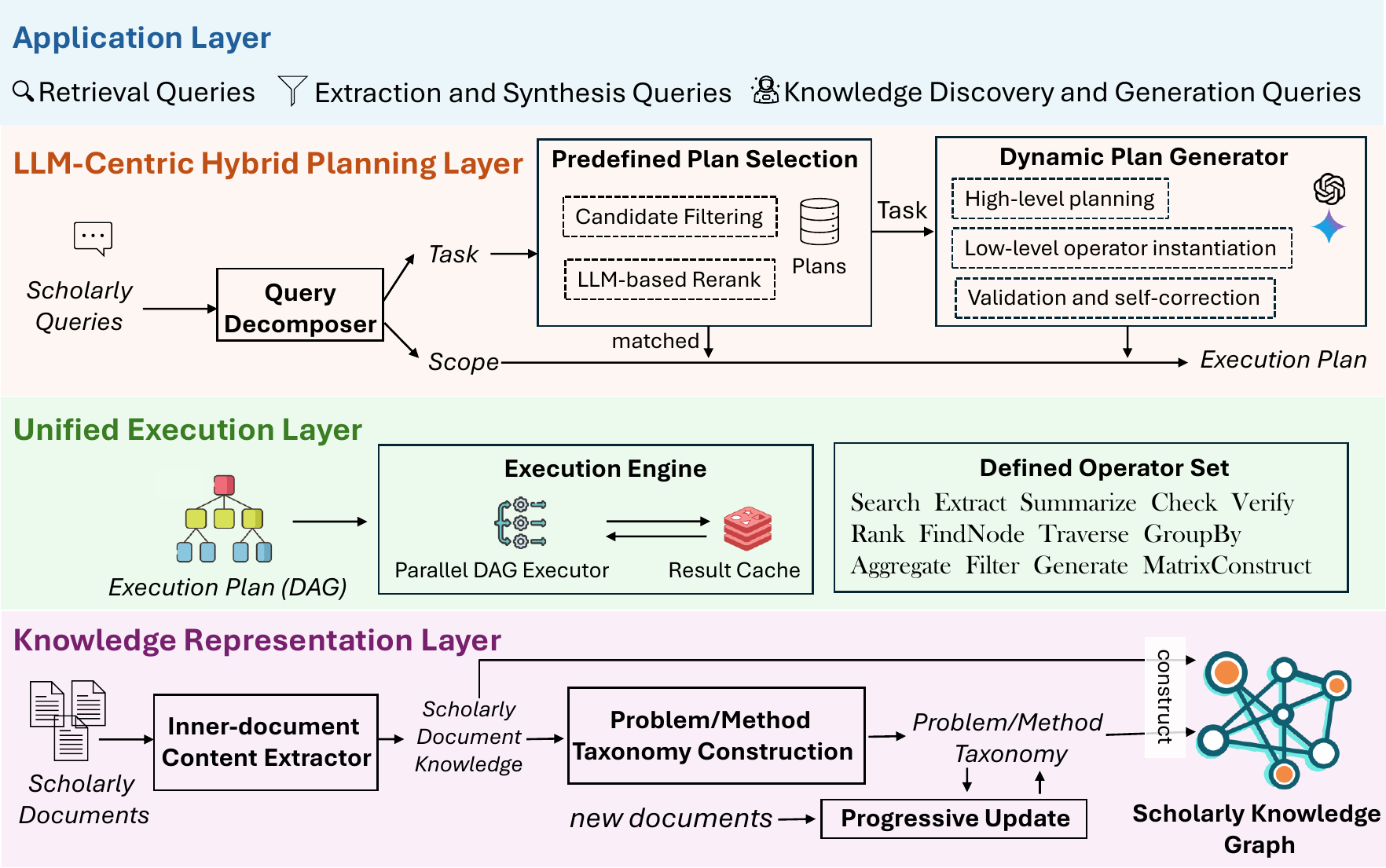}
\caption{System architecture of {\sysname}.}
\label{fig:sys-ov}
% \vspace{-0.2em}
\end{figure}

\smallskip\noindent\textbf{Challenge 3: Generality and Extensibility of the Execution Framework.} 
The space of scholarly queries is vast and continually evolving. As shown in Figure~\ref{fig:example}, a researcher's exploration often begins with an open question and progresses through diverse queries. A system built on hard-coded, query-specific pipelines would be brittle and incapable of adapting to new analytical needs. The challenge, therefore, is to design a framework that \emph{balances generality with extensibility}:
(1) its core set of analytical steps must be sufficiently expressive and composable to support a broad spectrum of queries without requiring custom implementations for each new one; and
(2) its architecture must allow new analytical steps and execution plans to be integrated seamlessly, enabling support for emerging query types without necessitating a full system redesign.

\smallskip\noindent\textbf{Challenge 4: Interpretability and Traceability.} For research-oriented data management, correctness alone is not enough -- researchers need to understand and trust how an answer was produced. 
The challenge lies in exposing interpretable reasoning chains that make internal decisions explicit, including applied analytical steps, intermediate results, and evidence propagation.
This requires reasoning processes to be materialized and inspectable, akin to query execution plans in DBMSs.
As illustrated in Figure~\ref{fig:example}, queries may include reasoning traces -- for instance, Q1 shows its execution plan, while Q4 cites supporting tables and figures.

\smallskip\noindent\textbf{Our Contributions -- A Holistic Architecture for Agentic Scholarly Data Management}.~
To address the above challenges collectively, we introduce {\sysname}, 
which embodies a holistic system architecture (Figure \ref{fig:sys-ov}) to unify knowledge representation, LLM-centric hybrid planning, and unified execution to support diverse scholarly queries.

\smallskip
\noindent$\bullet$~\textbf{Knowledge Representation Layer}
(Section~\ref{sec:storage}).
To address \emph{Challenge 1}, we introduce a \emph{\textbf{taxonomy-anchored knowledge graph}}
that unifies fine-grained document structure with an explicit conceptual organization of the research landscape.
Unlike existing academic knowledge graphs~\cite{WangSHWDK20,abs-2205-01833,WanZZT19} that primarily organize papers via bibliographic relationships,
our representation models papers together with their internal components
(e.g., sections, tables, figures, and experimental metadata), enabling fine-grained evidence access during query execution.
Crucially, the graph is anchored by \emph{\textbf{hierarchical taxonomies}} of research problems and methods,
which provide shared semantic abstractions and serve as the contextual backbone for topic-level analysis,
such as research trend analysis and research idea exploration.
To support this capability, we develop a novel algorithm to \emph{\textbf{automatically construct}} these taxonomies from the corpus
and \emph{\textbf{progressively update and refine}} them as new research outputs emerge.

\smallskip
\noindent$\bullet$ \textbf{LLM-Centric Hybrid Query Planning Layer} 
(Section~\ref{sec:planner}).
To tackle \emph{Challenge 2}, this layer places an \emph{\textbf{LLM at the center of query planning}} to interpret diverse user intents.
Unlike traditional query optimizers in DBMSs, the planner integrates
\emph{\textbf{LLM-driven semantic reasoning}} to interpret ambiguous natural language queries, decompose analytical intents, and reason about operator composition beyond relational algebra.
To reduce planning cost while ensuring reliable performance, the planner adopts a
\emph{\textbf{hybrid planning framework}} that combines reusable predefined plans
for common scholarly queries with dynamic plan generation for queries outside
this library.
\emph{\textbf{Explicit plan validation and self-correction}} are further incorporated to ensure plan executability.

\smallskip\noindent$\bullet$ \textbf{Unified Execution Layer} (Section~\ref{sec:execution_engine}). To address \emph{Challenge 3}, we introduce the Unified Execution Layer, which serves as the operational backbone of {\sysname}. First, we analyze typical scholarly queries in Figure~\ref{fig:query_set} and design a set of \emph{\textbf{general, reusable operators}} that can be composed to support a wide range of analytical goals (see Section~\ref{sec:designed_operators}). 
Second, we design a \emph{\textbf{parallel execution engine}} with \emph{\textbf{result caching}} to accelerate query processing by reusing (intermediate) results across queries (see Section~\ref{sec:executor}). The architecture remains extensible, allowing new operators to be seamlessly integrated as emerging query types arise. To ensure \emph{\textbf{interpretability and traceability}} (Challenge 4), the execution engine materializes the full query plan and data lineage for each result, and exposes them to users for inspection. Moreover, users can control the level of detail in the generated outputs through internal operator parameters -- for example, requesting a concise summary, a detailed explanation, or a fully evidence-backed reasoning trace. 

\smallskip\noindent$\bullet$ \textbf{System Implementation and Evaluation} (Section~\ref{sec:exp}). Together, these layers form a unified framework that addresses the four challenges outlined above and supports diverse, open-ended scholarly queries. We implement {\sysname} and conduct comprehensive experiments to evaluate its capabilities across retrieval, information extraction and synthesis, and knowledge discovery and generation queries. Furthermore, we present an end-to-end case study simulating realistic, open-ended scholarly exploration and perform ablation studies to quantify the impact of key architectural choices.
\section{Typical Queries and Related Work}\label{sec:preliminaies}

\subsection{Typical Scholarly Queries}
\label{sec:typical_query}

\begin{figure}[t]
\setlength{\abovecaptionskip}{0cm}
\setlength{\belowcaptionskip}{0cm}
\centering
\includegraphics[width=0.8\textwidth]{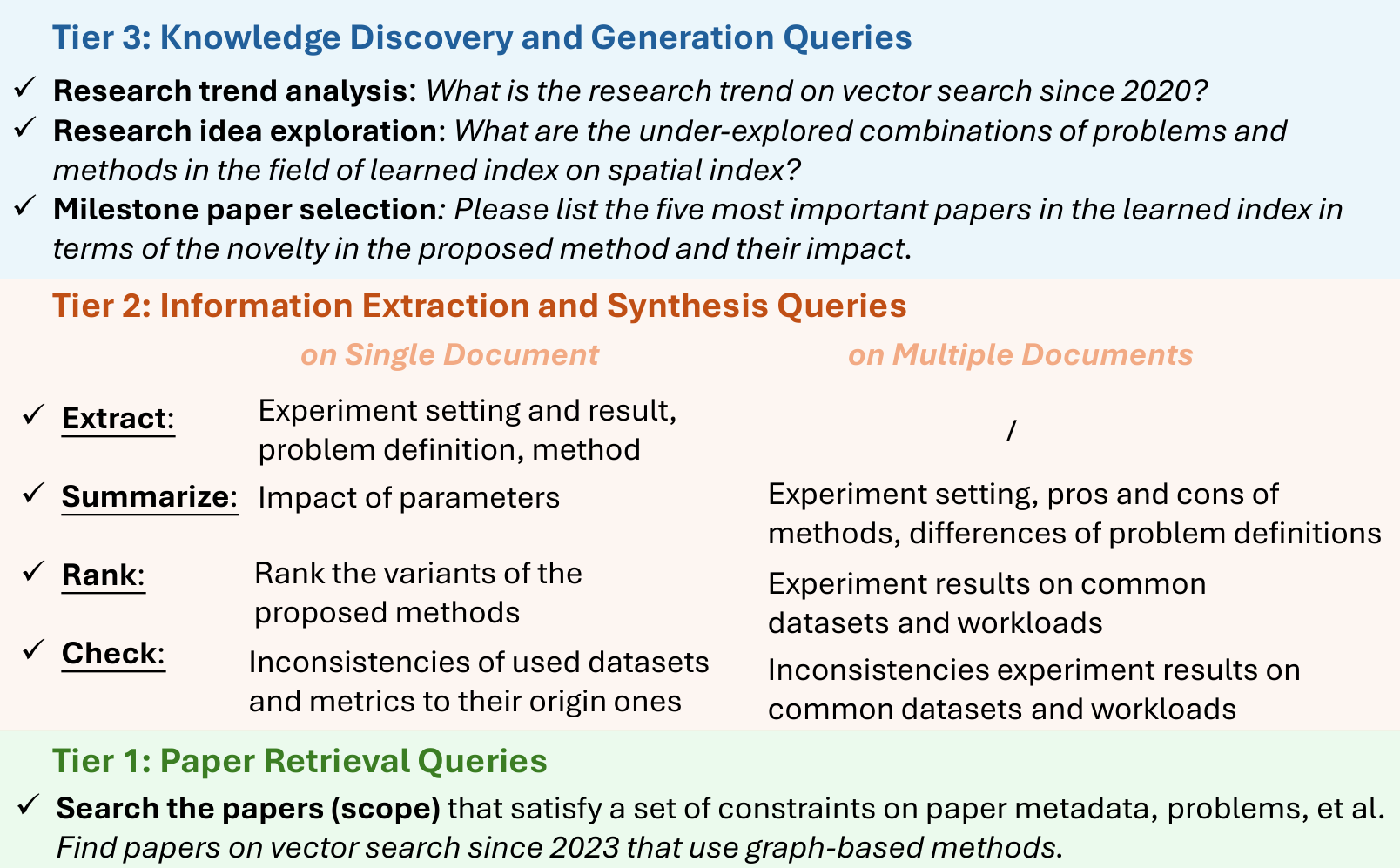}
\caption{Representative scholarly queries across three tiers.}
\label{fig:query_set}
\vspace{-0.2em}
\end{figure}

Figure~\ref{fig:query_set} shows representative scholarly queries in {\sysname}. 

\noindent\textbf{Tier 1: Paper Retrieval Queries.} This foundational tier focuses on locating papers matching specific criteria, serving as the basis for subsequent analytical tasks.

\noindent\textbf{Tier 2: Information Extraction and Synthesis Queries.} This tier involves fine-grained content analysis, covering both \textit{single-document extraction} to extract details of a paper and \textit{multi-document synthesis} to synthesize knowledge across papers.

\noindent\textbf{Tier 3: Knowledge Discovery and Generation Queries.} This advanced tier extends beyond retrieval or summarization to generate novel insights and actionable guidance.

\vspace{-1em}
\subsection{Related Systems}
\label{sec:related_system}

\setlength{\textfloatsep}{0pt}
\begin{table*}[t]
\setlength{\tabcolsep}{3pt}
\setlength{\abovecaptionskip}{0cm}
\setlength{\belowcaptionskip}{0cm}
\footnotesize
\centering
\begin{threeparttable}
\caption{Comparisons on related systems (\full~supported; \half~partially supported; \none~not supported).}
\label{tab:system}
\begin{tabular}{ccccccc}
\toprule
\multirow{3}{*}{System}
  & \multicolumn{1}{c}{Tier 1 Queries}
  & \multicolumn{2}{c}{Tier 2 Queries}
  & \multicolumn{3}{c}{Tier 3 Queries} \\
\cmidrule(lr){2-2} \cmidrule(lr){3-4} \cmidrule(lr){5-7}
& \makecell{Retrieval}
& \makecell{Single doc.}
& \makecell{Multiple doc.}
& \makecell{Trend  Analysis}
& \makecell{Idea Exploration}&\makecell{Milestone Selection} \\
\midrule
%PASA&$\blacktriangle$&--&--&--&--&--&--\\
Elicit~\cite{elicit}&\full&\full&\full&\full&\full&\full\\
Ai2 Asta$^1$~\cite{Asta}&\full&\none&\none&\none&\none&\none\\
ChatPDF~\cite{ChatPDF}&\none&\full&\full&\none&\none&\none\\
% OpenAI API~\cite{GPT}&\half&\full&\full&\full&\full&\full\\
% Gemini API~\cite{Gemini}&\half&\full&\full&\full&\full&\full\\
RAG~\cite{abs-2312-10997}&\none&\full&\full&\none&\none&\none\\
Gemini Deep Research~\cite{deep_research}&\half&\full&\full&\full&\full&\full\\
Tongyi Deep Research~\cite{tongyi}&\half&\full&\full&\full&\full&\full\\
AgenticData~\cite{abs-2508-05002}&\none&\full&\full&\none&\none&\none\\
\textbf{\sysname}&\full&\full&\full&\full&\full&\full\\
\bottomrule
\end{tabular}
\begin{tablenotes}[flushleft] 
\footnotesize
\item $^1$Ai2 Asta claims to support literature summarization but currently fails on complex Tier-2 and Tier-3 queries based on our test; hence marked as not supported.
\end{tablenotes}
\end{threeparttable}
% \vspace{-1.4em}
\end{table*}

Several systems support scholarly queries to a certain extent, as shown in Table~\ref{tab:system}. We categorize these systems into four classes according to their design focus and depth of reasoning.

\smallskip
\noindent\textbf{Scholarly-Centric Systems}. This class of systems~\cite{elicit,Asta} supports large-scale paper retrieval and literature summarization. Among them, Elicit~\cite{elicit} covers all tiers of scholarly queries but offers only limited query refinement -- restricted to improving existing summaries -- and often incurs longer response times. Other systems are designed for a single query tier. For example, ChatPDF~\cite{ChatPDF} allows users to upload papers for information extraction and synthesis, whereas PASA~\cite{he2025acl} targets paper retrieval by leveraging textual and citation-based signals to locate studies.

\smallskip
\noindent\textbf{LLM-based Tools}. Retrieval-Augmented Generation (RAG)~\cite{LewisPPPKGKLYR020, abs-2312-10997, guo2024lightrag, edge2024graphrag, luo2025hypergraphrag, WangZPG0025, JiangXGSLDYCN23} enhances LLMs by grounding their outputs in retrieved sources. This approach improves reliability in single-document extraction and multi-document synthesis, providing a scalable alternative to direct LLM APIs~\cite{GPT,Gemini,Mistral,Llama,Qwen}, which are constrained by token limits and high computational costs. However, most existing RAG implementations retrieve only a limited number of content chunks, resulting in poor source coverage and weak contextual integration. Consequently, they are less effective for open-ended queries that require progressive exploration and deep reasoning across corpora.

\smallskip
\noindent\textbf{Agentic Deep Research Systems.} Representative applications, such as Gemini Deep Research~\cite{deep_research} and Tongyi Deep Research~\cite{tongyi}, primarily target Tier-3 queries. Unlike standard LLM APIs, they autonomously plan multi-step reasoning -- decomposing complex questions, extracting real-time information, and synthesizing structured, citation-grounded reports. This enables higher-level capabilities like identifying research trends, milestone papers, and promising ideas. However, each presents trade-offs: Gemini Deep Research produces more comprehensive results but with longer response times, reducing interactivity, whereas Tongyi Deep Research may respond faster but is less effective for nuanced higher-tier queries.

\smallskip
\noindent\textbf{Semi-Structured Document Analytics Systems.} 
Recent systems~\cite{abs-2508-05002,PatelJPGAGZ25,abs-2405-04674,ShankarCSPW25,ChaiLDZYWC25,abs-2505-14661,WangF25,SunCDJGHYWC25,liu2025palimpzest} extend LLM-based reasoning to document analytics but mainly support Tier-2 queries. They struggle with Tier-3 queries, which are open-ended and involve evolving user intents, due to their reliance on rigid queryable forms such as manually written programs~\cite{PatelJPGAGZ25,ShankarCSPW25,liu2025palimpzest} or SQL-like queries~\cite{abs-2405-04674,SunCDJGHYWC25}. While systems such as AgenticData~\cite{abs-2508-05002} and Unify~\cite{WangF25} introduce natural language interfaces, their data models remain insufficient for multi-modal and context-aware representation, and their limited operator spaces restrict their ability to perform complex semantic reasoning, such as research idea exploration.

\section{Knowledge Representation Layer: From Multi-Modal Structure to Hierarchical Research Taxonomies}
\label{sec:storage}

\begin{figure}
\setlength{\abovecaptionskip}{0cm}
\setlength{\belowcaptionskip}{0cm}
    \centering
\includegraphics[width=0.45\linewidth]{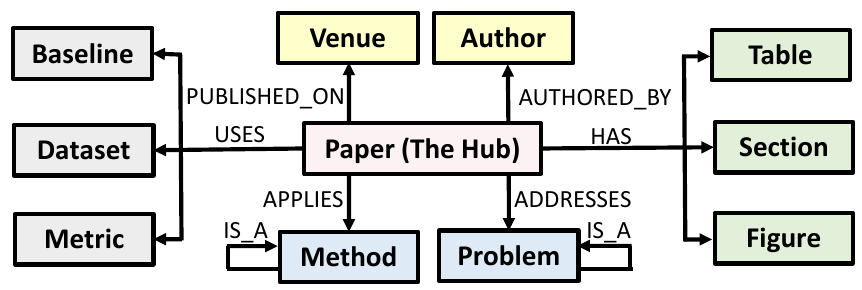}
\caption{The schema of the knowledge graph.}
\label{fig:schema}
\end{figure}

Supporting the diverse queries shown in Figure~\ref{fig:query_set}, especially Tier-3 queries that require reasoning over research topics, demands a storage representation that goes beyond document-centric storage. Such a representation must both expose the fine-grained, multi-modal structure of individual papers and anchor them within a shared, explicit conceptual organization of the research landscape.

Thus, the representation of a scholarly corpus must satisfy three key requirements: 
(1) provide fine-grained access to multi-modal content such as sections, tables, and figures; 
(2) maintain explicit links among related elements -- including datasets, metrics, baselines, and citations -- to capture experimental and bibliographic relationships; and 
(3) organize papers in structured hierarchies of problems and methods, enabling reasoning over research topics rather than isolated documents.

\smallskip
\noindent\textbf{Taxonomy-Anchored Knowledge Graph.}
To this end, \sysname{} introduces a \emph{Knowledge Representation Layer} that transforms the scholarly corpus into a structured knowledge graph (KG).
While prior academic KGs such as MAG~\cite{WangSHWDK20}, OpenAlex~\cite{abs-2205-01833}, and AMiner~\cite{WanZZT19} primarily model bibliographic relationships among papers, authors, venues, and citations, such document-centric representations are insufficient for supporting topic-level and compositional reasoning.
As a result, our KG is anchored by \emph{Taxonomy Nodes} -- specifically \texttt{Problem} and \texttt{Method} nodes.
These nodes define shared abstractions of research problems and algorithmic strategies, serving as first-class semantic primitives for reasoning.
For instance, when addressing Q5 in Figure~\ref{fig:example}, the system can directly infer unexplored problem--method combinations by traversing the problem and method taxonomies within the vector search domain, rather than aggregating evidence across individual papers, thereby enabling more efficient and scalable reasoning.

As illustrated in Figure~\ref{fig:schema}, our KG is structured \emph{around these taxonomy nodes} to capture the multi-modal structure of scholarly documents. Each \texttt{Paper} node acts as a central hub that links to Content Nodes (including \texttt{Section}, \texttt{Figure}, and \texttt{Table}), Experimental Context Nodes (including \texttt{Dataset}, \texttt{Metric}, and \texttt{Baseline}), and Bibliographic Nodes (including \texttt{Author} and \texttt{Venue}).
This joint modeling of taxonomy-level abstractions and document-level evidence enables \sysname{} to support complex Tier-3 queries that require topic-level reasoning grounded in document content.

As the KG instantiation follows engineering practices, we provide implementation details in 
% Appendix~\ref{app:construct_KB};
Appendix~B.1;
the following section presents our novel algorithm for taxonomy construction.

\vspace{-0.8em}
\subsection{Reference-enhanced Taxonomy Construction}
\label{sec:taxo}

Constructing high-quality taxonomies for a scholarly corpus remains a significant challenge. While various approaches exist, they face several limitations: (1) Corpus-driven methods~\cite{Chen2019CNProbaseAD,Huang2020CoRelST,kargupta2025taxoadapt,Shang2020TaxonomyCO,Shen2018HiExpanTT,Zhang2018TaxoGenUT,zhu2025context} build hierarchies directly from papers. While effective at capturing domain-specific nuances, they often suffer from data sparsity, yielding fragmented or overly narrow structures~\cite{lu2024tacl}. (2) LLM-driven methods~\cite{sun2024large,Chen2023PromptingOF,Zeng2024ChainofLayerIP,kargupta2025taxoadapt} leverage an LLM's internal knowledge to ensure broader conceptual coverage. However, they frequently fail to capture emerging topics that appear in the corpus but are absent from the model's pre-training data. (3) Crucially, both paradigms typically lack mechanisms for progressive refinement and incorporating newly papers may require reconstructing the entire taxonomy from scratch -- a prohibitive cost for rapidly evolving research fields.

\smallskip
\noindent\textbf{Our Novel Solution}. To overcome these limitations, we propose a \emph{Reference-enhanced Taxonomy Construction Framework}. The framework synergizes an LLM-derived scaffold with corpus-grounded evidence, ensuring that the resulting taxonomy is structurally comprehensive and faithful to the target corpus. Crucially, it supports \emph{progressive refinement}, allowing the taxonomy to evolve dynamically as new papers arrive, without full reconstruction. As outlined in Algorithm~\ref{alg:build}, the taxonomy construction proceeds in four stages:

\noindent$\bullet$ \underline{\emph{Structured Information Extraction}}: Extract key aspects that characterize paper’s problem or method.

\noindent$\bullet$ \underline{\emph{Cross-Paper Standardization}}: Synthesize extracted aspects across papers into standardized concepts by identifying recurring concepts and unifying synonyms.

\noindent$\bullet$ \underline{\emph{Reference Taxonomy Alignment}}: Align the extracted aspects of each paper and the standardized concepts with an LLM-generated reference taxonomy, constructed using topic labels from the corpus, to produce a taxonomy that integrates corpus-grounded evidence with domain knowledge.

\noindent$\bullet$ \underline{\emph{Progressive Update and Branch Refinement}}: 
Incorporating newly arriving papers by progressively refining the taxonomy structure through three operations: assigning them to existing nodes, creating new nodes, or introducing new branches.

\begin{figure}[t]
\setlength{\abovecaptionskip}{0cm}
\setlength{\belowcaptionskip}{0cm}
\centering
\includegraphics[width=0.7\textwidth] {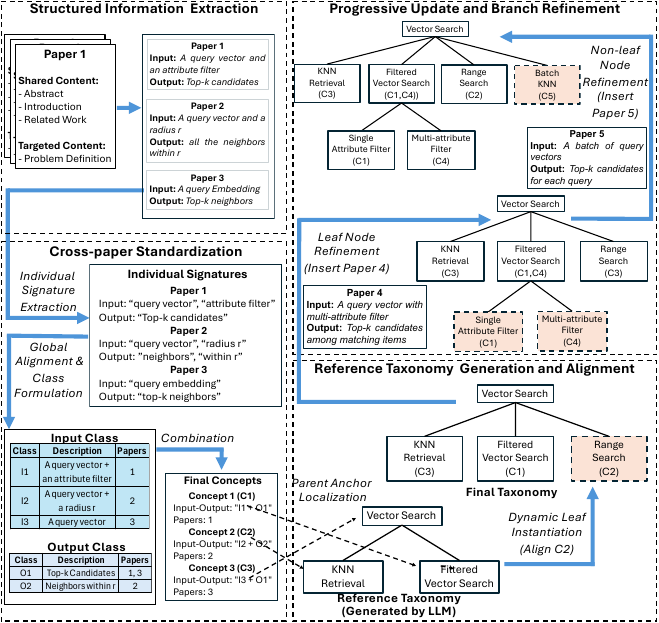}
\caption{Reference-enhanced taxonomy construction.}
\label{fig:taxonomy construction} 
\end{figure}

In the following, we describe each stage, referencing Algorithm~\ref{alg:build} and the illustrative example in Figure~\ref{fig:taxonomy construction}.
We use the construction of the problem taxonomy as a representative case, as the construction processes for problem and method taxonomies are analogous. Implementation details are provided in 
% Appendix~\ref{app:code_taxo}.
Appendix~B.2.

\setlength{\textfloatsep}{0pt}
\begin{algorithm}[t]
\footnotesize
\renewcommand{\algorithmicrequire}{\textbf{Input:}}
\renewcommand\algorithmicensure{\textbf{Output:}}
\caption{\textsc{TaxonomyConstruction}}
\label{alg:build}
\begin{algorithmic}[1]
\Require scholarly corpus $D$, optional new papers $D_{\mathrm{new}}$, refinement factor $\alpha$, match threshold $\tau_{\text{match}}$;
\Ensure taxonomy $T$;
\Statex \textbf{// Stage 1: Structured Information Extraction}
\State $\mathcal{A} \gets \{\}$; \Comment{map: paper $\mapsto$ problem/method aspects}
\State \textbf{for} {$p \in D$} \textbf{do} $\mathcal{A}[p] \gets \textsc{ExtractAspect}(p)$;
\Statex \textbf{// Stage 2: Cross-paper Standardization}
\State $\mathcal{C} \gets \textsc{CrossPaperStandardization}(\mathcal{A})$;
\Statex \textbf{// Stage 3: Reference Taxonomy Generation and Alignment}
\State $L \gets \textsc{InferTopicLabel}(D)$, $T_{\mathrm{ref}} \gets \textsc{GenerateReferenceTaxonomy}(L)$;
\State $T \gets \textsc{CopyTree}(T_{\text{ref}})$;
\For{$(C, D_C) \in \mathcal{C}$}
\State $T.u \gets \textsc{TopDownLocateParent}(T.\text{root}, C)$;
\State $T.v \gets \textsc{FindMatchingChild}(T.u, C, \tau_{\text{match}})$;
\State \textbf{if} {$T.v == NULL$} \textbf{then} $T.v \gets \textsc{AddChild}(T.u, C)$;
\State Assign papers $D_C$ to $T.v$;
\EndFor
\Statex \textbf{// Stage 4: Progressive Update and Branch Refinement }
\If{$D_{\mathrm{new}} \neq \emptyset$}
\State $\mathcal{A}_{\mathrm{new}} \gets \{\}$;
\For {$p \in D_{\mathrm{new}}$}
\State $A \gets \textsc{ExtractAspect}(p)$, $C \gets \textsc{MapToConceptsOrCreate}(A)$;
\State $T.u \gets \textsc{TopDownLocateParent}(T.\text{root}, C)$;
\State $T.v \gets \textsc{FindMatchingChild}(T.u, C,\tau_{\text{match}})$
\State \textbf{if} {$T.v == NULL$} \textbf{then} $T.v \gets \textsc{AddChild}(T.u, C)$;
\State Assign paper $p$ to $T.v$;
\State \textbf{if} $\textsc{NewPaperCount}(T.v) \ge \alpha$ \textbf{then} \textsc{RefineBranch}$(T.v)$
\EndFor
\EndIf
\State \Return $T$
\end{algorithmic}
\end{algorithm}

\smallskip
\noindent\textbf{Stage 1: Structured Information Extraction}. The first stage extracts essential aspects from each paper in the scholarly corpus to facilitate alignment (Line 2 in Algorithm~\ref{alg:build}). We adopt a context-aware extraction strategy that combines \emph{Shared Content} (e.g., Abstract, Introduction, and Related Work), which orients the LLM to the paper’s overall scope, with \emph{Targeted Content} (e.g., the Problem Definition for the problem taxonomy), enabling precise extraction of problem- or method-specific information. The extracted aspects differ by taxonomy type:

\noindent$\bullet$ For problem taxonomy, the aspects consist of formal specifications of the \emph{input} and \emph{output}.
As illustrated in Figure~\ref{fig:taxonomy construction}, this is achieved by decomposing each problem into concrete data constraints (e.g., ``Input: a query vector and an attribute filter'' in Paper 1) and objectives (e.g., ``Output: top-$k$ candidates'' in Paper 1). Such formalization enables fine-grained distinctions between closely related problems, e.g., differentiating KNN retrieval (Paper 2) from range search (Paper 3).

\noindent$\bullet$ For method taxonomy, the aspects include
\emph{key techniques} (e.g., core components, ordered workflows), and, crucially, the \emph{strengths and weaknesses}. While technical descriptions capture how a method operates, strengths and weaknesses explain why it was designed, enabling methods to be grouped not only by mechanics but also by shared optimization goals (e.g., trading accuracy for speed). This reveals implicit methodological families that are often missed by technical descriptions.

\smallskip
\noindent\textbf{Stage 2: Cross-Paper Standardization}. While the previous stage yields isolated aspects, this stage synthesizes them into coherent, corpus-level concepts (Line 3 in Algorithm~\ref{alg:build}). We employ a three-step process to standardize extracted aspects -- unifying diverse phrasings into canonical classes, synthesizing these classes into composite concepts, and assigning each paper to the concept, which serves as corpus-ground evidence for the next stage.

\noindent\emph{\underline{Step 1: Individual Signature Extraction.}}
We extract key noun phrases using an LLM, which serve as semantic signatures for the target aspect. As shown in Figure~\ref{fig:taxonomy construction}, when the input is the target aspect, Paper 1 yields signatures ``query vector'' and ``attribute filter''. These signatures capture essential semantics required for alignment.

\noindent\emph{\underline{Step 2: Global Alignment and Class Formation.}} We prompt an LLM to identify semantically equivalent signatures and group them under a canonical class. As shown in Figure~\ref{fig:taxonomy construction}, the distinct signatures ``top-k candidates'' (Paper 1) and ``top-k neighbors'' (Paper 3) are identified as synonyms and consolidated into Output Class $O_1$.

\noindent\emph{\underline{Step 3: Assignment and Concept Formulation.}} 
Canonical classes derived from aspects are combined to form \emph{concepts}, which represent abstractions of problem families
(e.g., pairing Input Class $I_1$ with Output Class $O_1$), and each paper is mapped to the most appropriate concept. As shown in Figure~\ref{fig:taxonomy construction}, Paper 1 is assigned to Concept $C_1$ (representing $I_1+O_1$), while Paper 2 is assigned to Concept $C_2$. Papers that do not match existing concepts form new subgroups that are recursively re-processed through \emph{Step 1} and \emph{Step 2}, ensuring comprehensive coverage.

\smallskip
\noindent\textbf{Stage 3: Reference Taxonomy Generation and Alignment}.
After standardizing corpus-grounded concepts, this stage integrates an LLM-derived reference taxonomy with these concepts to construct the final taxonomy, which forms the core of our framework. This integration are in three steps.

\smallskip
\noindent\underline{\emph{Step 1. Reference Taxonomy Generation.}} 
We first instruct an LLM to infer a topic label from the corpus and generate a skeletal reference taxonomy (Line 4 in Algorithm~\ref{alg:build}). In Figure~\ref{fig:taxonomy construction}, the LLM identifies topic ``Vector Search'' and generates child nodes ``KNN Retrieval'' and ``Filtered Vector Search'' based on general domain knowledge.

\smallskip
\noindent\underline{\emph{Step 2. Node Localization.}} 
We then map each standardized concept to the reference taxonomy through two sub-steps.

\noindent(1) \emph{Parent Anchor Localization:} 
We traverse the reference taxonomy in a top-down manner to identify the most specific parent node that subsumes the concept (Line 7 in Algorithm~\ref{alg:build}). In Figure~\ref{fig:taxonomy construction}, for Concept $C_2$, the system identifies the root node ``Vector Search'' as the parent anchor, since $C_2$ does not semantically fit under the existing ``KNN Retrieval'' or ``Filtered Vector Search'' nodes.

\noindent(2) \emph{Dynamic Leaf Instantiation:} 
If the concept does not match any existing child of the parent anchor, we instantiate a new node (Lines 8-9 in Algorithm~\ref{alg:build}). In Figure~\ref{fig:taxonomy construction}, because a specific ``Range Search'' node is absent from the reference taxonomy, the system dynamically creates a new node for $C_2$, prompts the LLM to generate a canonical name ``Range Search'', and attaches it to the taxonomy.

\smallskip
\noindent\underline{\emph{Step 3. Paper Assignment.}} 
Finally, all papers associated with the concept are linked to the corresponding taxonomy node (Line 10 in Algorithm~\ref{alg:build}). For example, Paper 2 is assigned to ``Range Search'' node, while Paper 1 is assigned to ``Filtered Vector Search'', ensuring that every document is reachable through the taxonomy hierarchy.

\smallskip
\noindent\textbf{Stage 4: Progressive Update and Branch Refinement}.
To accommodate the dynamic nature of research, our framework supports progressive update, allowing the taxonomy to evolve as papers arrive \emph{without requiring reconstruction from scratch} (Lines 11-20 in Algorithm~\ref{alg:build}). Specifically, new papers are first processed by extracting aspects following Stage 1 and mapping them to existing concepts or creating new concepts (Line 14 in Algorithm~\ref{alg:build}). They are then assigned to the most specific existing taxonomy node via Step 2 and Step 3 of Stage 3 (Lines 15-18 in Algorithm~\ref{alg:build}). To prevent taxonomy nodes from becoming overly broad due to concept drift, we further employ a branch refinement mechanism.

\smallskip
\noindent\underline{\emph{Step 1. Refinement Trigger.}} 
As shown in Line 19 of Algorithm~\ref{alg:build}, when the number of new papers linked to a node exceeds a predefined threshold $\alpha$, the node is flagged for refinement.

\smallskip
\noindent\underline{\emph{Step 2. Refinement Execution.}} 
The refinement strategy adapts based on the node’s position in the taxonomy and leverages corpus-grounded clustering to discover fine-grained structures that are often absent from the LLM’s pre-training data.

\noindent\emph{Case 1: Leaf Node Refinement.} 
This case captures the emergence of granular subtopics. As illustrated in Figure~\ref{fig:taxonomy construction}, the leaf node ``Filtered Vector Search'' accumulates new papers, including Paper 4, which introduces multi-attribute filtering. The system performs semantic clustering on these papers and splits the node into a more detailed subtree consisting of ``Single Attribute Filter'' ($C_1$) and ``Multi-attribute Filter'' ($C_4$).

\noindent\emph{Case 2: Non-leaf Node Refinement.} 
This case handles papers that fit a high-level category but do not align with any child nodes.
In Figure~\ref{fig:taxonomy construction}, Paper 5 maps to the root ``Vector Search'' but fits none of its children. The refinement process identifies this distinct cluster and instantiates a new sibling branch, ``Batch KNN'' ($C_5$), extending the taxonomy breadth without disrupting existing branches.

\section{LLM-Centric Hybrid Planning Layer: From Natural-Language Queries to Executable Plans}\label{sec:planner}

The Query Planning layer translates high-level natural-language queries into executable Directed Acyclic Graphs (DAGs) of operators. While this objective shares the high-level spirit of classical database query optimization~\cite{LeisGMBK015,SelingerACLP79}, our system operates in a fundamentally different setting.% beyond relational DBMSs.

\smallskip 
\noindent\textbf{Unique Challenges.}  
Traditional query optimizers operate over algebraic SQL queries with well-defined schemas and formally specified operator semantics. In contrast, our planner must reason over open-ended natural language scholarly queries, which introduces challenges absent in DBMS planning: (1) \emph{Lack of algebraic semantics:} Natural language queries do not admit a fixed algebraic representation, rendering rule-based transformations and cost-based enumeration ineffective. (2) \emph{Open-ended analytical intents:} Scholarly queries are often open-ended and may evolve into diverse analytical intents (Figure~\ref{fig:example}), leading to a vast plan space that extends well beyond relational algebra. (3) \emph{Verification complexity:} Unlike deterministic SQL compilation, generative planning requires explicit mechanisms to validate plan correctness and executability before execution.

\smallskip
\noindent\textbf{Our Planner}. To address these challenges, we propose an \emph{LLM-centric hybrid query planner} that places LLMs at the core of semantic reasoning while carefully constraining their role to ensure efficiency and reliability. Our planner is built on three key principles: 

\noindent$\bullet$ \underline{\emph{LLM-driven Semantic Reasoning.}} We leverage LLMs to interpret ambiguous natural language queries, decompose analytical intents, and reason about operator composition.

\noindent$\bullet$ \underline{\emph{Hybrid Planning for Efficiency and Generality.}} To reduce planning cost while ensuring reliable performance, we combine a library of predefined plans for common scholarly queries with dynamic plan generation for queries outside this library.

\noindent$\bullet$ \underline{\emph{Robustness through Validation and Self-correction.}} We introduce explicit plan validation and an iterative self-correction loop to mitigate the non-determinism and errors of LLM-generated plans.

Figure~\ref{fig:plan_example} illustrates the end-to-end process by which a natural language query is translated into a validated execution plan; the details are presented in the following subsections.

\subsection{LLM-Driven Query Decomposition}\label{sec:qo-decomp} 

To bridge the gap between ambiguous natural language and modular execution, our system first decomposes the analytical intent into two components: a \texttt{Scope} and a \texttt{Task}. The \texttt{Scope} specifies the entities or conceptual boundary of analysis, while the \texttt{Task} describes the analytical step to be performed over that scope. 

\begin{figure}[t]
\setlength{\abovecaptionskip}{0cm}
\setlength{\belowcaptionskip}{0cm}
\centering
\includegraphics[width=0.8\textwidth]{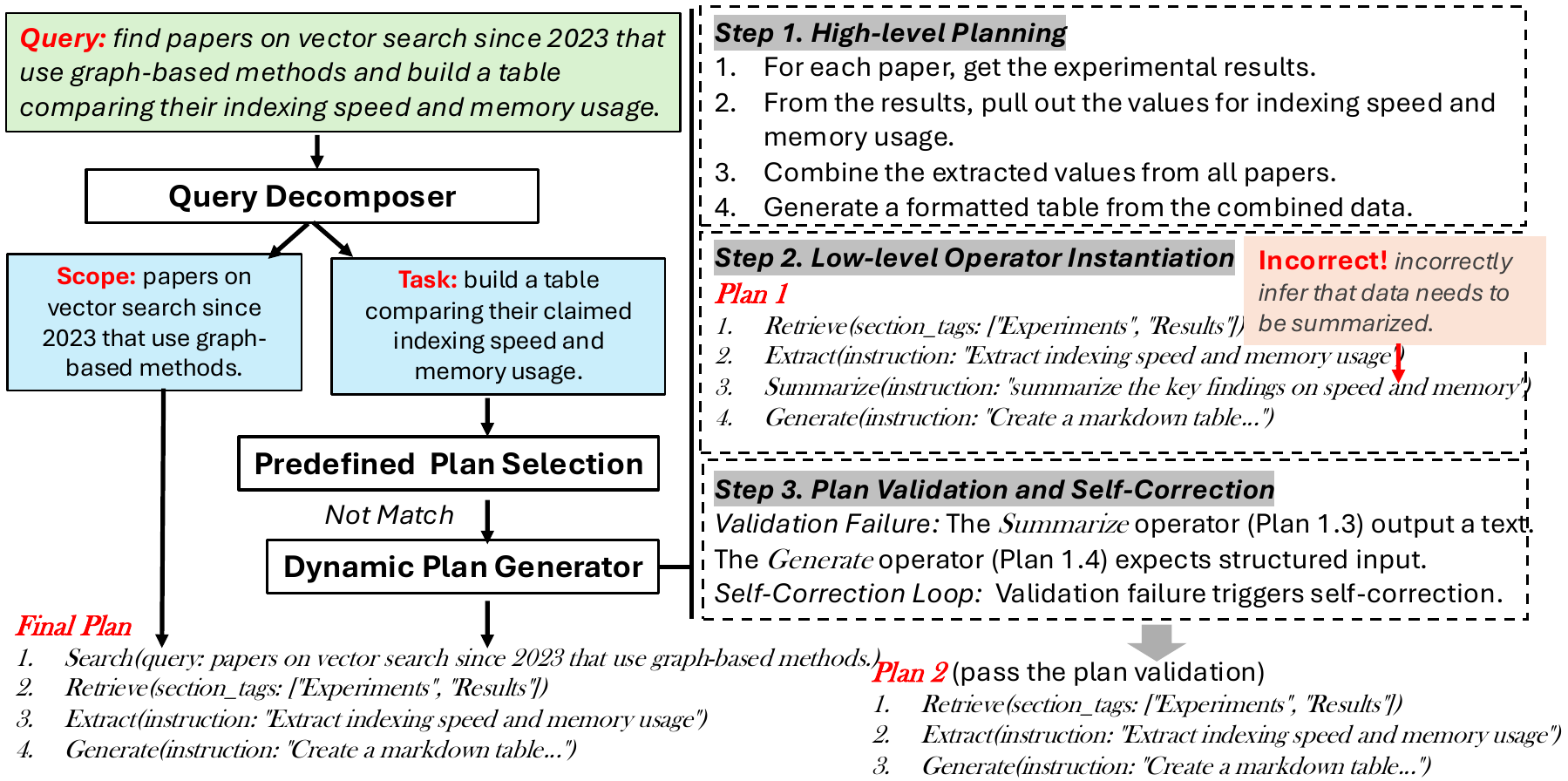}
\caption{From query to validated execution plan.}
\label{fig:plan_example}
\end{figure}

This decomposition is achieved by prompting an LLM with a structured instruction that explicitly separates ``what to analyze'' from ``how to analyze it''. As shown in Figure~\ref{fig:plan_example}, given the query ``Find papers on vector search since 2023 that use graph-based methods and build a table comparing their indexing speed and memory usage'', the LLM produces: (1) \texttt{Scope} -- ``papers on vector search since 2023 that use graph-based methods''; and (2) \texttt{Task} -- ``build a table comparing their indexing speed and memory usage''.

This decomposition yields a clearer and more modular representation of analytical intents, reducing ambiguity and enabling subsequent planning stages to operate more reliably and efficiently.

\subsection{Predefined Plan Selection}\label{sec:qo-predf}

For common and well-understood scholarly queries (Figure~\ref{fig:query_set}), invoking a fully generative planner is inefficient. To achieve low latency and high reliability for such queries, our system first attempts to match the query’s \texttt{Task} component to a plan in a \emph{Predefined Plan Library} 
% (Appendix~\ref{app:plan_libary})
(Appendix~C)
. This library plays a role analogous to cached query plans in traditional DBMSs~\cite{MoCWCB23,DursunBCK17,AlucDB12}; however, despite this resemblance, our design differs in two fundamental ways to accommodate the semantic nature of natural-language queries:

\noindent$\bullet$ \textbf{Natural-language-driven Parameters.} 
In our setting, plan ``parameters'' are not SQL-like constants (e.g., \texttt{WHERE id = ?}) but semantic intents expressed in natural language (e.g., ``Extract indexing speed and memory usage''). Consequently, plan selection requires semantic understanding rather than syntactic matching.

\noindent$\bullet$ \textbf{Reliability-oriented plan selection.}  
As the plan library is extensible and may grow large over time, directly prompting an LLM to select from all available plans is unreliable: long prompts with many candidates often lead to misselection. To ensure robust plan reuse, we adopt a two-stage selection strategy:

\noindent\emph{(1) Candidate Filtering:} A semantic search is first performed to compare the task description against the descriptions of all plans in the library, retrieving a small candidate set of plans.

\noindent\emph{(2) LLM-based Reranking and Validation:} The candidate plans are then evaluated by a reasoning LLM, which assesses their alignment with the analytical intent and assigns a confidence score.

A plan is selected only if its confidence score exceeds a high threshold (e.g., $>90\%$); otherwise, the query is treated as ad-hoc and passed to the dynamic plan generator. For example, the task \emph{``Extract indexing speed and memory usage''} directly matches a predefined extraction plan specialized for experiment sections.

This design enables our system to retain the low latency benefits of plan reuse while preserving the semantic flexibility and robustness required for natural language scholarly queries.

\subsection{Dynamic Plan Generation with Self-correction}
\label{sec:qo-dyn}

For queries that fall outside the predefined library, our system invokes the \emph{Dynamic Plan Generator}. To manage the combinatorial explosion of the plan space and reduce hallucinations, we employ a decomposed generation process: generating a high-level logical plan via few-shot demonstrations, followed by low-level operator instantiation. Crucially, we introduce a \textbf{self-correction loop} to ensure robustness.

\smallskip\noindent\textbf{Step 1. High-Level Planning with 
Demonstrations.}
To narrow the search space, our system first generates a high-level logical plan as an abstract sequence of steps. To guide the LLM, we retrieve representative \texttt{(query, plan)} demonstrations from an extensible library via semantic search and inject them into the LLM context.

\smallskip\noindent\textbf{Step 2. Low-Level Operator Instantiation.} 
Our system then proceeds to the low-level instantiation stage. It iterates through each abstract step of the logical plan and makes a series of focused, less complex calls to the LLM. For each step, the LLM's task is to select the single most appropriate operator from the operator library and fully parameterize it to accomplish that specific sub-task. This enables our system to ask the LLM to solve a much simpler problem at each step rather than a complex problem of generating an entire DAG at once, significantly increasing the success rate.

\smallskip\noindent\textbf{Step 3. Robustness through Plan Validation and Self-Correction.}
Generated plans are never trusted implicitly. Before execution, every plan is passed to a \emph{Plan Validator} that performs fast, deterministic checks for DAG validity, operator correctness. If a plan fails validation, 
or if a syntactically valid plan fails at runtime,
our system triggers a \emph{self-correction loop}. It is re-invoked with an augmented context that includes the failed plan and the specific error message. The prompt is shifted from ``planning'' to ``debugging'', instructing the LLM to analyze the error and propose a corrected plan. 
As shown in Figure~\ref{fig:plan_example}, the \texttt{Summarize} operator is ineffective because the subsequent \texttt{Generate} operator requires structured numeric evidence to construct the table. However, \texttt{Summarize} produces only high-level textual descriptions, which likely omit the detailed quantitative information needed for table generation.

\smallskip
\noindent\textbf{Final Plan Composition}. The final stage assembles the complete execution plan by composing the plan for the \texttt{Scope} component with the plan for the \texttt{Task} component. Our system chains these two parts together, ensuring the output of the initial operator (e.g., \texttt{Search} or \texttt{FindNode} from the scope analysis) is correctly fed as the input to the subsequent task plan (either the selected predefined plan or the dynamically generated one). The result is an DAG passed to the execution engine.
\section{Unified Execution Layer: Operator Orchestration and  Execution}\label{sec:execution_engine}

As established in Section~\ref{sec:typical_query}, scholarly queries span a wide spectrum from simple document retrieval to advanced knowledge discovery and generation. A system designed to support this diversity, along with future analytical needs, requires a \emph{unified and extensible} execution framework rather than a collection of hard-coded, query-specific pipelines. To achieve this, we first analyze the typical query patterns and our underlying knowledge representation to design a library of composable operators (Section~\ref{sec:designed_operators}). 
Following established principles from traditional database systems~\cite{mysql,postgresql,0001ZSYHJLWL21}, these operators serve as the building blocks for execution plans, which are structured as Directed Acyclic Graphs (DAGs). We build an execution engine that orchestrates these DAG-based plans to deliver accurate and interpretable results (Section~\ref{sec:executor}).

\setlength{\textfloatsep}{0pt}
\begin{table*}[t]
\setlength{\tabcolsep}{3pt}
\setlength{\abovecaptionskip}{0cm}
\setlength{\belowcaptionskip}{0cm}
\footnotesize
\centering
\begin{threeparttable}
\caption{Summary of Operator Library.}
\label{tab:operator}
\begin{tabular}{ccc}
\toprule
Category& Operator&Description\\
\midrule
\multirow{1}{*}{Knowledge Access}&\texttt{Search}&Locate documents from the knowledge graph\\
\hline
\multirow{3}{*}{Knowledge Navigation}&\texttt{FindNode}&Locate specific entry-point nodes within taxonomies\\
&\texttt{Traverse}&Traverse the knowledge graph from a set of starting nodes\\
&\texttt{Retrieve}&Fetch the raw content of specific sections from a document\\
\hline
\multirow{2}{*}{Knowledge Generation}&\texttt{MatrixConstruct}&Create a structured matrix summarizing problems and methods\\
&\texttt{Generate}&Generate new content based on inputs and specific instructions\\
\hline
\multirow{5}{*}{Semantic Content Processing}&\texttt{Extract}&Extract specific, structured information
from raw text content\\
&\texttt{Summarize}&Summarise textual information into a concise, coherent overview\\
&\texttt{Check}&Compare information across multiple peer documents\\
&\texttt{Verify}&Verify a specific claim against a designated source of evidence\\
&\texttt{Rank}&Order a set of entities based on a
specific, measurable criterion\\
\hline
\multirow{3}{*}{Relational-Style Data Manipulation}&\texttt{GroupBy}&Partition a list of entities
based on a specified attribute\\
&\texttt{Aggregate}&Compute summary statistics over a set
of entities\\
&\texttt{Filter}&Return a subset of entities that satisfy the condition\\
\bottomrule
\end{tabular}
\end{threeparttable}
% \vspace{-1.4em}
\end{table*}

\subsection{Operator Library}\label{sec:designed_operators}

The operator library in {\sysname} provides building blocks for execution plans and is summarized in Table~\ref{tab:operator}. 
We introduce novel operators tightly coupled with the taxonomy-anchored knowledge graph, enabling topic-level reasoning to support Tier-3 queries beyond retrieval and summarization. Representative operators are described below, with the full details in 
% Appendix~\ref{app:operator_detail}.
Appendix~D.

\smallskip\noindent\textbf{\texttt{Search}.} 
This operator serves as the main entry point for retrieving relevant scholarly documents from the knowledge graph. It supports natural language queries (e.g., Q2 in Figure~\ref{fig:example}) that combine precise constraints with abstract intent via a hybrid retrieval strategy balancing lexical precision and semantic recall.
Given a query, \texttt{Search} uses an LLM to decompose it into structured metadata constraints (e.g., title, authors, venue, year) and semantic intents over research aspects (e.g., topics, problems, methods, datasets). Metadata constraints are handled using BM25-based lexical retrieval~\cite{robertson2009probabilistic}, while semantic intents are matched via dense retrieval over BGE embeddings~\cite{xiao2023cpack} indexed by FAISS~\cite{faiss}. Retrieved candidates are then refined by an LLM-based reranker that jointly reasons over the query, paper metadata, and KG-derived aspect summaries to produce aligned results for downstream operators.

\smallskip\noindent\textbf{\texttt{FindNode}.} This operator identifies an entry-point taxonomy node for topic-level reasoning. It supports direct key-based lookup when an identifier is provided, or semantic matching over precomputed node description embeddings to select the best-matching concept.

\smallskip
\noindent\textbf{\texttt{Traverse}.} This operator enables controlled multi-hop navigation over the knowledge graph from given starting nodes. It uses a typed \textsf{traversal\_path} to constrain relationship and entity types at each hop, preventing unintended expansion. At runtime, \texttt{Traverse} compiles the path into a multi-hop graph query (e.g., Cypher~\cite{FrancisGGLLMPRS18}) and returns an entity list for downstream operators such as \texttt{Aggregate} (e.g., Q1 in Figure~\ref{fig:example}) and \texttt{Rank} (e.g., Q3 in Figure~\ref{fig:example}).

\smallskip\noindent\textbf{\texttt{MatrixConstruct}.} This operator supports research idea exploration (e.g., Q5 in Figure~\ref{fig:example}) by synthesizing a structured view of a research topic to reveal promising directions.
Given a topic, \texttt{MatrixConstruct} uses problem and method taxonomies to identify representative problems and methods, and organizes them into a problem-method matrix. Each cell in the matrix aggregates supporting evidence (e.g., representative papers and summaries). The matrix can be inspected directly or passed to \texttt{Generate} for evidence-grounded synthesis, enabling identification of sparse or missing problem–method combinations.

\smallskip
\noindent\underline{\textit{Remark on Extensibility and Generality.}} The operator library is designed to be both general and extensible: \emph{Generality} is achieved through two mechanisms: (1) a concise set of well-chosen operators and (2) flexible \textsf{instruction} parameters that allow each operator to adapt to a wide range of tasks without altering the system’s core logic.
\emph{Extensibility} arises from the modular architecture, which enables new operators to be added seamlessly as analytical intents evolve. 
Together, these properties ensure that the system can scale and adapt in tandem with the changing research landscape.

\subsection{Unified Query Execution Engine}\label{sec:executor}

The Unified Query Execution Engine realizes DAG-based query plans produced by the planner introduced in Section~\ref{sec:planner}. 
It instantiates operators, resolves their dependencies, and manages intermediate and final results during execution. Its design emphasizes scalable parallel execution, efficient result reuse, and end-to-end traceability to support reliable and interpretable query execution. Algorithm~\ref{alg:executor} sketches the end-to-end execution workflow.

\setlength{\textfloatsep}{0pt}
\begin{algorithm}[t]
\footnotesize
\renewcommand{\algorithmicrequire}{\textbf{Input:}}
\renewcommand\algorithmicensure{\textbf{Output:}}
\caption{\textsc{UnifiedQueryExecution}}
\label{alg:executor}
\begin{algorithmic}[1]
\Require query plan $P$;
\Ensure final result $R$;

\Statex \textbf{Step 1: Scope Execution and Plan Unfolding}
\State $S \leftarrow$ \textsc{ExecuteScopeOperators}$(P)$;
\State $G \leftarrow$ \textsc{UnfoldPlan}$(P, S)$; \Comment{Instantiate operators based on \textsf{execution\_mode}}

\Statex \textbf{Step 2: Dependency Resolution}
\For{operator $o \in G$}
  \State $\textit{dep\_count}[o] \leftarrow$ number of unresolved dependencies of $o$;
  \If{$\textit{dep\_count}[o] = 0$}
     enqueue $o$ into $Q$;
  \EndIf
\EndFor

\Statex \textbf{Step 3: Operator Dispatch and Parallel Execution}
\While{$Q$ is not empty}
  \State $o \leftarrow$ dequeue from $Q$ \Comment{May execute in parallel}
  \State $\textit{inputs} \leftarrow$ \textsc{FetchInputs}$(o, \textit{Buffer}, \textit{Cache}, \textit{Storage})$;
  \If{\textsc{CacheHit}$(o, \textit{inputs})$}
  $\textit{output} \leftarrow$ \textsc{LoadFromCache}$(o, \textit{inputs})$; 
  \Else \text{ }
   $\textit{output} \leftarrow$ \textsc{Execute}$(o, \textit{inputs})$, \textsc{StoreInCache}$(o, \textit{inputs}, \textit{output})$;
  \EndIf
  \State $\textit{Buffer}[o] \leftarrow \textit{output}$; 
  \State \textsc{RecordTrace}$(o, \textit{inputs}, \textit{output})$;

  \For{downstream operator $d$ of $o$}
    \State $\textit{dep\_count}[d] \leftarrow \textit{dep\_count}[d] - 1$;
    \If{$\textit{dep\_count}[d] = 0$}
      enqueue $d$ into $Q$;
    \EndIf
  \EndFor
\EndWhile

\State \Return outputs produced by terminal operators in $G$;
\end{algorithmic}
\end{algorithm}

\subsubsection{Execution Model} There are three key steps:

\noindent\textbf{Step 1. \texttt{Scope} Execution and Plan Unfolding.} The engine instantiates the query plan in two phases. Instead of materializing the entire DAG upfront, it first executes the operators in the query’s \texttt{Scope} (e.g., \texttt{Search}; Line 1 in Algorithm~\ref{alg:executor}) to obtain an initial result set (e.g., $N$ papers). It then enters a plan unfolding stage by inspecting each operator’s \textsf{execution\_mode} (Line 2 in Algorithm~\ref{alg:executor}). Operators marked as \textsf{instance} (e.g., \texttt{Extract}) are expanded into $N$ parallel instances, one per item, while \textsf{group} operators (e.g., \texttt{Summarize}) are instantiated once to aggregate across all results. This process transforms the compact plan into a data-aware execution graph that explicitly exposes parallelism.

\noindent\textbf{Step 2. Dependency Resolution.} Given the unfolded plan, the engine performs a topological analysis to determine execution order, enqueuing operators with no unresolved dependencies as ready for execution (Lines 3--5 in Algorithm~\ref{alg:executor}).

\noindent \textbf{Step 3. Operator Dispatch and Parallelism.} Operators are dispatched once all required inputs are available. Upon completion, the engine propagates outputs, updates dependency counters of downstream operators, and enqueues operators whose dependencies are satisfied (Lines 6--15 in Algorithm~\ref{alg:executor}). Multiple operators may become ready concurrently, enabling parallel execution while preserving dependency correctness.

\subsubsection{Result Management, Reuse, and Traceability} 

During execution, {\sysname} manages intermediate and final results through transient buffering, persistent caching, and execution traceability, enabling efficient reuse of computation while ensuring transparent and interpretable query execution.

\noindent\textbf{Transient Buffering.} Intermediate results are stored in a transient in-memory buffer keyed by operator IDs (Line 11 in Algorithm~\ref{alg:executor}). Upon completion, each operator materializes its output into the buffer, from which downstream operators retrieve inputs by referencing upstream IDs. Operators without dependencies read directly from the storage layer (Line 8 in Algorithm~\ref{alg:executor}), enabling efficient in-run reuse without redundant computation.

\noindent\textbf{Persistent Caching.} Beyond operator-level buffering, the engine maintains an execution cache to persist and reuse results across executions. Unlike transient buffers, the cache stores outputs of plans and subplans for reuse (Line~10 in Algorithm~\ref{alg:executor}). When identical or overlapping plans are reissued, cached results are retrieved instead of recomputed (Lines~8--9 in Algorithm~\ref{alg:executor}). Cached entries are persisted to disk, enabling durability across process lifetimes and accelerating iterative and incremental execution.

\noindent\textbf{Execution Traceability.} 
To support auditing and interpretation, all runtime artifacts are keyed by execution node IDs. The engine records per-node provenance and metadata, including dependencies, status transitions, and runtime statistics (Line 12 in Algorithm~\ref{alg:executor}), enabling reconstruction and inspection of the annotated execution DAG that produced each result.
\section{Experimental Study}\label{sec:exp}
We implement \sysname ~\cite{CODE} and conduct a comprehensive experimental study to evaluate its effectiveness, efficiency, and end-to-end usability. Our evaluation is grounded in a diverse corpus designed to assess the system's versatility across various scholarly domains, covering broad domains, specialized sub-fields, and emerging areas. A complete list of topics is in 
% Appendix~\ref{app:topics}.
Appendix~F.

Our evaluation begins by benchmarking the performance of \sysname's core capabilities: retrieval (Section~\ref{sec:exp_retrieval}), information extraction and synthesis (Section~\ref{sec:exp_qa}), and knowledge discovery and generation (Section~\ref{sec:exp_kb}). To demonstrate its practical usability, we then present an end-to-end case study that simulates a realistic open-ended scholarly exploration workflow (Section~\ref{sec:case_study}). Finally, we conduct a series of ablation studies (Section~\ref{sec:exp_ablation}) to validate the measurable benefits of our key architectural decisions.

\subsection{Evaluation of Retrieval}\label{sec:exp_retrieval}

We evaluate the effectiveness of {\sysname} in paper retrieval (Tier 1 in Figure~\ref{fig:query_set}) using 237 research papers spanning 10 topics.

\setlength{\textfloatsep}{0pt}
\begin{table}[t]
    \setlength{\tabcolsep}{4pt}
    \setlength{\abovecaptionskip}{0cm}
    \setlength{\belowcaptionskip}{0cm}
    \centering
    \small
    \caption{Effectiveness on Tier-1 queries.}\label{tab:search_prec_all_topic}
    \begin{tabular}{ccccc}
        \toprule
        \multirow{2}{*}{Methods} & \multicolumn{2}{c}{Factual Queries} & \multicolumn{2}{c}{Analytical Queries} \\ 
        \cmidrule(lr){2-3} \cmidrule(lr){4-5}
        & R-Precision & MAP & R-Precision & MAP \\ 
        \midrule
        BM25                 & 0.49 & 0.52 & 0.42 & 0.45 \\ 
        GTR         & 0.44 & 0.47 & 0.36 & 0.39 \\ 
        E5          & 0.51 & 0.54 & 0.46 & 0.48 \\ 
        Instructor        & 0.52 & 0.53 & 0.45 & 0.46 \\ 
        EmbeddingGemma       & 0.49 & 0.49 & 0.42 & 0.41 \\ 
        ColBERT           & 0.58 & 0.60 & 0.45 & 0.42 \\ 
        Qwen3-Embedding              & \underline{0.61} & \underline{0.65} & \underline{0.50} & \underline{0.49} \\ 
        \textbf{\sysname{}}  & \textbf{0.73} & \textbf{0.74} & \textbf{0.69} & \textbf{0.70} \\ 
        \bottomrule
    \end{tabular}
\end{table}

\smallskip\noindent\textbf{Query Generation.} 
We construct 220 queries of two types:
\emph{factual queries}, which target explicit paper metadata (e.g., title, authors, venue) and research aspects (e.g., topics, problems, methods, datasets); and \emph{analytical queries}, which require reasoning within or across topics. Each query is derived from paper metadata and aspect summaries and is paired with 1--10 manually verified ground-truth papers. Additional details are provided in 
% Appendix~\ref{app:retrieval_setup}.
Appendix~H.1.

\smallskip
\noindent\textbf{Baselines.} 
We compare \sysname{} with the classical lexical baseline \emph{BM25}~\cite{robertson2009probabilistic}, strong dense retrievers -- \emph{GTR}~\cite{ni2021large}, \emph{E5}~\cite{wang2022text}, \emph{Instructor}~\cite{su2022one}, \emph{EmbeddingGemma}~\cite{vera2025embeddinggemma}, and \emph{Qwen3-Embedding}~\cite{qwen3zhang} -- and the late-interaction retriever \emph{ColBERT}~\cite{colbertv22022santhanam}. \sysname{} and ColBERT have $\sim$0.1B parameters, while others range from $\sim$0.3B to $\sim$1.5B. Most models use 768-d embeddings, except E5 (1024), and ColBERT uses 128-d token embeddings. All methods are reranked with GPT-4.1 on the top-$k$ candidates for fair comparison~\cite{ajith2024litsearch}. We omit \emph{LLM-Direct} retrieval due to context-length limitations.

\smallskip\noindent\textbf{Evaluation Metrics.} We use R-Precision~\cite{Beitzel2009arp}, which is precision at $R$ (number of relevant papers), and Mean Average Precision (MAP)~\cite{Beitzel2009map}, which averages precision across relevant ranks. We omit Recall$@k$ and Precision$@k$ since fixed cutoffs are incomparable when queries have different numbers of relevant papers.

\smallskip\noindent\textbf{Results.}  Table~\ref{tab:search_prec_all_topic} shows an overall result 
% (see Appendix~\ref{app:search_eval_results} for results by topic). 
(see Appendix~H.2 for results by topic). 
{\sysname{}} consistently outperforms all baselines. This confirms its effectiveness, particularly in performing cross-modal synthesis that unifies precise lexical matching with the semantic generalization of dense encoders. For factual queries, this synthesis enables {\sysname{}} to capture both explicit facts and semantically related expressions often missed by single-modal methods. For analytical queries, LLM-driven query decomposition and reranking facilitate nuanced reasoning over abstract research aspect, going beyond simple text similarity. These strengths demonstrate that {\sysname{}} broadens retrieval coverage while enhancing contextual and analytical fidelity.

\subsection{Information Extraction and Synthesis}\label{sec:exp_qa}

We evaluate our system's ability to extract relevant information and synthesize it into coherent answers (Tier 2 in Figure~\ref{fig:query_set}). 

\smallskip
\noindent\textbf{Query Generation}. We consider four topics, each containing 10 papers, to construct two test sets.

\noindent$\bullet$ \emph{LLM-generated test set}. 
Using three topics, we employ an LLM to generate evaluation queries. Based on our plan library 
% (Appendix~\ref{app:plan_libary})
(Appendix~C)
, we generate 480 queries covering 16 
single-document plans and 72 queries for six multi-document plans with group sizes ranging from 3 to 10 papers. In addition, we generate 12 queries with group sizes ranging from 3 to 10 papers that require dynamic plan generation.

\noindent$\bullet$ \emph{Human-annotated test set}. We additionally construct a human-annotated test set consisting of 197 queries over two topics, derived from predefined single-document and multi-document plans that involve the \texttt{Rank} operator (see 
% Table~\ref{tab:prompts} in the 
Table~12 in the 
appendix), enabling quantitative evaluation. Specifically, we manually examine the experimental results reported in each paper to determine rankings of methods or datasets.
For single-document rank plans, methods or datasets are ranked according to commonly reported evaluation metrics within each paper. For multi-document rank plans, we group 3–10 papers and rank the evaluated methods based on shared datasets and metrics within each group.

\setlength{\textfloatsep}{0pt}
\begin{table}[t]
\setlength{\tabcolsep}{3pt}
\setlength{\abovecaptionskip}{0cm}
\setlength{\belowcaptionskip}{0cm}
\footnotesize
\centering
\caption{Effectiveness on Tier-2 queries.}
\label{tab:eff_tier2}
\begin{tabular}{lcccc}
\toprule
Method & NDCG@1&NDCG@3&NDCG@5&NDCG@7 \\
\midrule
\sysname&0.606&0.629&0.644&0.655\\
RAG&0.411&0.482&0.491&0.494\\
\bottomrule
\end{tabular}
\end{table}

\begin{figure}[t]
\setlength{\abovecaptionskip}{0cm}
\setlength{\belowcaptionskip}{0cm}
\centering
\subfigtopskip=0pt
\subfigbottomskip=0pt
\subfigcapskip=-6pt
\subfigure[Input Tokens]{\label{subfig:LI_input} \includegraphics[height=28mm, width=36mm]{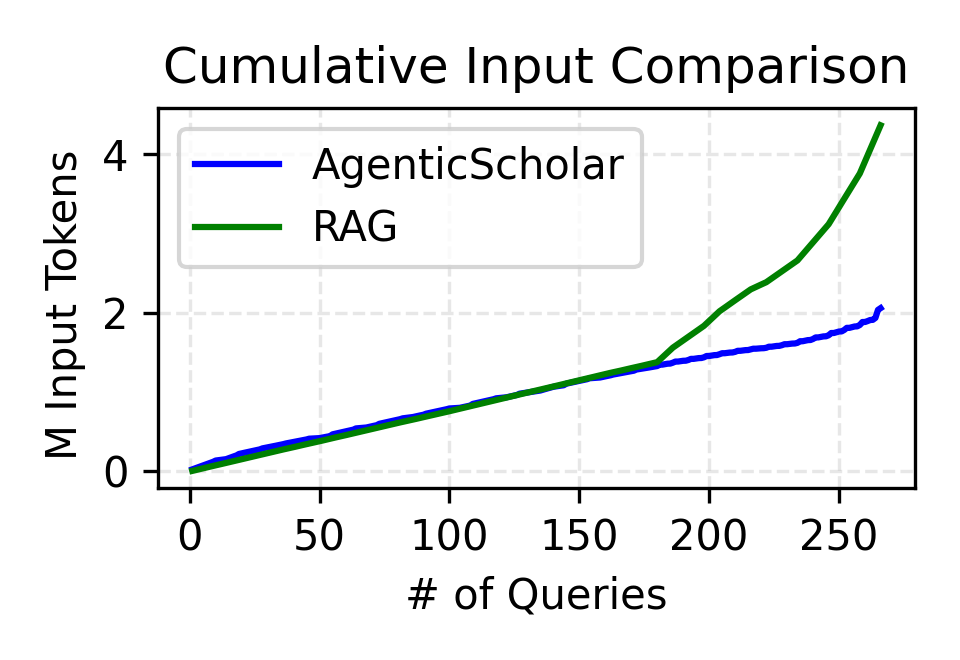}}
\subfigure[Output Tokens]{\label{subfig:LI_output} \includegraphics[height=28mm, width=36mm]{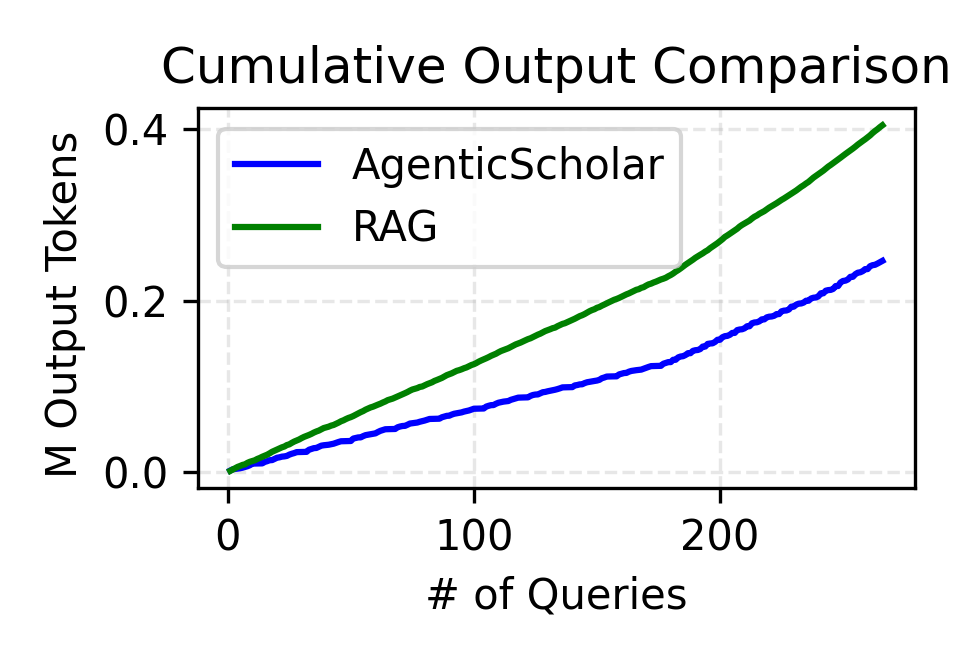}}
\subfigure[Response Time]{\label{subfig:LI_time} \includegraphics[height=28mm, width=36mm]{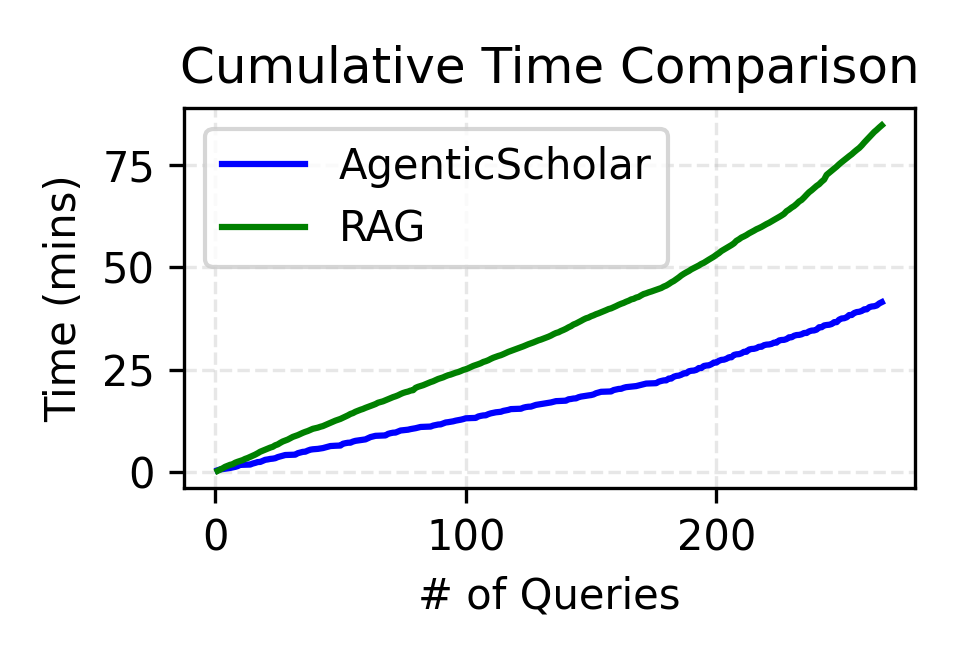}}
\caption{\label{fig:qa}{Cost and efficiency on Tier-2 queries.}
}
\end{figure}

\smallskip
\noindent\textbf{Baselines}. We compare \sysname{} with an \emph{RAG-based} method implemented using LangChain~\cite{LangChain}. The method retrieves relevant document chunks and synthesizes answers using \texttt{GPT-4.1 nano}. We set the chunk size to $1000$ to capture broader scholarly context and retrieve $8$ chunks per document for each query.

\smallskip
\noindent\textbf{Evaluation Metrics.} We evaluate effectiveness on the human-annotated test set using NDCG@$k$~\cite{JarvelinK02}, a widely used ranking metric.
Cost and efficiency are measured on the LLM-generated test set in terms of input/output tokens and response time.

\noindent\textbf{Results}. We observe the following:

\noindent\underline{Effectiveness}. Table~\ref{tab:eff_tier2} reports NDCG@k for $k$ ranging from 1 to 7. \sysname{} consistently outperforms RAG across all $k$ values, indicating that it accurately extracts relevant information from papers and effectively executes the \texttt{Rank} operator to place correct results at the top of the ranked list.

\noindent\underline{Cost and Efficiency}. Figure~\ref{fig:qa} reports cumulative input tokens, output tokens, and response time for the \emph{learned index} topic using the LLM-generated test set. Results on other topics show similar trends and are provided in 
% Appendix~\ref{sec:app_qa}. 
Appendix~I. 
Across all metrics, \sysname{} consistently outperforms the standard RAG framework, using fewer input and output tokens and achieving lower cumulative latency as query volume increases.

\setlength{\textfloatsep}{0pt}
\begin{table*}[t]
\setlength{\tabcolsep}{3pt}
\setlength{\abovecaptionskip}{0cm}
\setlength{\belowcaptionskip}{0cm}
\footnotesize
\centering
\caption{Results on Tier-3 queries. Elicit is the baseline; other entries report relative improvements (green, $\uparrow$) or degradations (red, $\downarrow$). \textit{Efficiency} shows percentage runtime change (positive = faster).}
\label{tab:tier3_queries}
\begin{tabular}{p{1.0cm}P{1.2cm}P{1.3cm}P{1.3cm}P{1.2cm}P{1.2cm}P{1.9cm}P{1.9cm}P{1.9cm}}
\toprule
\textbf{Metric} & Elicit & \makecell[tc]{Gemini DR} & \makecell[tc]{Tongyi DR} & \makecell[tc]{Open DR} & SmolAgent & \sysnameC & \sysnameR & \textbf{\sysname} \\
\midrule
\multicolumn{9}{l}{\textit{Research Trend Analysis --- Problem-oriented}} \\
Correctness & 5.84 & \textcolor{green!50!black}{+21.2\%} & \textcolor{red!70!black}{-10.3\%} & \textcolor{green!50!black}{+25.0\%} & \textcolor{green!50!black}{+26.7\%} & \textcolor{green!50!black}{+32.7\%} & \textcolor{green!50!black}{+30.7\%} & \textcolor{green!50!black}{+34.8\%} \\
Relevance & 5.83 & \textcolor{green!50!black}{+45.5\%} & \textcolor{green!50!black}{+41.2\%} & \textcolor{green!50!black}{+56.1\%} & \textcolor{green!50!black}{+47.5\%} & \textcolor{green!50!black}{+52.1\%} & \textcolor{green!50!black}{+53.5\%} & \textcolor{green!50!black}{+54.7\%} \\
Diversity & 5.42 & \textcolor{green!50!black}{+26.0\%} & \textcolor{green!50!black}{+17.5\%} & \textcolor{green!50!black}{+24.0\%} & \textcolor{green!50!black}{+11.8\%} & \textcolor{green!50!black}{+27.9\%} & \textcolor{green!50!black}{+26.8\%} & \textcolor{green!50!black}{+29.3\%} \\
Specificity & 5.23 & \textcolor{green!50!black}{+54.7\%} & \textcolor{green!50!black}{+25.4\%} & \textcolor{green!50!black}{+53.7\%} & \textcolor{green!50!black}{+23.1\%} & \textcolor{green!50!black}{+59.1\%} & \textcolor{green!50!black}{+62.9\%} & \textcolor{green!50!black}{+61.8\%} \\
Efficiency & 1{,}276\,s & \textcolor{green!50!black}{+40.2\%} & \textcolor{green!50!black}{+68.1\%} & \textcolor{green!50!black}{+71.6\%} & \textcolor{green!50!black}{+18.7\%} & \textcolor{green!50!black}{+68.9\%} & \textcolor{green!50!black}{+70.5\%} & \textcolor{green!50!black}{+69.6\%} \\
\midrule
\multicolumn{9}{l}{\textit{Research Trend Analysis --- Method-oriented}} \\
Correctness & 6.02 & \textcolor{green!50!black}{+6.5\%} & \textcolor{red!70!black}{-6.8\%} & \textcolor{green!50!black}{+27.1\%} & \textcolor{green!50!black}{+24.6\%} & \textcolor{green!50!black}{+26.7\%} & \textcolor{green!50!black}{+29.7\%} & \textcolor{green!50!black}{+28.9\%} \\
Relevance & 7.40 & \textcolor{green!50!black}{+13.9\%} & \textcolor{green!50!black}{+2.7\%} & \textcolor{green!50!black}{+21.6\%} & \textcolor{green!50!black}{+6.1\%} & \textcolor{green!50!black}{+18.4\%} & \textcolor{green!50!black}{+15.4\%} & \textcolor{green!50!black}{+20.1\%} \\
Diversity & 4.83 & \textcolor{green!50!black}{+41.6\%} & \textcolor{green!50!black}{+36.2\%} & \textcolor{green!50!black}{+49.1\%} & \textcolor{green!50!black}{+24.2\%} & \textcolor{green!50!black}{+50.3\%} & \textcolor{green!50!black}{+49.1\%} & \textcolor{green!50!black}{+51.6\%} \\
Specificity & 5.83 & \textcolor{green!50!black}{+21.3\%} & \textcolor{green!50!black}{+16.6\%} & \textcolor{green!50!black}{+37.9\%} & \textcolor{green!50!black}{+17.5\%} & \textcolor{green!50!black}{+41.2\%} & \textcolor{green!50!black}{+36.0\%} & \textcolor{green!50!black}{+41.9\%} \\
Efficiency & 1{,}432\,s & \textcolor{green!50!black}{+52.9\%} & \textcolor{green!50!black}{+69.7\%} & \textcolor{green!50!black}{+73.8\%} & \textcolor{red!70!black}{-47.2\%} & \textcolor{green!50!black}{+75.1\%} & \textcolor{green!50!black}{+71.8\%} & \textcolor{green!50!black}{+74.2\%} \\
\midrule
\multicolumn{9}{l}{\textit{Research Idea Exploration}} \\
Novelty & 6.38 & \textcolor{green!50!black}{+17.1\%} & \textcolor{red!70!black}{-21.6\%} & \textcolor{green!50!black}{+16.0\%} & \textcolor{green!50!black}{+12.7\%} & \textcolor{green!50!black}{+22.7\%} & \textcolor{green!50!black}{+26.0\%} & \textcolor{green!50!black}{+25.1\%} \\
Feasibility & 7.18 & \textcolor{red!70!black}{-5.0\%} & \textcolor{red!70!black}{-1.3\%} & \textcolor{green!50!black}{+2.4\%} & \textcolor{red!70!black}{-2.5\%} & \textcolor{green!50!black}{+6.5\%} & \textcolor{green!50!black}{+6.5\%} & \textcolor{green!50!black}{+8.4\%} \\
Impact & 7.82 & \textcolor{green!50!black}{+0.3\%} & \textcolor{red!70!black}{-10.9\%} & \textcolor{green!50!black}{+1.0\%} & \textcolor{green!50!black}{+5.5\%} & \textcolor{green!50!black}{+3.2\%} & \textcolor{green!50!black}{+3.2\%} & \textcolor{green!50!black}{+5.8\%} \\
Efficiency & 1{,}723\,s & \textcolor{green!50!black}{+66.7\%} & \textcolor{green!50!black}{+77.3\%} & \textcolor{green!50!black}{+80.7\%} & \textcolor{green!50!black}{+91.9\%} & \textcolor{green!50!black}{+75.3\%} & \textcolor{green!50!black}{+77.4\%} & \textcolor{green!50!black}{+78.0\%} \\
\midrule
\multicolumn{9}{l}{\textit{Milestone Paper Selection}} \\
Impact & 6.67 & \textcolor{green!50!black}{+19.9\%} & 0.0\% & \textcolor{green!50!black}{+7.9\%} & \textcolor{red!70!black}{-21.3\%} & \textcolor{green!50!black}{+12.4\%} & \textcolor{green!50!black}{+19.9\%} & \textcolor{green!50!black}{+19.9\%} \\
Coverage & 5.67 & \textcolor{green!50!black}{+20.5\%} & 0.0\% & \textcolor{green!50!black}{+12.9\%} & \textcolor{red!70!black}{-41.6\%} & \textcolor{green!50!black}{+12.4\%} & \textcolor{green!50!black}{+11.6\%} & \textcolor{green!50!black}{+11.6\%} \\
Coherence & 6.33 & \textcolor{green!50!black}{+23.7\%} & \textcolor{green!50!black}{+5.4\%} & \textcolor{green!50!black}{+16.9\%} & \textcolor{red!70!black}{-40.8\%} & \textcolor{green!50!black}{+10.5\%} & \textcolor{green!50!black}{+15.7\%} & \textcolor{green!50!black}{+13.1\%} \\
Efficiency & 213\,s & \textcolor{red!70!black}{-209.9\%} & \textcolor{red!70!black}{-36.6\%} & \textcolor{red!70!black}{-48.4\%} & \textcolor{red!70!black}{-333.8\%} & \textcolor{green!50!black}{+99.6\%} & \textcolor{green!50!black}{+99.6\%} & \textcolor{green!50!black}{+99.6\%} \\
\bottomrule
\end{tabular}
% \vspace{-1.4em}
\end{table*}

\subsection{Knowledge Discovery and Generation}\label{sec:exp_kb}

We evaluate how effective {\sysname} supports knowledge discovery and generation queries (Tier-3 in Figure~\ref{fig:query_set}), i.e., research trend analysis, idea exploration, and milestone paper selection. The implementation details are provided in Appendix E.

\smallskip
\noindent\textbf{Query Generation.} We use five research topics spanning broad domains, specialized sub-fields, and emerging areas 
% (Appendix~\ref{app:topics}). 
(Appendix~F). 
For each topic, we manually design Tier-3 queries of three types -- \textit{research trend analysis}, \textit{research idea exploration}, and \textit{milestone paper selection} -- following a unified template that specifies the topic, query type, and output format.
Query examples and prompt templates are provided in 
% Appendix~\ref{app:gen_query_kb}.
Appendix~J.

\smallskip
\noindent\textbf{Baselines.} 
We compare \sysname{} against three closed-source systems -- \elicit~\cite{elicit}, \qwen~\cite{tongyi}, and \gdr~\cite{deep_research} -- and two open-source systems, \smol~\cite{smolagents} and \open~\cite{opendeepresearcher}. We also evaluate two variants of \sysname: \sysnameR, which adopts LLM-generated reference taxonomies, and \sysnameC, which derives taxonomies solely from the corpus. For fairness, all open-source baselines and \sysname{} use GPT-5 as the backbone LLM.

\smallskip
\noindent\textbf{Evaluation Metrics.} We use the \emph{LLM-as-a-judge protocol}~\cite{gu2024llm_as_a_judge1,li2024llm_as_a_judge2,raina2024llm_as_a_judge3}, with GPT-5 serving as an impartial reviewer to score each method on specific dimensions (1–10 scale).

\begin{itemize}[leftmargin=*, noitemsep]
\item \textbf{Research Trend Analysis:}
(1) \textit{Correctness} -- factual accuracy, logical coherence, and temporal consistency;
(2) \textit{Relevance} -- topical alignment with the target domain;
(3) \textit{Diversity} -- coverage of distinct, non-overlapping trends; and
(4) \textit{Specificity} -- concreteness and technical precision.
\item \textbf{Research Idea Exploration:}
(1) \textit{Feasibility} -- technical soundness and practical achievability;
(2) \textit{Novelty} -- originality relative to existing research directions; and
(3) \textit{Impact} -- potential significance for future studies.
\item \textbf{Milestone Paper Selection:}
(1) \textit{Impact} -- influence and contribution to the research community;
(2) \textit{Coverage} -- representativeness across key subtopics; and
(3) \textit{Coherence} -- consistency and thematic alignment within the selected set.
\end{itemize}
Additionally, for each query type, we report \textit{efficiency}, measured as average inference time per query, as well as \emph{token} and \emph{monetary} costs for \sysname{} and open-source baselines.

\smallskip
\noindent\textbf{Results.} We observe the following:

\smallskip
\noindent\underline{Effectiveness}. As shown in Table~\ref{tab:tier3_queries}, {\sysname} achieves the best overall performance across nearly all dimensions. 

\noindent$\bullet$ \emph{Research trend analysis.} \sysname{} and its variants perform strongly across all evaluation dimensions. High \textit{correctness} and \textit{relevance} stem from its knowledge-grounded reasoning pipeline, which decomposes topics into structured subtopics and retrieves evidence via semantically anchored nodes (e.g., \texttt{Problem}, \texttt{Method}). They also achieve superior \textit{specificity} and \textit{diversity}, reflecting the robustness of its taxonomy in modeling hierarchical problem–method relations. Notably, \open attains comparable -- and occasionally higher -- \textit{relevance}, which we attribute to its strong online search capability for dynamically retrieving up-to-date sources.

\begin{figure}[t]
\setlength{\abovecaptionskip}{0cm}
\setlength{\belowcaptionskip}{0cm}
    \centering
    \includegraphics[width=\linewidth]{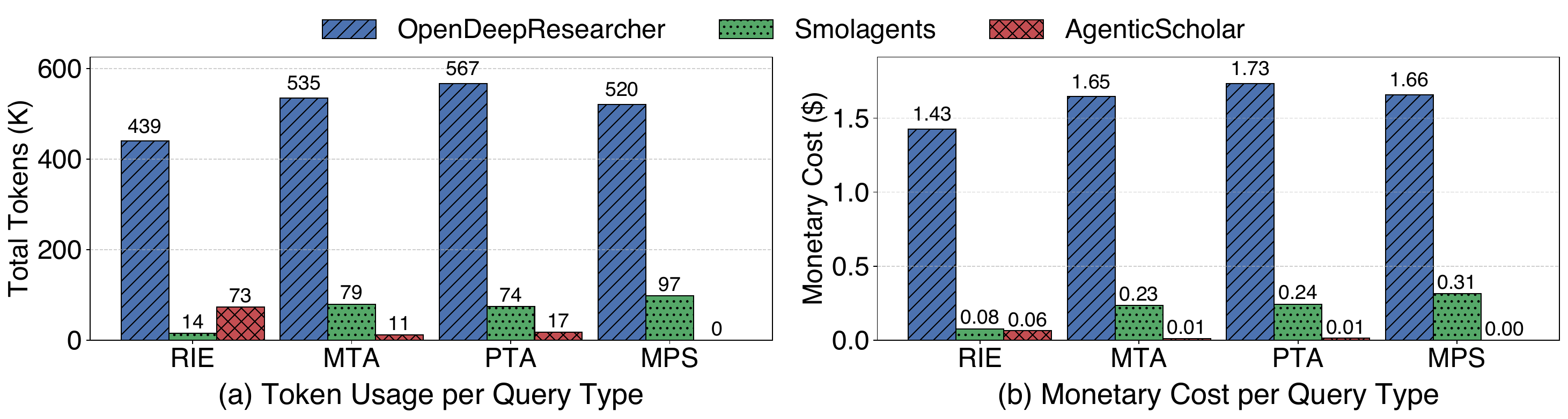}
    \caption{The comparison of token and monetary cost on Tier-3 Queries (RIE: Research Idea Exploration; MTA: Method-oriented Trend Analysis; PTA: Problem-oriented Trend Analysis; MPS: Milestone Paper Selection).}
    \label{fig:cost_comparison}
\end{figure}

\setlength{\textfloatsep}{0pt}
\begin{table}[t]
\setlength{\tabcolsep}{3pt}
\setlength{\abovecaptionskip}{0cm}
\setlength{\belowcaptionskip}{0cm}
\footnotesize
\centering
\caption{Breakdown of token cost on Tier-3 queries.}
\label{tab:token_breakdown}

\begin{threeparttable}
\begin{tabular}{lcccc}
\toprule
Query type & Planning$^1$ & Retrieval & Reasoning & Total Tokens \\
\midrule
Research Idea Exploration & 410 & 0 & 73{,}245 & 73{,}519 \\
Method Trend Analysis    & 410 & 0 & 11{,}345 & 11{,}649 \\
Problem Trend Analysis   & 410 & 0 & 17{,}463 & 17{,}935 \\
Milestone Paper Selection & 410 & 0 & 0 & 372 \\
\bottomrule
\end{tabular}

\begin{tablenotes}[flushleft]
\footnotesize
\item $^1$ As these queries directly match plans in our predefined plan library, their token cost in the planning stage remains fixed.
\end{tablenotes}
\end{threeparttable}
\end{table}

\noindent$\bullet$ \emph{Research idea exploration.}
Consistent with trend analysis, the \sysname{} family produces research ideas that are \emph{novel}, \emph{feasible}, and \emph{impactful}, driven by its structured discovery of underexplored problem–method intersections. This grounding enables forward-looking ideas while avoiding the speculative outputs common in baselines. Notably, \sysnameC achieves slightly higher \emph{Novelty}, suggesting that removing reference-guided constraints reduces anchoring bias and encourages broader hypothesis generation. However, this increased exploration comes at the cost of lower \emph{Impact} and \emph{Feasibility}.

\begin{figure*}[t]
\setlength{\abovecaptionskip}{0cm}
\centering
\subfigure[Research Trend]{\label{subfig:case_study_trend} \includegraphics[height=28mm, width=51mm]{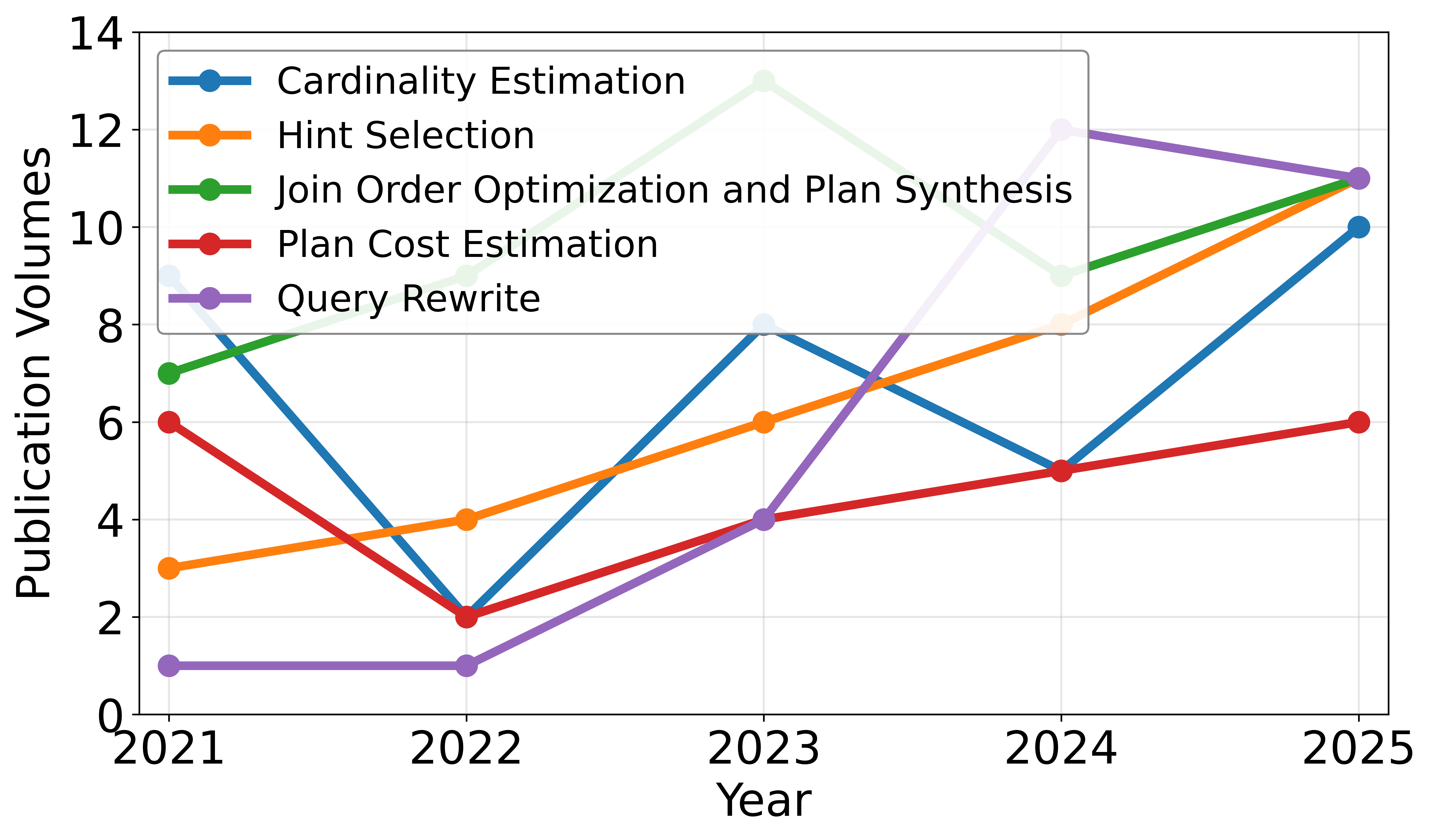}}
\subfigure[Problem and Method Summaries for Milestone Papers]{\label{subfig:case_study_summary} \includegraphics[height=32mm, width=74mm]{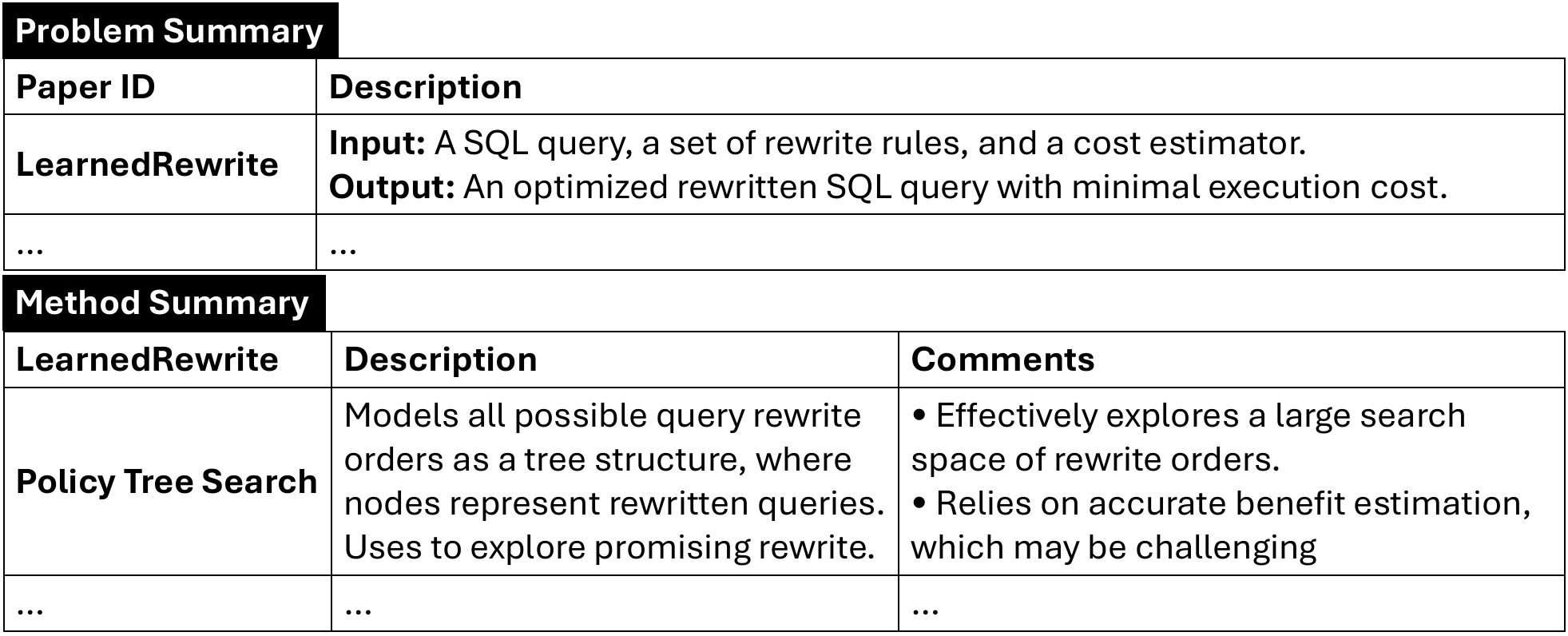}}
\subfigure[Explored Research Ideas]{\label{subfig:case_study_idea} \includegraphics[height=25mm, width=88mm]{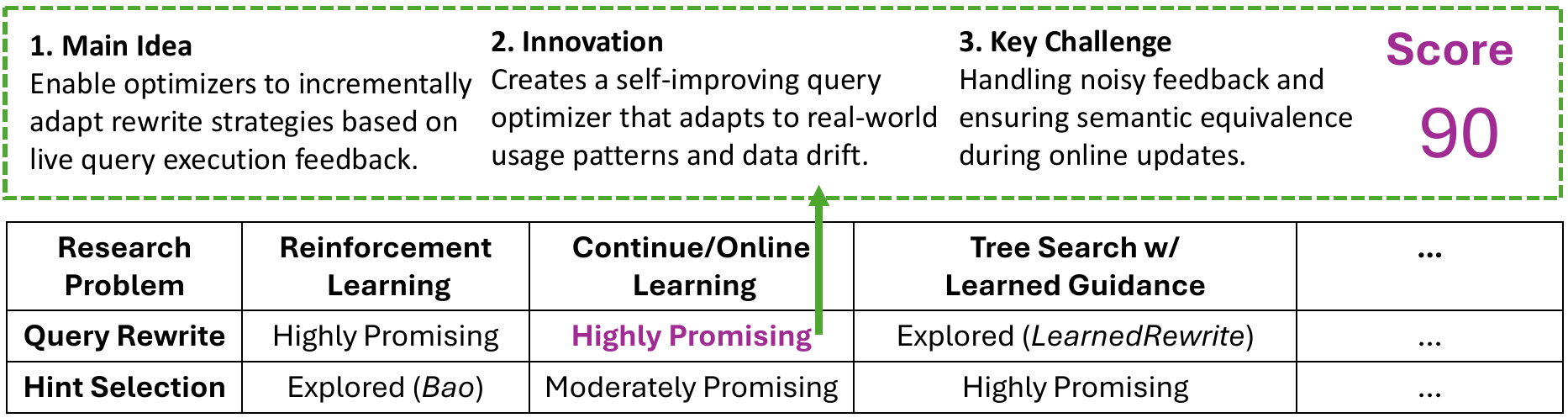}}
\caption{\label{fig:case_study}{\sysname{}'s results for case study on learning-based query optimization.}
}
\end{figure*}

\noindent$\bullet$ \emph{Milestone paper selection.} 
\sysname{} and its variants remain highly competitive, achieving strong \emph{Impact} scores. Their relatively lower \emph{Coverage} and \emph{Coherence} compared to \gdr{} and \open{}. This may stem from the latter’s web search capabilities, which enable broader and more up-to-date evidence retrieval.

\smallskip
\noindent\underline{Efficiency.} \sysname{} consistently delivers strong efficiency gains across Tier-3 tasks, with substantial speedups on \emph{Milestone Paper Selection}. The only exception is \emph{Research Idea Exploration}, where it is slightly slower than \open, likely due to evaluating a larger space of problem–method combinations for verification and scoring. Overall, these results indicate that structure-guided planning and targeted retrieval reduce unnecessary browsing and stabilize tool-call overhead, with the largest benefits emerging in focused selection tasks.

\smallskip
\noindent\underline{Token and Monetary Costs.} Figure~\ref{fig:cost_comparison} compares token and monetary costs against open-source baselines, with all systems using a unified GPT-5 backbone. 
Owing to its internal knowledge layer, \sysname{} minimizes token consumption across most query types, except research idea exploration. This efficiency stems from its lightweight LLM-based planner and reliance on internal knowledge for search and reasoning, which substantially reduces token overhead in the planning and retrieval stage. 
For instance, the average token usage for \open and \smol during planning and retrieval is 237,057 and 12,056, respectively, whereas \sysname{} consumes only 355 tokens on average. Table~\ref{tab:token_breakdown}  further provides a stage-level breakdown, 
\sysname{} consistently maintains minimal token consumption in planning and retrieval

\subsection{Case Study on Learning-based Query Optimization}
\label{sec:case_study}

To illustrate the end-to-end capabilities of \sysname{} in a realistic research setting, we present a case study simulating a typical researcher’s workflow. The task is to explore the rapidly evolving area of \textit{learning-based query optimization} and identify underexplored research directions -- an objective that challenges existing systems due to multi-step semantic reasoning over large scholarly corpora. The case study proceeds as a four-step progressive query session, demonstrating how \sysname{} enables iterative intent refinement and reasoning toward novel research ideas.

\smallskip
\noindent\textbf{Step 1: Understanding the Research Landscape (Tier 3)}. The researcher begins with an open-ended question to gain a high-level overview of the field and issues the query:
\emph{Q1: ``What are the overall research trends in applying learning-based methods to query optimization since 2021?''} As shown in Figure~\ref{subfig:case_study_trend}, {\sysname} generates a trend chart depicting publication volumes over time across key sub-topics. This goes beyond simple keyword search -- {\sysname}’s \emph{Knowledge Representation Layer} has already organized the corpus into a hierarchical taxonomy of research problems in query optimization. Its \emph{Execution Layer} then applies \texttt{Traverse} and \texttt{Aggregate} operators to quantify and visualize the evolution of these research focus areas, allowing users to understand how interest shifts across sub-topics. 

\smallskip
\noindent\textbf{Step 2: Identifying Foundational Studies Across Key Sub-fields (Tier 3)}. The trend analysis from Step 1 provides a crucial insight: a clear upward trend in \textit{query rewrite} and \textit{hint selection}, indicating that the community is actively exploring these techniques to build more robust query plans under estimation errors. To further understand this development, the researcher decides to investigate foundational studies across these two areas and issue the query \emph{``Can you identify three milestone papers for `query rewrite' and `hint selection'?''} {\sysname} returns the three most influential papers for each of the two requested topics: \emph{`learned query rewriting'}: SIA~\cite{SIA}, LLM-R$^2$~\cite{LLM-R2}, and LearnedRewrite~\cite{MCTS_QR}; \emph{`optimizer hint selection'}: Bao~\cite{Bao}, Steering~\cite{Steering}, and COOOL~\cite{COOOL}. This highlights {\sysname}’s ability to conduct fine-grained, taxonomy-guided knowledge discovery that identifies representative contributions within specific sub-fields.

\smallskip
\noindent\textbf{Step 3: Summarizing Core Problems and Methods (Tier 2)}. 
The researcher seeks a concise overview of each group of milestone papers to understand the central problems and methodological paradigms. The researcher issues the query:
\emph{``Could you summarize the core problems and methods for each of these three groups of papers?''}
As shown in Figure~\ref{subfig:case_study_summary}, {\sysname} produces structured summaries for each paper (e.g., LearnedRewrite), distilling both the problem formulation and proposed method. This step demonstrates how {\sysname}’s \texttt{Extract} and \texttt{Summarize} operators transform unstructured scholarly text into interpretable, structured insights -- laying the foundation for higher-level reasoning.

\smallskip
\noindent\textbf{Step 4: Identifying Novel Research Opportunities (Tier 3)}. With a clear understanding of the milestone papers' problems and methods, the researcher is now ready to identify new research directions and issues the query: \emph{``Okay, what would be some promising research ideas or directions to explore in query rewriting and hint selection?''}. As shown in Figure~\ref{subfig:case_study_idea}, {\sysname} traverses its knowledge graph to generate problem–method combinations representing underexplored. It then ranks these ideas to provide concise descriptions of the most promising ideas for future research.

\subsection{Validating Architectural Design}\label{sec:exp_ablation}

\subsubsection{The Effectiveness of Reference-enhanced Taxonomy Construction}\label{sec:exp_taxonomy}

To assess the quality of taxonomies produced by our reference-enhanced construction method, we follow~\cite{zhu2025context} and build a benchmark using gold-standard taxonomies manually curated from seven authoritative surveys published in ACM Computing Surveys\footnote{\url{https://dl.acm.org/journal/csur}} 
% (see Appendix~\ref{ap:kb_benchmark} for details). 
(see Appendix~K.1.1 for details). 
As baselines, we include two state-of-the-art taxonomy construction methods, \emph{TaxoAdapt}~\cite{kargupta2025taxoadapt} and \emph{\matht}~\cite{zhu2025context}, along with two variants of our approach, \sysnameC{} and \sysnameR{}.
We evaluate taxonomy quality using \emph{soft} Precision, Recall, and F1~\cite{franti2023soft}, which account for semantic similarity between concept labels and partial structural alignment, and further assess structural quality via an LLM-as-judge protocol 
% (See Appendix~\ref{app:ab_metric_defs} for details). 
(See Appendix~K.1.2 for details). 
As shown in Table~\ref{tab:taxonomy_quality}, \sysname{} consistently achieves the best overall performance across soft Precision, Recall, and F1 at nearly all taxonomy levels. These results indicate improvements in both concept identification (node quality) and hierarchical organization (edge quality), showing that anchoring taxonomies in corpus evidence while leveraging LLM knowledge yields more coherent and fine-grained structures.

\setlength{\textfloatsep}{0pt}
\begin{table}[t]
\setlength{\tabcolsep}{3pt}
\setlength{\abovecaptionskip}{0cm}
\setlength{\belowcaptionskip}{0cm}
\footnotesize
\centering
\caption{Level-wise taxonomy evaluation.}
\label{tab:taxonomy_quality}
\begin{tabular}{c l ccc ccc}
\toprule
 &  & \multicolumn{3}{c}{Node quality (soft)} & \multicolumn{3}{c}{Edge quality (soft)} \\
Level & Methods & Pre. & Rec. & F1. & Pre. & Rec. & F1. \\
\midrule
\multirow{5}{*}{1}
& TaxoAdapt         & 0.632          & 0.561          & 0.594          & 0.498          & 0.421          & 0.456 \\
& \matht             & 0.661          & 0.598          & 0.628          & 0.523          & 0.462          & 0.491 \\
& \sysnameC         & 0.694          & 0.640          & 0.677          & 0.562          & 0.504          & 0.531 \\
& \sysnameR         & 0.688          & 0.676          & 0.682          & \textbf{0.586} & 0.512          & 0.547 \\
& \textbf{\sysname} & \textbf{0.718}          & \textbf{0.684} & \textbf{0.703} & 0.574          & \textbf{0.538} & \textbf{0.555} \\
\midrule
\multirow{5}{*}{2}
& TaxoAdapt         & 0.598          & 0.532          & 0.563          & 0.458          & 0.392          & 0.422 \\
& \matht             & 0.620          & 0.558          & 0.587          & 0.487          & 0.424          & 0.453 \\
& \sysnameC         & 0.673          & 0.587          & 0.628          & 0.515          & 0.456          & 0.484 \\
& \sysnameR         & 0.649          & 0.632          & 0.640          & 0.541          & 0.466          & 0.501 \\
& \textbf{\sysname} & \textbf{0.684} & \textbf{0.642} & \textbf{0.657} & \textbf{0.549} & \textbf{0.487} & \textbf{0.507} \\
\midrule
\multirow{5}{*}{3}
& TaxoAdapt         & 0.543          & 0.472          & 0.505          & 0.402          & 0.327          & 0.361 \\
& \matht             & 0.568          & 0.495          & 0.529          & 0.428          & 0.345          & 0.382 \\
& \sysnameC         & \textbf{0.621} & 0.533          & 0.574          & 0.452          & 0.396          & 0.422 \\
& \sysnameR         & 0.593          & 0.574          & 0.583          & \textbf{0.477} & 0.406          & 0.439 \\
& \textbf{\sysname} & 0.608          & \textbf{0.582} & \textbf{0.595} & 0.466          & \textbf{0.432} & \textbf{0.448} \\
\bottomrule
\end{tabular}
\end{table}

\subsubsection{The Effectiveness of Hybrid Planning}
\noindent

\smallskip\noindent\textbf{Efficiency of Predefined Plan Selection}
We generate $50$ Tier-2 queries which can hit our predefined plans pool on the  \textit{vector search} topic and compare two variants: enable predefined plan selection (Enable) and disable it (Disable). Table~\ref{table:predefined_opt} shows the average \textit{number of input/output tokens} and \textit{planning latency} per query of each variant. We can observe: with predefined plan selection enabled, \our{} reduces both input/output tokens and planning time by more than 90\%, indicating that Predefined Plan Selection can effectively reduce token overhead and planning latency.

\setlength{\textfloatsep}{0pt}
\begin{table}[t]
\setlength{\tabcolsep}{3pt}
\setlength{\abovecaptionskip}{0cm}
\setlength{\belowcaptionskip}{0cm}
\footnotesize
\centering
\caption{Effectiveness of Predefined Plan Selection.}\label{table:predefined_opt} 
\begin{tabular}{ccccc}
\toprule
Variant& Input Tokens (K) & Output Tokens (K)  & Planning Time (s) \\ 
\midrule
Enable & 0.41 & 0.013 & 0.96 \\
Disable & 4.08 & 0.77 & 15.17\\
\bottomrule
\end{tabular}
\end{table}

\smallskip\noindent\textbf{Robustness through Self-Correction}
We designed $16$ complex Tier-2 queries on the  \textit{vector search} topic that genuinely reflect our daily research needs. We record the frequency and cumulative percentage of different repair attempts in Table~\ref{table:correction_opt}. We observe that in most cases ($\geq 60\%$), {\sysname} is already able to generate perfect plans, and with a reasonable small self-correction overhead ($\leq$ 3 times).

\setlength{\textfloatsep}{0pt}
\begin{table}[t]
\centering
\setlength{\abovecaptionskip}{0cm}
\setlength{\belowcaptionskip}{0cm}
\captionsetup{margin={2em,0pt}}
\caption{Robustness through Self-Correction.}\label{table:correction_opt} 
\setlength{\tabcolsep}{4pt}
\footnotesize
\begin{tabular}{lcccc}
\toprule
 & 0 & 1 & 2 & 3 \\
\midrule
Frequency & 10 & 4 & 1 & 1 \\
Cum. Pct. & 62.5\% & 87.5\% & 93.75\% & 100\% \\
\bottomrule
\end{tabular}
\end{table}

\subsubsection{The Impact of Execution Performance Optimizations}

We generated $266$ Tier-2 queries on the \textit{learned indexes} topic  and compare three variants: disable cache (W/O Cache), disable cache and parallel processing (W/O Cache and Para. Proc.), and enable all of them (Enable All). Table~\ref{table:execution_opt} shows the average \textit{number of input/output tokens} and \textit{latency} per query of each variant. We can observe: with caching and parallel processing enabled, \our{} uses about one-third of the input tokens and achieves the lowest latency ($9.3$ s). Disabling caching increases token usage and latency, while removing both optimizations roughly doubles the runtime.

\setlength{\textfloatsep}{0pt}
\begin{table}[t]
\centering
\setlength{\abovecaptionskip}{0cm}
\setlength{\belowcaptionskip}{0cm}
\captionsetup{margin={2em,0pt}}
%\vspace{-1.5em}
\caption{Impact of Performance Optimizations.}\label{table:execution_opt} 
\setlength{\tabcolsep}{3pt}
\footnotesize
\begin{tabular}{ccccc}
\toprule
Variant&  Input Tokens (K) & Output Tokens (K)  & Time (s) \\ 
\midrule
Enable All & 7.7 & 0.9 & 9.3 \\
W/O Cache & 23.0 & 2.8 & 13.7\\
W/O Cache and Para. Proc. & 22.8 & 2.8 & 23.3 \\
\bottomrule
\end{tabular}
\end{table}

\subsubsection{Impact of Backbone LLMs on End-to-End Performance}
We study how different backbone LLMs affect the end-to-end performance of \sysname. We instantiate \sysname{} with eight representative backbones from different providers, including GPT-5 and GPT-5 mini~\cite{openai_gpt5,openai_gpt5mini}, Claude Sonnet 4.5 and Claude Opus 4.5~\cite{anthropic_sonnet45,anthropic_opus45}, Gemini 3 Flash Preview and Gemini 3 Pro Preview~\cite{google_gemini3flash,google_gemini3pro}, and Grok-4.1 Fast variants without and with explicit reasoning (NR/R)~\cite{xai_grok41fast}. For fair comparison, we fix the agentic pipeline, prompts, and all non-LLM components, varying only the backbone. Table~\ref{tab:backbone_llms} reports results for research idea exploration 
% (see Appendix~\ref{ap:backbone} 
(see Appendix~K.3 
for research trend analysis). Following~\cite{abs-2511-01716}, we categorize backbone LLMs into Better, Average, and Worse based on relative performance (top, middle, and bottom 33\%). We observe that: (1) backbone choice has limited impact on effectiveness but substantially influences efficiency and cost; (2) GPT-5 achieves the strongest effectiveness, followed by Gemini-3.0-Pro; (3) Grok-4.1 Fast variants deliver the best efficiency and lowest cost; and (4) among the evaluated backbones, Gemini-3.0-Flash offers the best accuracy–efficiency trade-off.

\setlength{\textfloatsep}{0pt}
\begin{table}[t]
\setlength{\tabcolsep}{3pt}
\setlength{\abovecaptionskip}{0cm}
\setlength{\belowcaptionskip}{0cm}
\footnotesize
\centering
\caption{The impact of backbone LLMs for research idea exploration. The colors express relative values (\belowavg{worse than average}, \avg{average}, and \aboveavg{better than average}).}
\label{tab:backbone_llms}
\begin{tabular}{lcccccc}
\toprule
\textbf{Metric} &\makecell{Token \\ cost}&\makecell{Monetary \\cost}&Efficiency&Novelty&Feasibility&Impact\\
\midrule
GPT-5 & \avg{73,245}& \avg{\$0.640}& \avg{379.1s} & \aboveavg{8.22} & \aboveavg{7.92} & \aboveavg{8.42} \\
GPT-5  Mini&\avg{66,531}&\avg{\$0.116} & \avg{338.1s}& \belowavg{7.60} & \belowavg{7.36} & \belowavg{7.66} \\
Claude 4.5 Sonnet&\avg{71,017}&\belowavg{\$0.957} & \belowavg{468.9s}&\avg{7.98}& \avg{7.66} & \avg{8.02} \\
Claude  4.5 Opus &\belowavg{77,001}&\belowavg{\$1.737} & \belowavg{673.5s}& \avg{8.12}& \avg{7.78}  & \avg{8.18}   \\
Gemini-3.0 Flash&\aboveavg{60,240}& \avg{\$0.160}& \avg{305.2s} & \avg{7.72}& \avg{7.44}& \avg{7.78}  \\
Gemini-3.0 Pro&\belowavg{74,094}&\belowavg{\$0.793}& \belowavg{433.7s} & \aboveavg{8.36} & \aboveavg{7.84} & \aboveavg{8.34}  \\
Grok-4.1 Fast (NR)& \aboveavg{54,429} & \aboveavg{\$0.026}& \aboveavg{147.4s}& \belowavg{7.28} & \belowavg{7.14} & \belowavg{7.34} \\
Grok-4.1 Fast (R)& \belowavg{93,794}& \aboveavg{\$0.046}& \aboveavg{233.6s}& \belowavg{7.38}& \belowavg{7.22}& \belowavg{7.44}\\
Average     & 71,294 & \$0.559 & 372.4s & 7.83 & 7.55 & 7.90 \\
Standard deviation & $\pm$16.6\% & $\pm$107.2\% & $\pm$43.0\% & $\pm$5.1\% & $\pm$4.0\% & $\pm$5.2\% \\
\bottomrule
\end{tabular}
\end{table}

\section{Lessons Learned and Future Work}
\label{sec:limitations}

\noindent\textbf{Knowledge Graph Construction and Maintenance}.
Our experience shows that the quality of structured knowledge is pivotal to the success of {\sysname}. While the automated pipeline for constructing knowledge graphs and taxonomies is effective, it remains prone to occasional errors in entity extraction, relation detection, and semantic labeling that can propagate through downstream components. Heavy reliance on large models also incurs substantial computational and financial cost. Future efforts can be made to incorporate human-in-the-loop validation and cost-efficient hybrid pipelines that combine lightweight NLP components with compact fine-tuned models.

\smallskip\noindent\textbf{Multi-Modal Data Extraction}. Figures in scholarly documents are essential for high-quality scholarly synthesis, yet quantitative extraction remains a challenge. Recovering accurate numerical values from complex plots, such as spider charts or line charts, remains unreliable, even with SOTA multi-modal LLMs. This insight motivates the integration of specialized vision–language models and structured figure-parsing frameworks to improve numerical fidelity and support richer evidence-grounded analysis.

\smallskip\noindent\textbf{Model Selection and Monetary Cost in Complex Analysis}. A key lesson is that scholarly queries often require reasoning across multiple papers and synthesizing experimental evidence. This creates an important trade-off among model efficiency, reliability, and cost-effectiveness. Lightweight models such as \texttt{GPT-4.1-nano} offer faster inference and lower monetary cost, making them attractive for cost-sensitive large-scale use, but they are also more prone to hallucinations than stronger models such as \texttt{GPT-5} or \texttt{Gemini 2.5 Pro}. Future work could therefore explore adaptive model selection strategies that dynamically choose models based on task complexity and the desired cost-quality balance. One promising direction is to decompose complex analysis into smaller subtasks that can be handled by lightweight, more cost-effective models, while reserving more powerful and expensive models for reasoning-intensive stages. Such a design could improve overall cost-effectiveness while maintaining a strong balance among accuracy, speed, and monetary cost.

\smallskip\noindent\textbf{Domain-Aware Planning and Reasoning}. Finally, our study underscores the value of domain sensitivity in agentic reasoning. While {\sysname} already generalizes across disciplines, domain-agnostic planning can miss field-specific reasoning conventions -- such as performance-driven evaluation in computer science, proof-centric validation in physics, or evidence hierarchies in medicine. Future efforts will explore domain-aware planning that integrates domain-specific reasoning and analytical workflows, enabling results that are both precise and contextually aligned with expert reasoning practices.
\section{Conclusion}
\label{sec:conclusion}
In this paper, we have developed {\sysname}, an agentic data management system for scholarly corpora. It unifies knowledge representation, hybrid query planning, and operator-based execution to support diverse retrieval, synthesis, and discovery tasks. To the best of our knowledge and evaluation, {\sysname} is the first system to execute agentic reasoning over multi-modal scholarly data through DAG-based plans, offering a practical foundation for future research in agentic data management. While {\sysname} demonstrates strong performance on several scholarly domains, our current evaluation focuses primarily on research papers from a limited set of fields in computer science. Supporting broader disciplines may require domain-aware planning and reasoning mechanisms to better capture field-specific structures and analytical conventions.

% \begin{acks}
% This project is supported in part by ARC FT240100832 and DP240101211. 
% % Yuwei Peng is supported in part by the National Key Research and Development Program of China under Grant No. 2023YFB4503604. Yuwei Peng is the corresponding author.
% \end{acks}

% \received{October 2025}
% \received[revised]{January 2026}
% \received[accepted]{February 2026}

\clearpage

% \begin{acks}
% \end{acks}

\bibliographystyle{ACM-Reference-Format}
\bibliography{sample-base}

\clearpage
\appendix
\section*{Appendix}

\section{Demonstration of \sysname} \label{app_sys_ui}
\sysname{} supports a spectrum of scholarly query intents within a unified, interactive workflow. 
In the following, we will demonstrate how a researcher can issue queries at multiple tiers and engage with the system’s interfaces.

\begin{figure*}[t]
\centering
\includegraphics[width=\linewidth]{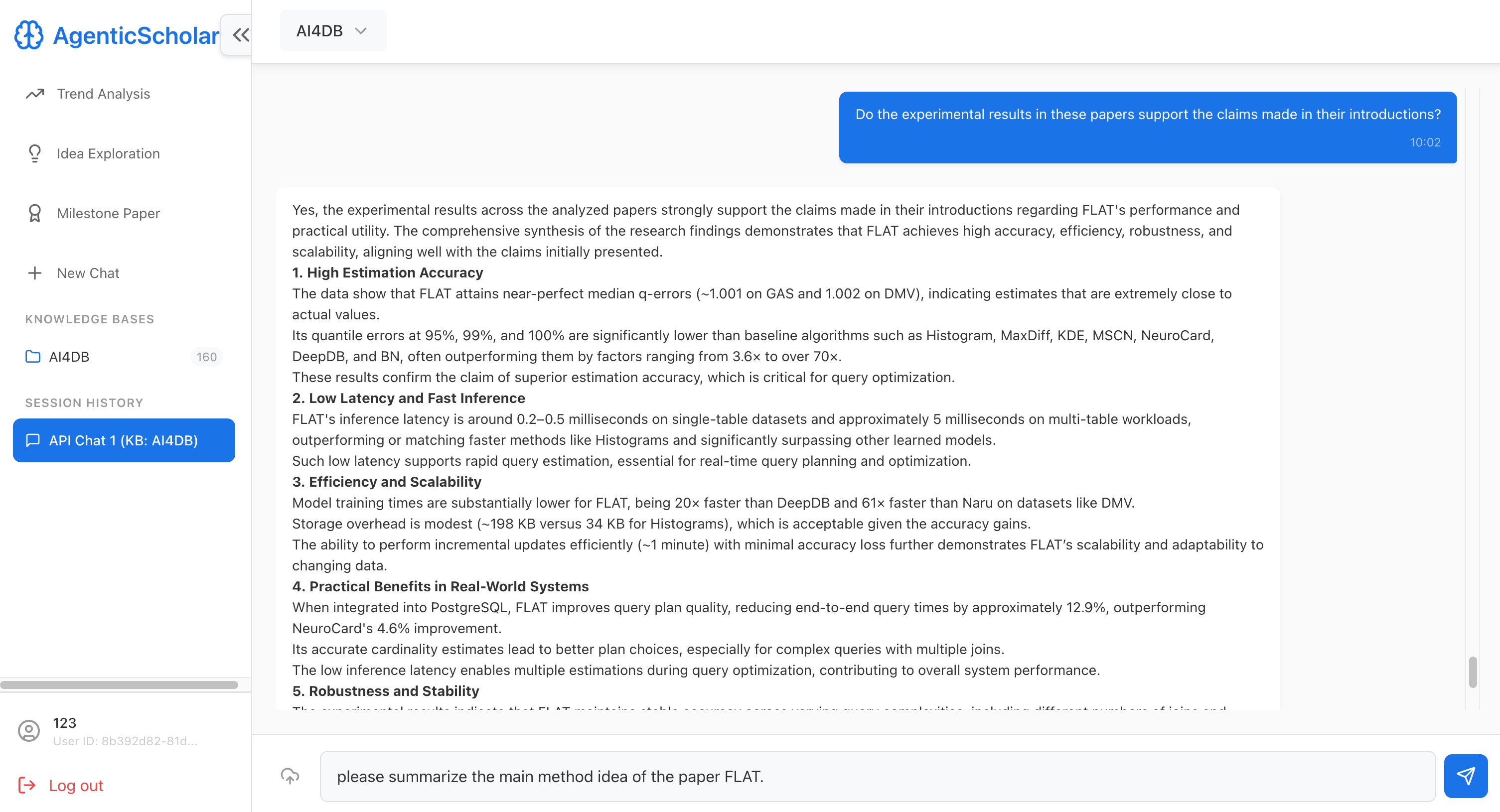}
\caption{Information Extraction and Synthesis.}
\label{fig:ui_ie}
\end{figure*}

\noindent\textbf{Scenario.} Consider a database researcher exploring the topic ``AI4DB''.
\noindent
(1) \emph{Information Extraction and Synthesis.} 
The researcher aims to determine whether the paper FLAT’s experimental results substantiate the claims made in its introduction.
As illustrated in Figure \ref{fig:ui_ie}, the researcher opens a new session via \texttt{+ New Chat}, submits the query ``Do the experimental results in these papers support the claims made in their introductions?'', and \sysname{} returns a structured, evidence-backed answer.

\begin{figure*}[t]
\centering
\includegraphics[width=0.9\linewidth]{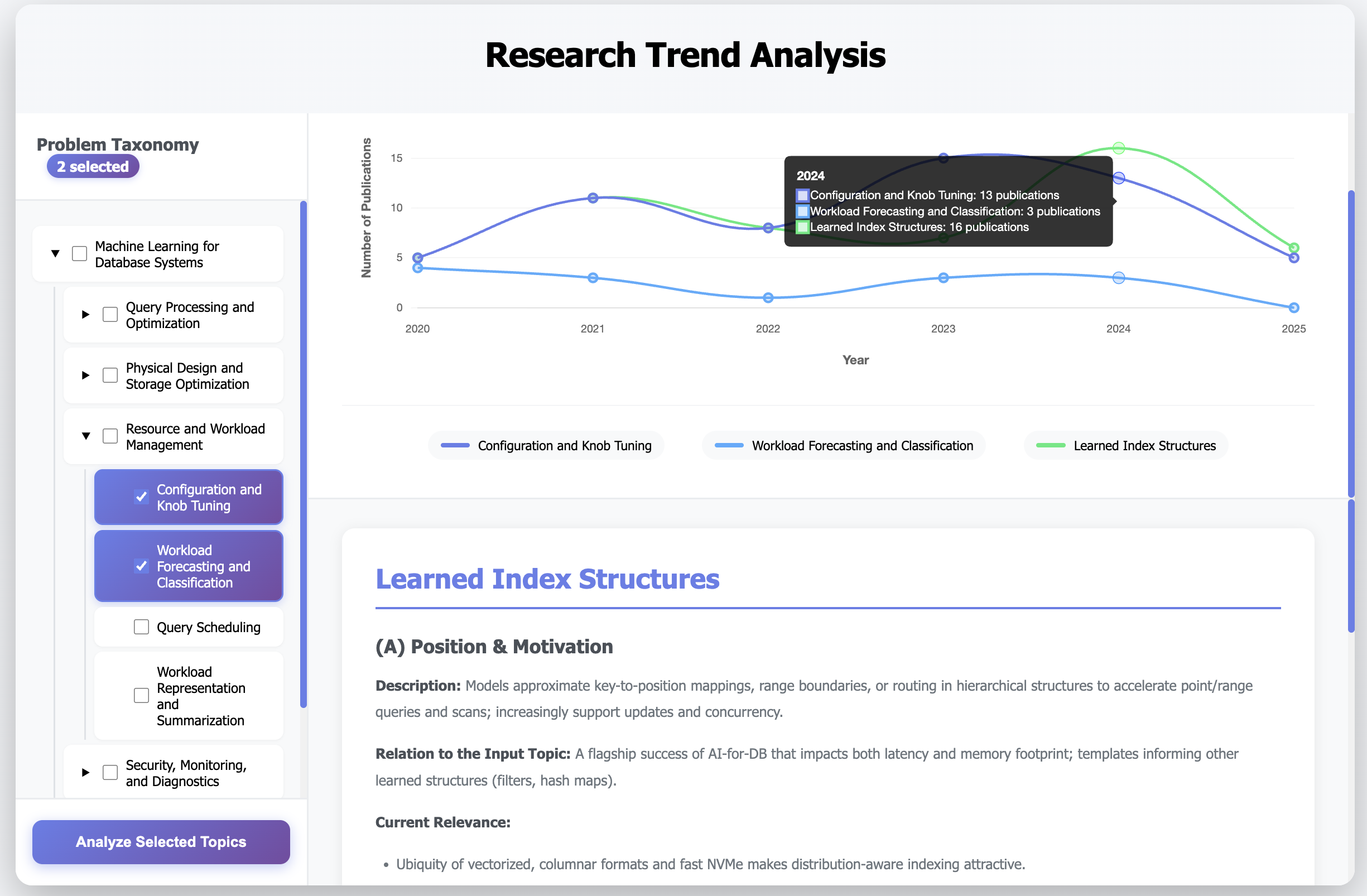}
\caption{Research trend analysis.}
\label{fig:ui_trend}
\end{figure*}

\noindent
(2) \emph{Knowledge Discovery and Generation.} The researcher moves from single-paper analysis to corpus-level exploration of a recently collected ``AI4DB'' paper set, proceeding through three queries, such as research trend analysis, research idea exploration, and milestone paper selection:

\noindent$\bullet$ \textit{Research trend analysis.} As shown in Figure~\ref{fig:ui_trend}, the left panel displays the hierarchical tree view of the problem taxonomy. Users can select one or more nodes within the taxonomy for trend analysis. The right panel then searches relevant papers online for these selected topics and visualizes their publication trends from 2020 to 2025 as line charts.
When a user clicks on a specific subtopic represented by a line in the chart—e.g., \emph{learned index structures}—a descriptive panel appears in the lower-right corner of \sysname{}, providing a concise definition of the subtopic and highlighting representative works.

\begin{figure*}[t]
\centering
\includegraphics[width=0.9\linewidth]{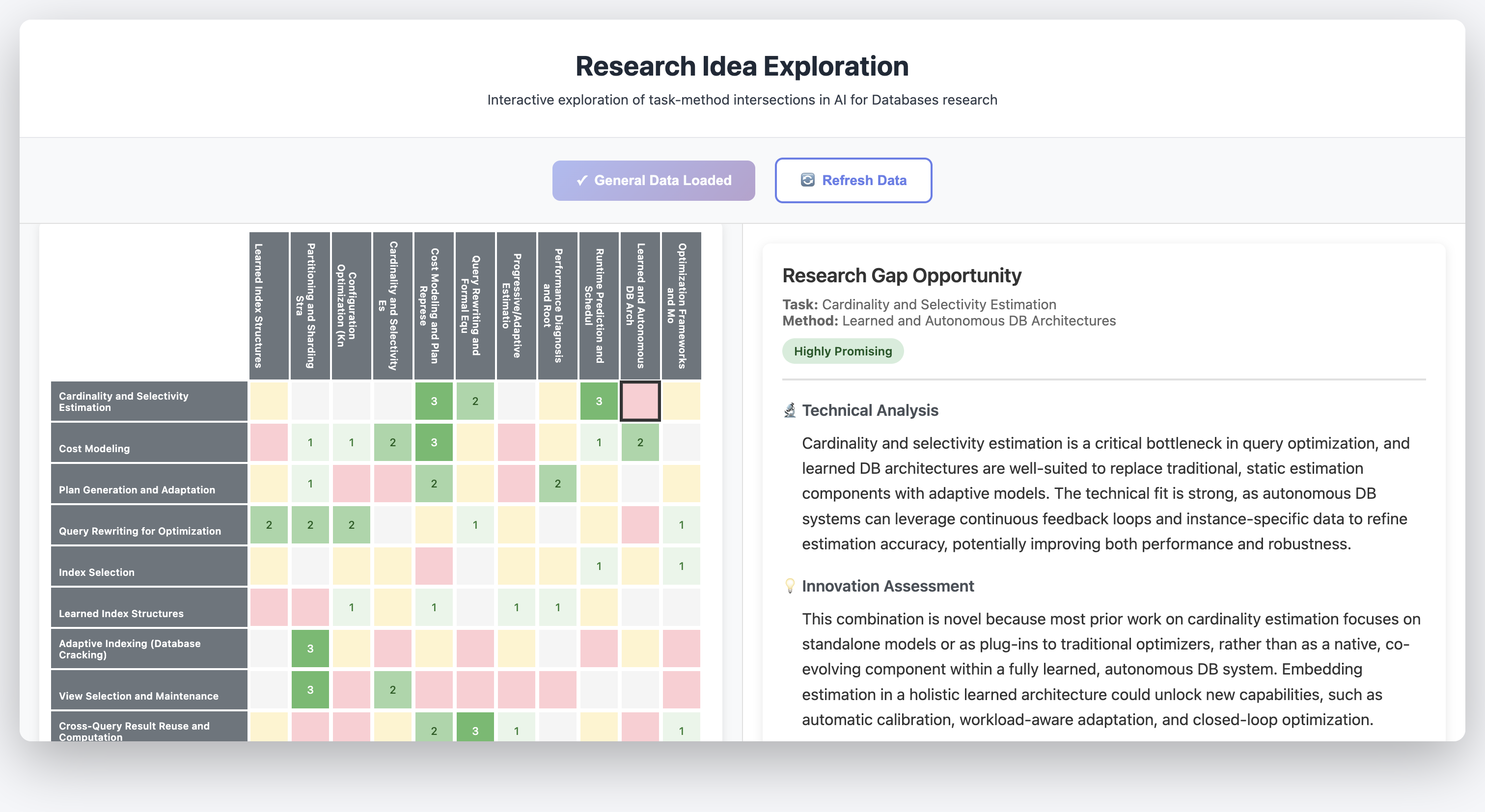}
\caption{Research idea exploration.}
\label{fig:ui_idea}
\end{figure*}

\noindent$\bullet$ \textit{Research idea exploration.} The researcher then requests research idea exploration. \sysname{} analyzes the corpus against problem and method taxonomies, aligns papers to these categories, and constructs a matrix whose rows are problem categories and columns are method categories (Figure~\ref{fig:ui_idea}). Each cell displays the number of matched papers. Interacting with a sparse (or empty) cell—e.g., the problem \emph{Cardinality and Selectivity Estimation} crossed with the method \emph{Learned and Autonomous DB Architectures}—opens a right-side panel that synthesizes the gap and surfaces promising research directions.

\begin{figure*}[t]
\centering
\includegraphics[width=0.9\linewidth]{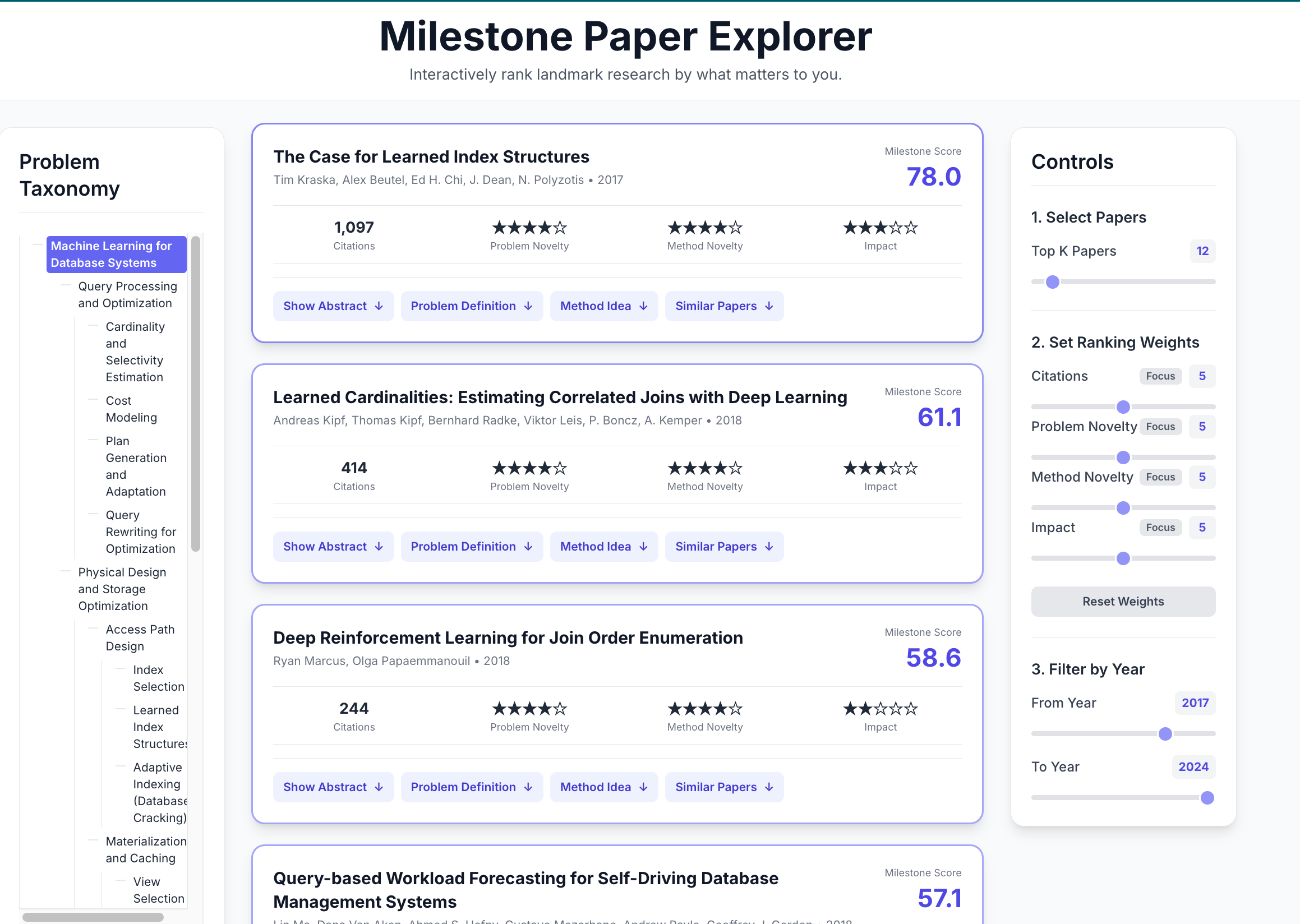}
\caption{Milestone paper selection.}
\label{fig:ui_milestone}
\end{figure*}

\noindent$\bullet$  \textit{Milestone paper selection.} Finally, the researcher issues a milestone identification request. As shown in Figure \ref{fig:ui_milestone}, they choose relevant subtopics from the problem taxonomy on the left, and configures ranking granularity on the right (e.g., top-$k$, citation window). The central panel lists the top-$k$ papers ranked by a milestone score, together with brief justifications, facilitating rapid triage for deeper study.

\section{Implementation Details of Knowledge Representation Layer}
\label{app:implement_KB}

\subsection{Knowledge Representation Layer Construction Pipeline}\label{app:construct_KB}

The knowledge graph construction pipeline forms the foundation of the Knowledge Representation Layer, transforming raw scholarly documents into a structured, semantically enriched graph that supports fine-grained retrieval and reasoning in {\sysname}. The pipeline integrates \textit{OCR-based text extraction}, \textit{LLM-assisted segmentation and tagging}, \textit{entity and relation extraction}, and \textit{graph materialization}. The overall workflow proceeds through five main stages.

\vspace{0.5em}
\noindent
\textbf{Stage 1: Document Parsing and Section-Level Chunking.}
For each research paper in PDF format, we first employ the \textit{Mistral OCR API} \cite{Mistral} to rapidly convert the document into Markdown format for text analysis, and then apply \textit{MinerU} \cite{mineru} to detect visual content and extract high-resolution tables and figures with preserved layout. 
We adopt the \textit{Mistral OCR API} and \textit{MinerU} due to their strong efficiency in processing individual paper PDFs and their effectiveness in accurately extracting tables and figures, as evidenced by the experimental results in Appendix~\ref{appendix_eval_doc_parsing}.
To facilitate accurate downstream processing, we adopt a \textit{section-level chunking strategy} that segments each paper by its natural sections and groups them into semantic units (e.g., Abstract, Introduction, Method, Experiment).
We prioritize section boundaries because research papers are structurally self-contained and queries typically target specific parts (e.g., problem definition or experimental setup), making this approach more effective than fixed windows or embedding-based topic segmentation~\cite{zhang2025sage} for scholarly analytics.
To robustly detect sections across heterogeneous formats, we first match standardized patterns with regular expressions (numeric, Roman, alphabetic, or style-only) and extract the section titles.
We then utilize an LLM as the classifier to infer the semantic label of each section title and maps it to standardized labels such as \textit{Abstract}, \textit{Introduction}, \textit{Related Work}, \textit{Problem Formulation}, \textit{Methodology}, and \textit{Experiments}.
Sections sharing the same label are consolidated into the corresponding semantic unit.
This strategy enables efficient, precise retrieval and aggregation for diverse queries.

\vspace{0.5em}
\noindent
\textbf{Stage 2: Bibliographic Information Search.}
After text extraction, we retrieve bibliographic information for each paper using its title as a query to the \textit{Semantic Scholar Paper Search API} \cite{semanticscholar}.
The returned metadata—such as \textit{authors} and \textit{venue}—is incorporated into the knowledge graph as \textit{bibliographic nodes}.
This enrichment ensures consistency across the corpus and enables reasoning over scholarly provenance, co-authorship, and venue-specific trends.

\vspace{0.5em}
\noindent
\textbf{Stage 3: Entity Extraction and Normalization.}
After obtaining the processed semantic units across sections, we feed them into LLM-based extractors to construct different categories of entities:
(1)~\textit{Taxonomy nodes:} sections labeled as \textit{Problem Formulation} and \textit{Methodology} are used to extract formalized problem definitions and methodological strategies, forming the \textit{Problem} and \textit{Method} taxonomies described in Section~\ref{sec:taxo};
(2)~\textit{Experimental context nodes:} sections labeled as \textit{Experiments} or \textit{Evaluation} are used to extract datasets, metrics, and baselines, forming the corresponding \texttt{Dataset}, \texttt{Metric}, and \texttt{Baseline} nodes.
Since different papers may refer to the same dataset or metric using variant names, we perform an LLM-based normalization step to unify semantically equivalent nodes across papers.
This normalization ensures global consistency and enables cross-paper comparisons in higher-tier analytical queries.

\vspace{0.5em}
\noindent
\textbf{Stage 4: Relation Construction and Schema Population.}
The extracted entities are linked to their corresponding \texttt{Paper} nodes through typed relations defined in the schema (Figure~\ref{fig:schema}), such as \texttt{ADDRESSES}, \texttt{APPLIES}, \texttt{USES}, and \texttt{HAS}.
These relations capture both intra-document and inter-document dependencies, situating each paper within a broader scholarly ecosystem.

\vspace{0.5em}
\noindent
\textbf{Stage 5: Graph Materialization and Physical Storage.}
Finally, the complete structure is materialized as a \textit{property graph} in \textit{Neo4j} \cite{neo4j}.
Each node and edge stores (i) symbolic attributes (e.g., names, descriptions, formal specifications) for rule-based filtering and graph traversal, and (ii) dense text embeddings for semantic similarity search.
This hybrid representation supports both efficient traversal queries and embedding-driven retrieval.
The Neo4j backend serves as the persistent substrate for higher-level components in {\sysname}, including the hybrid query planner and the unified execution engine.

\begin{algorithm}[t]
\caption{\textsc{BuildTaxonomy} (End-to-End Pipeline)}
\label{alg:build_app}
\begin{algorithmic}[1]
\small
\Require Corpus $\mathcal{D}$; taxonomy type $\kappa$ (determines $m$ aspects and extraction schema);
optional reference taxonomy $\mathcal{T}_{\text{ref}}$ (or topic label $l$ to generate it);
match threshold $\tau_{\text{match}}$; refinement factor $\alpha$; max branches $K_{\max}$
\Ensure Instantiated taxonomy $\mathcal{T}$; per-paper templates $S[\cdot]$; canonical aspect-class sets $\{\mathcal{A}_i\}_{i=1}^{m}$

\State $m \gets \textsc{NumAspects}(\kappa)$
\State $S \gets \emptyset$

\ForAll{$p \in \mathcal{D}$}
  \State $S[p] \gets \textsc{ExtractTemplate}(p,\kappa)$
\EndFor

\State $\{\mathcal{A}_i\}_{i=1}^{m} \gets \textsc{CrossPaperStandardization}(\{S[p]\}_{p\in\mathcal{D}})$

\If{$\mathcal{T}_{\text{ref}} = \texttt{null}$}
  \State $l \gets \textsc{InferTopicLabel}(\mathcal{D})$
  \State $\mathcal{T}_{\text{ref}} \gets \textsc{GenerateRefTaxonomy}(l,\kappa)$
\EndIf
\State $\mathcal{T} \gets \textsc{AlignAndInstantiate}(\mathcal{T}_{\text{ref}}, S, \{\mathcal{A}_i\}_{i=1}^{m})$

\State \textsc{RegisterUpdateHandler}$(\mathcal{T},\kappa,\{\mathcal{A}_i\}_{i=1}^{m},\alpha,\tau_{\text{match}},K_{\max})$

\State \Return $(\mathcal{T}, S, \{\mathcal{A}_i\}_{i=1}^{m})$
\end{algorithmic}
\end{algorithm}

\subsection{Pseudo-code of Taxonomy Construction}
\label{app:code_taxo}

This section presents the detailed pseudo code for our taxonomy construction algorithm, which forms the foundation for the hierarchical organization of scholarly knowledge in {\sysname}. The pseudo code is presented in Algorithm~\ref{alg:build_app}.

The algorithm performs in four stages.
(1) \textit{Structured Information Extraction}: \textsc{ExtractTemplate} extracts key aspects of the target concept into a structured template, leveraging relevant sections of the document.
(2) \textit{Cross-paper Standardization}: \textsc{CrossPaperStandardization} aligns the extracted aspects from templates to form unified canonical aspect-class sets $\{\mathcal{A}_i\}_{i=1}^{m}$ (where $m$ depends on the taxonomy type).
(3) \textit{Reference Taxonomy Generation and Alignment}: A soft-guidance reference taxonomy $\mathcal{T}_{\text{ref}}$ is generated from the inferred topic label $l$ and aligned with corpus-derived aspect signatures through \textsc{AlignAndInstantiate}.
(4) \textit{Progressive Update and Branch Refinement}: \textsc{UpdateTaxonomy} incrementally integrates newly arriving papers $\mathcal{D}_{\mathrm{new}}$ under a size-controlled policy ($\alpha$): each paper is routed to the most specific node and attached when similarity exceeds $\tau_{\mathrm{match}}$, otherwise a new branch is created; once a node becomes overpopulated ($\ge \alpha \cdot \textsc{BaseSize}$), \textsc{RefineBranch} applies corpus-grounded \textsc{LLM\_ClusterAndLabel} to split leaf nodes into finer subtopics or extend non-leaf nodes by adding a new sibling branch, and then reassigns papers accordingly.
In the next, we will present the detailed algorithms for each stage.

\begin{algorithm}[t]
\caption{\textsc{ExtractTemplate} (Structured Information Extraction)}
\label{alg:extract}
\begin{algorithmic}[1]
\small
\Require Document $p$; taxonomy type $\kappa$ (defines aspect schema $\{(\text{name}_i,\text{instr}_i)\}_{i=1}^{m}$); LLM $\mathcal{M}$
\Ensure Structured template $S[p]=\langle \text{desc}_p, \text{aspect}^{(1)}_p,\ldots,\text{aspect}^{(m)}_p\rangle$

\State $m \gets \textsc{NumAspects}(\kappa)$
\State $\mathcal{U} \gets \textsc{SelectSections}(p,\kappa)$
\State $ctx \gets \textsc{BuildContext}(p,\mathcal{U})$  \Comment{concatenate/trim to fit budget}

\State $\text{desc}_p \gets \textsc{LLM\_Summarize}(\mathcal{M}, ctx)$

\For{$i=1$ to $m$}
  \State $q_i \gets \textsc{BuildAspectPrompt}(\kappa,i)$
  \State $\text{aspect}^{(i)}_p \gets \textsc{LLM\_ExtractAspect}(\mathcal{M}, ctx, q_i)$
  \State $\text{aspect}^{(i)}_p \gets \textsc{PostProcess}(\text{aspect}^{(i)}_p)$
\EndFor

\State $S[p] \gets \langle \text{desc}_p, \text{aspect}^{(1)}_p,\ldots,\text{aspect}^{(m)}_p\rangle$
\State \Return $S[p]$
\end{algorithmic}
\end{algorithm}

\begin{algorithm}[t]
\caption{\textsc{CrossPaperStandardization}}
\label{alg:standardize}
\begin{algorithmic}[1]
\small
\Require Set of per-paper templates $\{ \langle \text{aspect}^{(1)}_p, \ldots, \text{aspect}^{(m)}_p \rangle \}_{p \in \mathcal{D}}$
\Ensure Canonical aspect-class sets $\{\mathcal{A}_i\}_{i=1}^{m}$
\For{$i=1$ to $m$}
  \State $\mathcal{E}_i \gets \emptyset$
\EndFor
\ForAll{$p \in \mathcal{D}$}
  \For{$i=1$ to $m$}
    \State $\mathcal{E}_i \gets \mathcal{E}_i \cup \textsc{ExtractEntities}(\text{aspect}^{(i)}_p)$
  \EndFor
\EndFor
\For{$i=1$ to $m$}
  \State $\mathcal{V}_i \gets \textsc{ComputeEmbeddings}(\mathcal{E}_i)$
  \State $\mathcal{A}_i \gets \textsc{ClusterAndLabel}(\mathcal{V}_i)$
\EndFor
\State \Return $\{\mathcal{A}_i\}_{i=1}^{m}$
\end{algorithmic}
\end{algorithm}

\begin{algorithm}[t]
\small
\caption{\textsc{AlignAndInstantiate}}
\label{alg:align}
\begin{algorithmic}[1]
\Require Reference taxonomy $\mathcal{T}_{\text{ref}}$, signatures $S$, aspect classes $\{\mathcal{A}_i\}_{i=1}^{m}$
\Ensure Instantiated taxonomy $\mathcal{T}$
\State $\mathcal{T} \gets \textsc{CopyTree}(\mathcal{T}_{\text{ref}})$
\State $\mathcal{P} \gets \textsc{DistinctTuples}\!\Big(\textsc{MapToClasses}\big(S,\{\mathcal{A}_i\}_{i=1}^{m}\big)\Big)$
\ForAll{$\mathbf{a} \in \mathcal{P}$} \Comment{$\mathbf{a}=(a_1,\ldots,a_m)$}
  \State $u \gets \textsc{TopDownLocateParent}(\mathcal{T}.\text{root}, \mathbf{a})$
  \State $v \gets \textsc{BestChildSim}(u.\text{children}, \mathbf{a})$
  \If{$\textsc{Sim}(v,\mathbf{a}) \ge \tau_{\text{match}}$}
    \State $target \gets v$
  \Else
    \State $(name,desc) \gets \textsc{LLM\_NameAndDescr}(\mathbf{a})$
    \State $target \gets \textsc{AddChild}(u, \texttt{Node}(name,desc,\mathbf{a}))$
  \EndIf
  \ForAll{$p \in \textsc{PapersWith}(S,\mathbf{a})$}
    \State $target.\text{papers} \gets target.\text{papers} \cup \{p\}$
  \EndFor
\EndFor
\State \Return $\mathcal{T}$
\end{algorithmic}
\end{algorithm}

\begin{algorithm}[t]
\small
\caption{\textsc{UpdateTaxonomy}}
\label{alg:update}
\begin{algorithmic}[1]
\Require Taxonomy $\mathcal{T}$, refinement factor $\alpha$, match threshold $\tau_{\text{match}}$
\Ensure Updated taxonomy $\mathcal{T}$
\State \textbf{on} \textsc{NewPaper}$(p)$:
\State \quad $sig \gets \textsc{ExtractSignature}(p)$
\State \quad $\mathbf{a} \gets \textsc{MapToClassesOrCreate}(sig)$ \Comment{$\mathbf{a}=(a_1,\ldots,a_m)$}
\State \quad $u \gets \textsc{TopDownLocateParent}(\mathcal{T}.\text{root}, \mathbf{a})$
\State \quad $v \gets \textsc{FindMatchingChild}(u,\mathbf{a},\tau_{\text{match}})$
\State \quad $target \gets v \text{ if } v \neq \text{null else } \textsc{AddChild}(u,\texttt{NodeFrom}(\mathbf{a}))$
\State \quad $target.\text{papers} \gets target.\text{papers} \cup \{p\}$
\State \quad \textbf{if} $\textsc{NewPaperCount}(target) \ge \alpha \cdot \textsc{BaseSize}(target)$ \textbf{then} \textsc{RefineBranch}$(target)$
\end{algorithmic}
\end{algorithm}

\begin{algorithm}[t]
\small
\caption{\textsc{RefineBranch} (Progressive Update: Branch Refinement)}
\label{alg:refine}
\begin{algorithmic}[1]
\Require Taxonomy node $u$; per-paper templates $S[\cdot]$; refinement threshold $\alpha$; match threshold $\tau_{\text{match}}$; max branches $K_{\max}$
\Ensure Refined taxonomy structure around $u$
\If{$\textsc{NewPaperCount}(u) < \alpha$} \State \Return \EndIf \Comment{Step 1: Trigger}

\State $P \gets \textsc{DirectPapers}(u)$ \Comment{papers assigned to $u$ (excluding descendants)}
\If{$|P| \le 1$} \State \Return \EndIf

\State $K \gets \min(\textsc{EstimateSubtopics}(P), K_{\max})$
\If{$K \le 1$} \State \Return \EndIf

\State $\{(\mathcal{C}_1,\ell_1),\ldots,(\mathcal{C}_K,\ell_K)\} \gets
\textsc{LLM\_ClusterAndLabel}(\{S[p]\}_{p\in P}, K)$
\Comment{Step 2: Execution via corpus-grounded LLM clustering}

\For{$k=1$ to $K$}
  \For{$i=1$ to $m$}
    \State $\mathcal{A}_i^{(k)} \gets \textsc{Canonicalize}\big(\{S[p].\text{aspect}^{(i)} \mid p \in \mathcal{C}_k\}\big)$
  \EndFor
  \State $\mathbf{a}^{\star} \gets \textsc{DominantTuple}\big(\{\mathcal{A}_i^{(k)}\}_{i=1}^{m}\big)$
  \If{$\textsc{IsLeaf}(u)$} \Comment{Case 1: Leaf node refinement (split into a deeper subtree)}
    \State $v \gets \textsc{AddChild}(u, \texttt{NodeFrom}(\ell_k, \mathbf{a}^{\star}))$
  \Else \Comment{Case 2: Non-leaf refinement (add new branch under $u$)}
    \State $v \gets \textsc{FindMatchingChild}(u.\text{children}, \mathbf{a}^{\star}, \tau_{\text{match}})$
    \If{$v = \texttt{null}$}
      \State $v \gets \textsc{AddChild}(u, \texttt{NodeFrom}(\ell_k, \mathbf{a}^{\star}))$
    \EndIf
  \EndIf
  \State \textsc{MovePapers}$(\mathcal{C}_k, v)$
\EndFor

\State \textsc{ResetNewPaperCount}$(u)$
\end{algorithmic}
\end{algorithm}

\vspace{0.5em}
\noindent\textbf{Stage 1: Structured Information Extraction.}
This stage extracts key aspects from each paper and represents them in a structured template, including (i) a concise description and (ii) a taxonomy-dependent set of structured aspect fields $\{\text{aspect}^{(i)}\}_{i=1}^{m}$ that capture salient dimensions for hierarchical organization. We leverage an LLM to analyze sections relevant to the target concept (e.g., abstract, introduction, and the most relevant definition/method sections). The procedure is detailed in Algorithm~\ref{alg:extract}.

\vspace{0.5em}
\noindent\textbf{Stage 2: Cross-Paper Standardization.}
This stage unifies heterogeneous representations across papers by standardizing the extracted aspects. As shown in Algorithm~\ref{alg:standardize}, the procedure takes the collection of structured templates from Stage~1 and applies semantic canonicalization to each aspect dimension. Specifically, \textsc{ExtractEntities} identifies candidate terms from each aspect field; \textsc{ComputeEmbeddings} encodes them into a shared vector space; and \textsc{ClusterAndLabel} groups semantically similar terms to derive the canonical aspect-class sets $\{\mathcal{A}_i\}_{i=1}^{m}$. These sets ensure cross-paper consistency and enable downstream taxonomy alignment.

\vspace{0.5em}
\noindent\textbf{Stage 3: Reference Taxonomy Generation and Alignment.}
This stage integrates canonicalized aspect classes into a hierarchical structure guided by a draft reference taxonomy. As shown in Algorithm~\ref{alg:align}, the procedure first duplicates the reference tree via \textsc{CopyTree} and maps each paper to a canonical aspect-signature tuple $\mathbf{a}=(a_1,\ldots,a_m)$ using \textsc{MapToClasses}, where $a_i \in \mathcal{A}_i$. For each distinct tuple, the algorithm locates the most relevant parent node through \textsc{TopDownLocateParent}, identifies the best-matching child using \textsc{BestChildSim}, and either reuses an existing node or invokes \textsc{LLM\_NameAndDescr} and \textsc{AddChild} to instantiate a new one. Finally, all associated papers are linked to their corresponding taxonomy nodes.

\vspace{0.5em}
\noindent\textbf{Stage 4: Progressive Update and Refinement.}
This stage enables the taxonomy to evolve dynamically as new papers arrive. As shown in Algorithm~\ref{alg:update}, each incoming paper is converted into a structured signature by \textsc{ExtractSignature}, mapped to a canonical tuple $\mathbf{a}$ via \textsc{MapToClassesOrCreate}, and routed to the most specific node by \textsc{TopDownLocateParent} and \textsc{FindMatchingChild}. If no match exists, \textsc{AddChild} instantiates a new node and the paper is attached. When a node accumulates enough new papers ($\ge \alpha\cdot \textsc{BaseSize}$), we invoke \textsc{RefineBranch} (Algorithm~\ref{alg:refine}). Refinement partitions linked papers via \textsc{LLM\_ClusterAndLabel} and, for each cluster, derives a dominant canonical tuple $\mathbf{a}^{\star}$ by aggregating evidence across all aspect dimensions. The taxonomy is then refined by adding or matching branches and reassigning papers accordingly, keeping the structure adaptive under concept drift.

\section{Predefined Plan Library}
\label{app:plan_libary}

Table~\ref{tab:prompts} presents our predefined plan library, which comprises 16 plans for executing the \texttt{extract}, \texttt{summarize}, \texttt{rank}, and \texttt{check} operators within individual scholarly documents, and 6 plans for applying the \texttt{summarize}, \texttt{rank}, and \texttt{check} operators across multiple documents.

\begin{table*}[t]
\setlength{\abovecaptionskip}{0cm}
\setlength{\belowcaptionskip}{0cm}
\small
\centering
\caption{Predefined plan library}
\label{tab:prompts}
\begin{tabular}{ccccc}
\toprule
Type&Operator&\multicolumn{2}{c}{Plan description}\\
\midrule
\multirow{18}{*}{Single doc.}&\multirow{12}{*}{Extract}&1&Extract experimental settings\\
&&2&Extract experimental datasets \\
&&3&Extract evaluation metrics \\
&&4&Extract compared baselines \\
&&5&Extract the experimental results of comparison with baselines \\
&&6&Extract the experimental results of parameter study\\
&&7&Extract the experimental results of ablation study\\
&&8&Extract the problem definition\\
&&9&Extract the input of the problem\\
&&10&Extract the output of the problem\\
&&11&Extract the goal of the problem\\
&&12&Extract the proposed method\\
\cline{2-4}
&\multirow{1}{*}{Summarize}&13&\makecell{Summarize the impact of parameters based \\on experimental results}\\
\cline{2-4}
&\multirow{2}{*}{Rank}&14&\makecell{Rank variants of the proposed method \\based on experimental results} \\
&&15&Rank experimental datasets based on experimental results\\
\cline{2-4}
&\multirow{1}{*}{Check}&16&\makecell{Check inconsistencies between the used \\evaluation metrics and the original evaluation metrics}\\
\hline
\multirow{6}{*}{Multiple doc.}&\multirow{4}{*}{Summarize}&17&Summarize common settings in experiments\\
&&18&Summarize missing settings in experiments\\
&&19&Summarize pros and cons across different methods\\
&&20&Summarize differences among the problem definition\\
\cline{2-4}
&Rank&21&\makecell{Rank experiment results of different methods on \\a common dataset and a common evaluation metric}\\
\cline{2-4}
&Check&22&\makecell{Check inconsistencies in experiment results on \\a common dataset and a common evaluation metric}\\
\hline
\multirow{3}{*}{Topic}&\multirow{3}{*}{\makecell{Multiple \\operators}}&23&Research trend analysis\\
&&24&Research idea exploration\\
&&25&Milestone paper selection\\
\bottomrule
\end{tabular}
\end{table*}

\section{Details for Operators}
\label{app:operator_detail}

In this section, we present the parameters and detailed implementation of each operator. All operators -- except \texttt{Search}, \texttt{FindNode}, and \texttt{Traverse} -- expose an \textsf{execution\_mode} parameter (\textsf{instance} or \textsf{group}) that specifies whether the operator is applied to each item individually or to the entire collection as a whole. This design enables fine-grained control over execution behavior.

\subsection{Knowledge Access}

\smallskip\noindent\textbf{\texttt{Search}.} The \texttt{Search} operator performs ranked retrieval of scholarly documents from our knowledge graph.
        
\noindent\textit{\underline{Parameters.}} It takes a natural language query $q$ as its primary input. The operator behavior is controlled by two main parameters: \textsf{candidates\_per\_query}, which specifies the number of initial candidates retrieved by each retrieval strategy to ensure high recall (default 20), and \textsf{top\_k}, which defines the maximum number of the final ranked results returned to the user (default 10). 
\noindent\textit{\underline{Implementation.}} The search operator is implemented as a five-stage pipeline that progressively refines the candidate set of papers to balance lexical precision and semantic relevance.
\begin{enumerate}[leftmargin=*]
\renewcommand{\labelenumi}{\roman{enumi}.}
\item \emph{Query decomposition}. The input query is first parsed by an LLM-based component into structured attribute–value pairs, identifying which metadata fields and research aspects are relevant. We use five metadata fields = \{\emph{title}, \emph{authors}, \emph{affiliations}, \emph{publication year}, \emph{venue}\}, and six research aspects = \{\emph{research topic}, \emph{problem formulation}, \emph{proposed method}, \emph{experimental datasets}, \emph{experimental baselines}, \emph{experimental results}\}.
\item \emph{Temporal filtering}. Publication year constraints are applied as a strict preliminary filter to prune temporally invalid papers, reducing downstream computation.
\item \emph{Hybrid candidate retrieval}. (1) Lexical matching on metadata: BM25 keyword search is applied to precise terms such as ``authors'' or ``venue''; (2) Semantic search on research aspects: Cosine similarity search over pre-computed dense embeddings is used for descriptive research aspects such as ``approximate nearest neighbor search'' or ``attribute/range filtering in high-dimensional vector spaces''.
\item \emph{Score aggregation}. Scores from the two retrieval strategies are normalized and averaged within each group (BM25 for metadata, cosine similarity for research aspects), then combined with equal weight into a unified relevance score to prevent bias toward either lexical or semantic matches.
\item \emph{LLM-based reranking}. The top candidates from score aggregation are passed to an LLM-based reranker. This component jointly considers the original query together with the candidate's metadata and research aspects, refining the final ranking and discarding results that are semantically plausible but contextually irrelevant. 
\end{enumerate}

\subsection{Knowledge Navigation}

\smallskip\noindent\textbf{\texttt{FindNode}.} The \texttt{FindNode} operator is designed to locate specific \emph{entry-point} nodes within the taxonomy nodes. The identified node will serve as the starting point for a more complex analytical plan.

\noindent\textit{\underline{Parameters.}} Its behavior is primarily controlled by one of two mutually exclusive parameters: \textsf{node\_id}, which provides a specific identifier for direct and unambiguous retrieval, or \textsf{node\_description}, a natural language phrase used to perform a semantic search to find the most relevant node in the taxonomy.

\noindent\textit{\underline{Implementation.}} If a \textsf{node\_id} is provided, the system performs a direct, high-speed key lookup against the graph database. If a \textsf{node\_description} is provided instead, the system performs a semantic similarity search over the embeddings of the taxonomy nodes' descriptions to identify and return the best match.

\smallskip\noindent\textbf{\texttt{Traverse}.} The \texttt{Traverse} operator is the primary navigation function of the knowledge graph. It enables movement from a given set of starting nodes to new sets of connected nodes by following specified relationships.

\noindent\textit{\underline{Parameters.}} The operator takes a collection of \textsf{start\_nodes} (produced by a preceding operator) as input. Its behavior is governed by the \textsf{traversal\_path} parameter -- a structured list that defines the sequence of relationship types to follow and the target entity types to retrieve at each step along the graph.

\noindent\textit{\underline{Implementation.}} It takes the \textsf{traversal\_path} and translates it into a single, efficient, multi-hop query in the native language of the underlying knowledge graph engine (e.g., a Cypher query with multiple \textsf{MATCH} clauses). This query is then executed against the database, and the results are formatted into a standard list of entities for consumption by subsequent operators.

\smallskip\noindent\textbf{\texttt{Retrieve}.} This operator fetches the raw content of specific sections from a document. 

\noindent\textit{\underline{Parameters.}} The operator is controlled by a \textsf{document\_id} and \textsf{section\_tags} (e.g., ``Introduction'', ``Methodology'') to precisely locate and return the required text for downstream processing.

\subsection{Semantic Content Processing}

\noindent\textbf{\texttt{Extract}.} This operator takes raw text content (typically from a \textsf{Retrieve} operator) and extracts specific, structured information from it. It is controlled by an \textsf{instruction} parameter, which provides details to guide LLM's extraction logic, and \textsf{detail\_level}, including \textsf{short}, \textsf{detailed}, or \textsf{detailed with evidence}.
Unlike a simple retrieval function that returns raw text, the  \texttt{Extract} operator interprets a user's query to identify and return precise details, such as experimental settings.

\noindent\textit{\underline{Parameters.}} The operator is invoked with a natural language query, $q$, and requires identifiers for the target document, \textsf{document\_id}, and the specific \textsf{section\_tag}s from which to extract information. Its behavior is controlled by an optional \textsf{extract\_instruction} parameter, which provides a detailed template to guide the LLM's extraction logic and output format. {\sysname} offers a library of predefined plans (See Table~\ref{tab:prompts}) for common scholarly queries, such as extracting experimental settings or problem definitions, while also allowing users to supply custom instructions for their own extraction needs.

\noindent\textit{\underline{Implementation.}} The operator's execution follows a three-stage pipeline:
\begin{enumerate}[leftmargin=*]
     \renewcommand{\labelenumi}{\roman{enumi}.}
\item \emph{Section retrieval}: The operator first retrieves the full text content of the section specified by \textsf{section\_tag}s from the document identified by \textsf{document\_id}.
\item \emph{Prompt construction}: A dynamic prompt is then constructed by integrating the retrieved section content, the user query $q$, and the \textsf{extract\_instruction}. The user query $q$ provides high-level intent, while the \textsf{extract\_instruction} provides a low-level template that directs the LLM on precisely what to extract and how to structure it (e.g., as a JSON object with predefined keys).
\item \emph{LLM generation}: The composed prompt is dispatched to an LLM. The model processes the input and generates a structured response containing the specific information requested by the user.
\end{enumerate}

\smallskip\noindent\textbf{\texttt{Summarize}.} This operator is designed to condense textual information from one or more inputs into a coherent and concise overview. Its behavior is controlled by an \textsf{instruction} parameter, which guides the LLM's summarization logic.
Its primary purpose is to reduce lengthy content, optimize context for subsequent processing, and distill key insights/aspects tailored to a user's specific focus.

\noindent\textit{\underline{Parameters.}} The operator is invoked with the original user query, q, and requires a collection of \textsf{input\_data} from a preceding operator(s) as its primary input. Its summarization behavior is controlled by two main parameters: \textsf{detail\_level} and \textsf{focus}. The \textsf{detail\_level} parameter dictates the output's length and depth, with options for a short overview (\textsf{short}), a detailed summary (\textsf{detailed}), or a summary that includes supporting evidence from the source texts (\textsf{detailed\_with\_evidence}). The {\textsf{focus}} parameter provides a specific thematic instruction that guides the summary. Same as \texttt{Extract}, {\sysname} offers a library of predefined plans (See Table~\ref{tab:prompts}) for common scholarly queries, such as focusing on the ``pros and cons across different methods'', while also allowing users to supply custom instructions to meet their own summarization needs.

\noindent\textit{\underline{Implementation.}}  The implementation is in a straightforward manner: the system constructs an appropriately parameterized prompt over the operator’s input data and issues a single LLM call to obtain the result. In the remainder of this paper, we omit explicit implementation details for operators that are realized in this same manner.

\smallskip\noindent\textbf{\texttt{Check}.} This analytical operator compares information across multiple peer documents to identify and report on inconsistencies or disagreements, guided by \textsf{instruction} parameter. 

\noindent\textit{\underline{Parameters.}} The operator is invoked with the original user query, $q$, and requires a list of two or more contents from  preceding operator(s) as its primary input to serve as the peer sources for comparison. Its behavior is controlled by the \textsf{check\_instruction} parameter, which defines the instructions to guide the checking process.

\smallskip\noindent\textbf{\texttt{Verify}.} This operator fact-checks a specific \textsf{claim} against a designated source of \textsf{evidence}, operating either within a single paper for internal consistency or across papers under the guidance of the \textsf{instruction} parameter.

The \texttt{Verify} operator is an analytical function designed to fact-check a specific claim against a designated source of evidence. It is a versatile tool that can operate both within a single paper to check for internal consistency and across multiple papers to validate citations and reported findings.

\noindent\textit{\underline{Parameters.}} The operator is invoked with the original user query, $q$, and requires two direct string inputs: a \textsf{claim} and the \textsf{evidence} generated from preceding operator(s).

\smallskip\noindent\textbf{\texttt{Rank}.} This operator is a specialized function designed to evaluate and order a set of entities from the \textsf{input\_data} parameter based on a specific, measurable criterion defined in the \textsf{instruction} parameter, producing a structured, ordered list as its primary output.

The \texttt{Rank} operator is a specialized function designed to evaluate and order a set of entities based on a specific, measurable criterion, producing a structured, ordered list as its primary output. It can be used within a single paper, such as ranking variants of a proposed method based on ablation study results, or it can be applied across multiple papers, such as ranking different approaches based on their reported performance on a common benchmark.

\noindent\textit{\underline{Parameters.}} The operator is invoked with the original user query $q$ and a collection of \textsf{input\_data} produced by preceding operators. Its ranking behavior is governed by the parameter \textsf{rank\_instruction}, which may either (i) directly specify the entities to be ranked and the ranking criterion, or (ii) provide instructions on how these entities or criteria should be identified from the input. In addition, users may supply auxiliary aspects to annotate the ranking results, enabling richer comparative outputs, and ranking\_criterion. The entities\_to\_rank parameter specifies the set of items to be ordered, such as ``variants of the proposed method'' or ``datasets used in the same method''. The ranking\_criterion parameter defines the specific metric for the comparison, for instance, ``experimental results on a common dataset and a common evaluation metric''.

\subsection{Relational-Style Data Manipulation}

\smallskip\noindent\textbf{\texttt{GroupBy}.} This operator partitions a list of entities in \textsf{input\_data} based on a specified attribute. Its \textsf{grouping\_key} specifies the attribute or attributes to be used for partitioning the data.

The \texttt{GroupBy} operator takes a list of entities and restructures it into a nested, grouped format based on specified attributes.

\noindent\textit{\underline{Parameters.}} The operator takes a collection of \textsf{input\_data} from a preceding operator as its primary input. Its behavior is controlled by the \textsf{grouping\_key} parameter, which specifies the attribute or attributes to be used for partitioning the data.

\smallskip\noindent\textbf{\texttt{Aggregate}.} 
This operator computes summary statistics over a set of entities from \textsf{input\_data}. The \textsf{instruction} specifies the function (e.g., COUNT, MAX) and the target of the function (e.g., a node's attributes), which may require semantic interpretation.

\noindent\textit{\underline{Parameters.}} The operator takes a collection of \textsf{input\_data} from a preceding operator as its primary input. Its behavior is controlled by the \textsf{aggregation\_instruction} parameter, which is a structured object that specifies the exact calculation to be performed. This instruction defines the \textsf{function} (e.g., COUNT, MAX, AVG), the \textsf{target} of the function (e.g., a node's properties or its relationships), and any required other specifics. 

\smallskip\noindent\textbf{\texttt{Filter}.} 
This operator returns a subset of entities from a collection that satisfies a logical condition. The condition is provided via a \textsf{filter\_instruction} parameter, which is evaluated semantically by an LLM for each entity in the \textsf{input\_data}.

The \texttt{Filter} operator is a selection function to take a list of entities and return a subset that satisfies a specific, logical criterion.

\noindent\textit{\underline{Parameters.}} The operator takes a collection of \textsf{input\_data} from a preceding operator as its primary input. Its behavior is controlled by the \textsf{filter\_instruction} parameter, which describes the logical condition to be applied.

\subsection{Knowledge Generation and Synthesis}

\smallskip\smallskip\noindent\textbf{\texttt{Generate}.} The \texttt{Generate} operator is a versatile, LLM-powered function designed to produce new, coherent text based on the \textsf{input\_data} and a specific \textsf{instruction}.

\noindent\textit{\underline{Parameters.}} The operator takes a collection of \textsf{input\_data} from preceding operators as its primary contextual input. Its behavior is controlled by the \textsf{generation\_instruction} parameter, which is a detailed, task-specific prompt or template.

\smallskip\noindent\textbf{\texttt{MatrixConstruct}.} A specialized operator designed for research idea exploration. It synthesizes information on problems and methods from a research area into a structured matrix, enabling the systematic discovery of under-explored combinations.

\noindent\textit{\underline{Parameters.}} The operator takes a collection of input\_data from preceding operators as its primary input. It automatically parses this input to identify the problem and method taxonomies, mapping problems to matrix rows and methods to matrix columns. This organization forms the grid structure required to identify potential research gaps.

\section{Implementation of Query Processing Pipelines}
 This section details the implementation of query processing pipelines for complex analytical queries (Tier 3), which are constructed using the proposed operators.

\subsection{Research Trend Analysis}
The research trend analysis query processing pipeline integrates taxonomy navigation, external retrieval, statistical aggregation, and LLM-based synthesis to construct a comprehensive view of emerging research dynamics. Specifically, with the \texttt{Traverse} operator, each leaf node in the problem or method taxonomy is first identified to capture its contextual hierarchy within the knowledge graph. The \texttt{Search} operator then issues structured API queries to external scholarly sources, Semantic Scholar, retrieving publications that match the node name or its semantically expanded variants. Retrieved metadata, including publication year and citation count, is subsequently processed by the \texttt{Aggregate} operator to compute temporal distributions and citation growth metrics across recent years. 

Unlike traditional numeric ranking, {\sysname} employs an LLM-centric strategy for trend evaluation. The aggregated statistics are passed directly to the \texttt{Generate} operator, where an LLM interprets the multi-dimensional trend data -- combining publication growth, citation velocity, and topical coherence -- to both \emph{rank} subtopics and \emph{summarize} their significance. This unified semantic reasoning enables a more nuanced prioritization of emerging subfields, highlighting the top-$k$ trends within each topic together with concise, evidence-grounded explanations of what drives their recent prominence.

\smallskip
\subsection{Research Idea Exploration}
The research idea exploration query processing pipeline supports systematic discovery of under-explored research directions by reasoning over the cross-product space of problem and method taxonomies. The \texttt{MatrixConstruct} operator first builds a 2D matrix where each cell corresponds to a possible (method, problem) pairing. The \texttt{Filter} operator examines the knowledge graph to detect existing publications aligned with these pairs and removes those already explored, yielding a candidate set of unexplored intersections.

To evaluate and prioritize these intersections, {\sysname} introduces a \emph{two-stage LLM-driven assessment process}. In the first stage, the LLM performs a lightweight scoring over all unexplored combinations, guided by compact descriptors of the method’s capability, e.g., strengths and weaknesses, and the problem’s characteristics, e.g., the input and output. The LLM estimates the plausibility and potential impact of applying the given method to the task, outputting a normalized relevance score that enables efficient ranking and selection of the top-$k$ promising combinations. In the second stage, each selected pair undergoes a deeper generative reasoning phase: the LLM further produces a structured, insight-rich research proposal outlining motivation, feasibility, novelty, and potential contributions. 

This hierarchical design balances scalability and analytical depth: the first stage provides broad exploration across a large combinatorial space, while the second stage offers focused, high-fidelity synthesis. Collectively, this two-phase pipeline transforms sparse knowledge signals into actionable research hypotheses, supporting a principled and data-driven workflow for identifying novel problem–method intersections within {\sysname}.

\subsection{Milestone Paper Selection}
\label{ap:milestone}
The milestone paper identification processing pipeline focuses on recognizing highly influential works that have significantly shaped a research field. It integrates bibliometric data, taxonomic novelty signals, and temporal impact dynamics to assign a composite milestone score to each candidate paper. Specifically, {\sysname} retrieves papers within the target topic through \texttt{Search} and aggregates relevant metadata—such as publication year, citation history, and taxonomy alignment—via the \texttt{Aggregate} operator. 

Each paper is evaluated across four primary dimensions: (i) \emph{citations}, reflecting normalized citation volume and trajectory; (ii) \emph{problem novelty}, measuring how early and distinctively the paper introduced a new problem formulation in the taxonomy; (iii) \emph{method novelty}, capturing the originality of algorithmic techniques and their contextual innovation; and (iv) \emph{impact}, quantifying both early influence and sustained citation momentum. Additionally, a delayed recognition adjustment identifies ``sleeping beauty" papers that gained influence belatedly and applies an appropriate novelty boost. After normalization, the results are ranked via the \texttt{Rank} operator, and the top-$k$ milestone papers are synthesized by the \texttt{Generate} operator into concise summaries outlining their novelty, influence, and long-term significance.

\begin{table}[t]
\centering
\small
\begin{threeparttable}
\captionsetup{skip=0pt}
\caption{Topics used in our evaluation.}
\label{tab:topic-strata}
\begin{tabular}{p{0.22\linewidth} p{0.72\linewidth}}
\toprule
\textbf{Topic Type} & \textbf{Topics} \\
\midrule
Broad Domains &
AI for Databases \\ \midrule

Specialized Sub-fields &
Query Optimization; Database Design \& Execution; Cross-Modal Retrieval; Dataset Discovery; Entity Resolution; Health IR; Learned Indexes; Recommendation; Time-Series Forecasting; Time-Series Similarity Search; Trajectory Search \\ \midrule

Emerging Areas &
Vector Search; LLMs for Structured Data Analysis \\
\bottomrule
\end{tabular}
\end{threeparttable}
\end{table}

\section{Selected Topics for Experiments}
\label{app:topics}

Our experimental design adopts a stratified topic selection to ensure a comprehensive evaluation of \sysname’s versatility across diverse scholarly domains. As shown in Table~\ref{tab:topic-strata}, we categorize topics into three types: (1) \emph{broad domains}, representing overarching areas that encompass multiple sub-fields; (2) \emph{specialized sub-fields}, focusing on specific technical problems within a domain; and (3) \emph{emerging areas}, capturing rapidly evolving research frontiers. This stratification enables coverage from large, foundational domains to focused, mature sub-fields. Importantly, including emerging areas allows us to assess \sysname’s ability to reason over novel, corpus-specific information that may not yet be well-represented in an LLM’s internal knowledge.

We tailor the number of topics to each evaluation’s objective. For foundational retrieval (Tier 1), we use all 10 specialized sub-field topics (see Table~\ref{tab:search_prec_per_topic} for per-topic results) to ensure robust benchmarking. For information extraction and synthesis (Tier 2) and knowledge discovery and generation (Tier 3), we use a smaller but more diverse set of five topics spanning all three categories. This setup rigorously tests the system’s generality and reasoning ability across the full spectrum of scholarly domains. Specifically, the \emph{broad domain} includes AI for Databases; the \emph{specialized sub-fields} comprise Query Optimization and Database Design and Execution; and the \emph{emerging areas} cover Vector Search and LLMs for Structured Data Analysis, reflecting the growing intersection between database systems and modern AI techniques.

\begin{figure*}[t]
    \centering
    \includegraphics[width=0.4\textwidth]{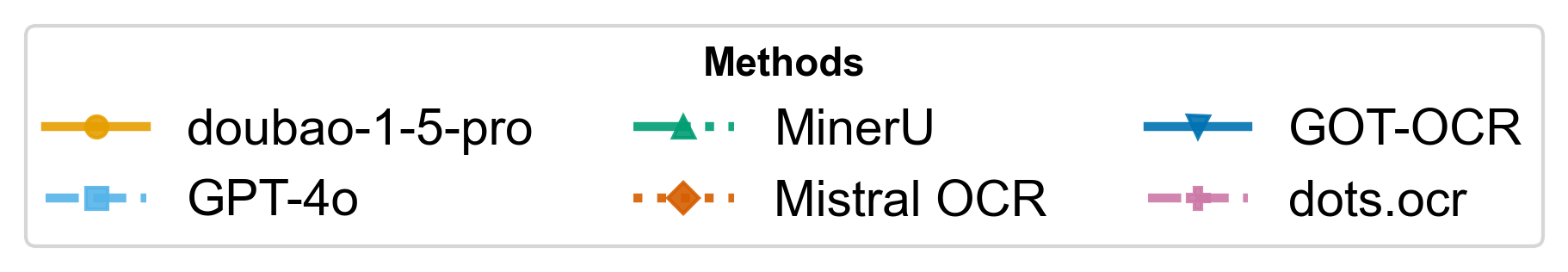}

    \vspace{0.6em}

    \begin{minipage}[t]{0.48\textwidth}
        \centering
        \includegraphics[width=\linewidth]{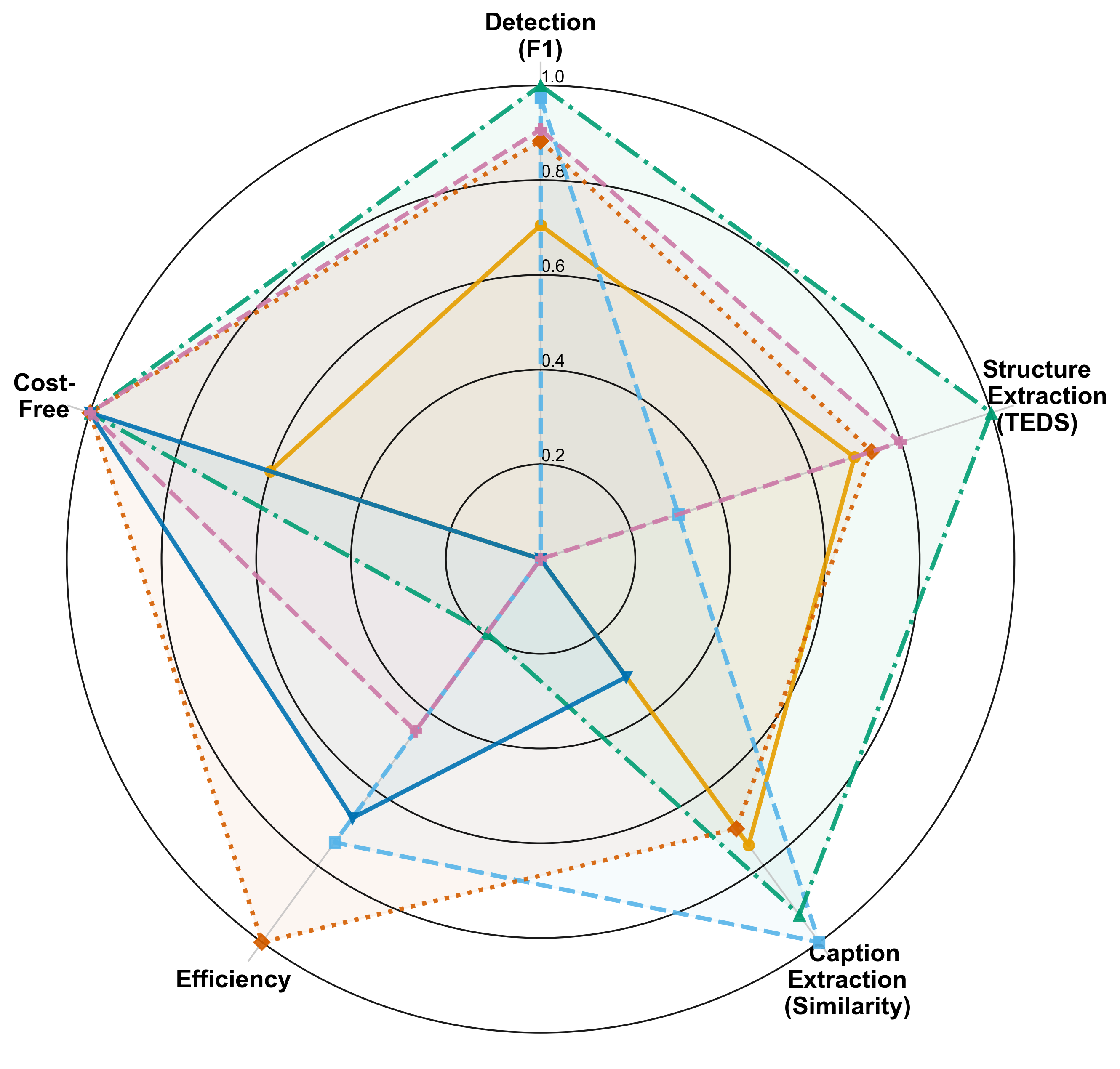}
        \caption*{\small (a) Table identification and extraction.}
    \end{minipage}\hfill
    \begin{minipage}[t]{0.48\textwidth}
        \centering
        \includegraphics[width=\linewidth]{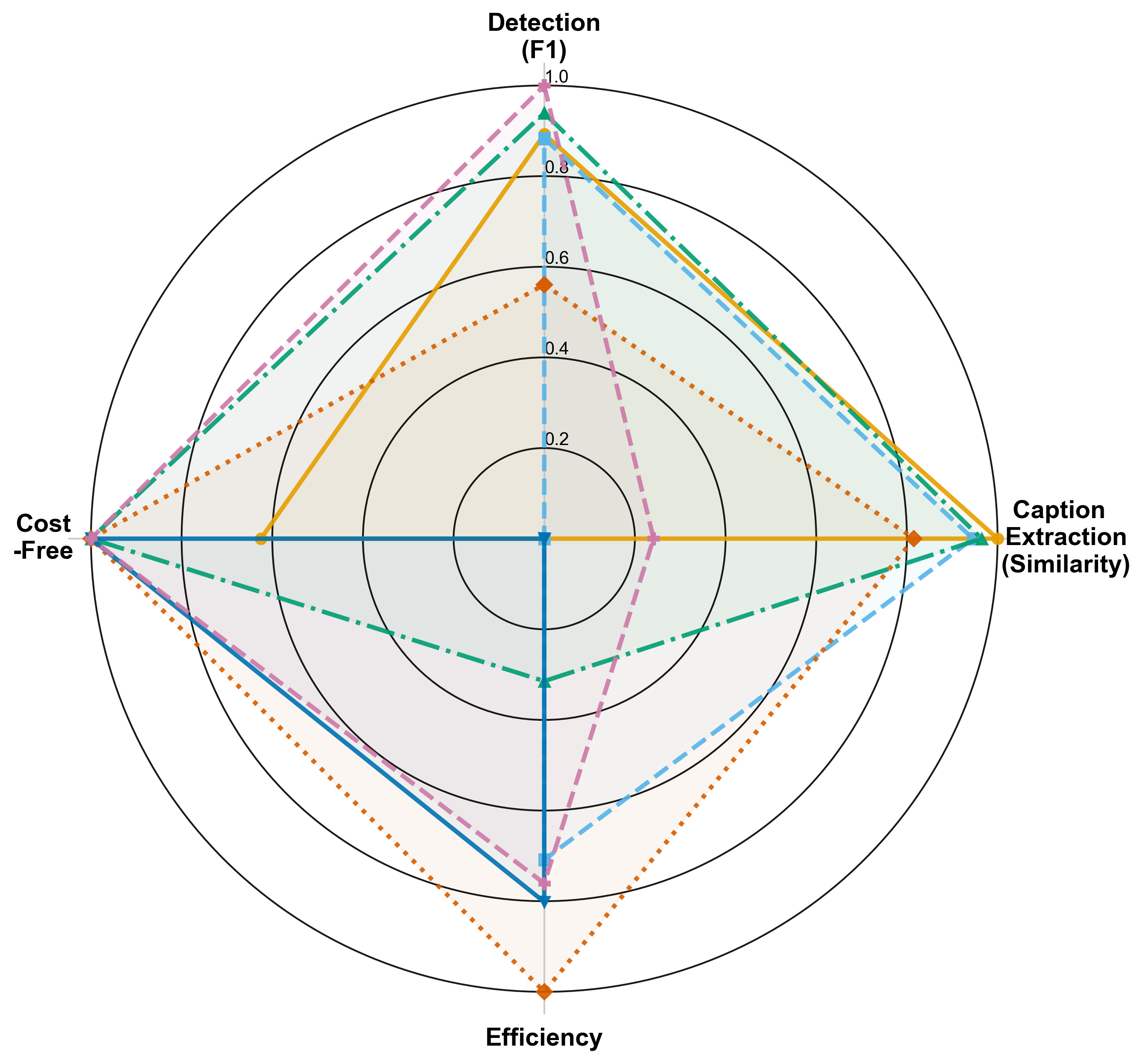}
        \caption*{\small (b) Figure identification and extraction.}
    \end{minipage}

    \caption{Comparison of document parsing tools across effectiveness, efficiency, and monetary cost. ``Cost-Free'' denotes the normalized monetary-cost axis (higher is better, with free tools achieving the maximum score).}
    \label{appendix_doc_parse_radar}
\end{figure*}

\section{Evaluation of Document Parsing Tools} \label{appendix_eval_doc_parsing}

In this section, we evaluate different document parsing tools based on their \emph{effectiveness} in correctly identifying and extracting tables/figures and their \emph{efficiency} and \emph{monetary cost}  in processing individual paper PDFs.

\subsection{Experimental Setup} 
\noindent\textbf{Dataset.}
We use 94 research papers from the \texttt{academic\_literature} split of OmniDocBench~\cite{DBLP:conf/cvpr/OuyangQZZZL0ZJ025} that contain at least one table or figure. For each paper, OmniDocBench provides a sampled page with tables/figures and fine-grained page-level ground-truth annotations, including: (i) table bounding boxes, table HTML, and table attributes (e.g., layout type, cell spans, formulas); (ii) figure bounding boxes; and (iii) captions and footnotes for both tables and figures. We treat these annotations as ground truth.

\noindent\textbf{Compared Methods.}
We compare representative tools from three categories: (i) a pipeline-style document analysis tool, MinerU~\cite{mineru}; (ii) specialized OCR tools, Mistral OCR~\cite{Mistral}, GOT-OCR \cite{DBLP:journals/corr/abs-2409-01704}, and dots.ocr \cite{li2025dotsocrmultilingualdocumentlayout}; and (iii) general-purpose VLMs, doubao-1.5-thinking-vision-pro (doubao-1-5-pro) \cite{guo2025seed1} and GPT-4o \cite{hurst2024gpt}.

\noindent\textbf{Implementation.}
Pipeline/OCR tools take the raw PDF page as input, whereas VLMs take the rendered page image from the PDF page.
All outputs are normalized into a structured format that contains, for each detected element, (i) the table/figure region (bounding box), (ii) table HTML  capturing structure and cell content, and (iii) captions and footnotes.

\noindent\textbf{Evaluation Metrics.}
Following OmniDocBench~\cite{DBLP:conf/cvpr/OuyangQZZZL0ZJ025}, we assess \textit{effectiveness} for table/figure extraction via both detection and extraction quality: 
(i) \textit{Detection:} we compute table/figure F1 by matching predictions to ground-truth instances with IoU $\ge 0.5$;
(ii) \textit{Structure and Caption Extraction:} for matched elements, we evaluate table structure extraction quality using TEDS (Tree-Edit-Distance–based Similarity)  between predicted and ground-truth HTML, and caption extraction quality for both tables and figures using embedding-based cosine similarity between predicted and ground-truth captions.
For \textit{efficiency}, we report the average runtime per paper page. For API-based methods, we additionally report the average \textit{monetary cost} per paper page based on logged token usage and provider pricing.

\subsection{Evaluation Results}
Figure~\ref{appendix_doc_parse_radar} compares document parsing tools in terms of effectiveness, efficiency, and monetary cost, based on a normalized radar chart (higher is better on all axes).
Specifically, for effectiveness metrics, we use standard min--max scaling, $\mathrm{Score}(x)=\frac{x-x_{\min}}{x_{\max}-x_{\min}}$; for efficiency and monetary cost, we use the inverted form, $\mathrm{Score}(x)=\frac{x_{\max}-x}{x_{\max}-x_{\min}}$, so lower latency and lower cost yield higher scores.
From the results, we can observe that MinerU achieves the strongest overall effectiveness for identifying and extracting tables and figures, but at a noticeably higher runtime, whereas Mistral OCR offers the best efficiency--effectiveness trade-off, delivering competitive extraction quality with substantially lower latency.
In contrast, general-purpose VLMs (e.g., GPT-4o and doubao-1-5-pro) incur non-trivial API costs without providing proportional effectiveness gains. 
Based on these observations, we adopt a hybrid preprocessing strategy: we use the \textit{Mistral OCR API} for rapid, low-latency conversion and deploy \textit{MinerU} as the default backend when high-fidelity, open-source table/figure extraction is required.

\begin{figure*}[t]
\centering

\begin{evbox}
You are an expert assistant for generating evaluation data for a paper search system.
You are given a set of papers with structured attributes.
Your task is to generate 20 realistic natural language queries that a human researcher might ask to find these papers.

## Attributes Available for Query Construction: 
Metadata
- "title": Full title of the paper.
- "authors": List of author names.
- "venue": Institutions or organizations corresponding to the authors.
- "publication_year": Year of publication.
- "venue": The conference, workshop, or journal where the paper appeared.

Aspects
- "research_topic": The broad area or domain of study (e.g., "time series similarity search", "graph neural networks").
- "problem_formulation": The precise problem formulation addressed by the paper, including how it is defined or formalized.
- "proposed_method": The main algorithm, model, or framework introduced by the paper, including its name (if available) and a brief description of its core idea or mechanism.
- "experimental_datasets": Datasets used in experiments.
- "experimental_baselines": Methods, models, or algorithms that the paper explicitly compares against in experiments (excluding ablations or internal variants).
- "experimental_results": A concise but specific summary of the reported outcomes, including metrics and improvements where available.

## Query Requirements:
1. Natural phrasing: Queries must be expressed in fluent, researcher-style language (not keyword lists, Boolean operators, or artificial phrasing).
2. Aspect focus: At least 80\% of queries must be centered on aspects (problem_formulation, proposed_method, datasets, baselines, results, research_topic). Metadata (title, authors, affiliations, year, venue) may be included but primarily as secondary constraints.  
3. Ground truth size: Each query must correspond to at least one paper, with an ideal ground truth set of 2-10 matching papers. Do not generate queries with no valid matches or queries that would match nearly all papers.
4. Query type diversity: Queries must span three distinct types:  
    - Single-attribute queries (4 queries): Target exactly one attribute. At least 3 should focus on aspects (e.g., "Which papers use the UCR dataset?", "Papers that use attention mechanisms in graph neural networks"). Metadata-only queries are rare but allowed.  
    - Multi-attribute queries (10 queries): Combine multiple attributes across aspects and optionally metadata (e.g., "Find papers that introduce transformer-based models for time series forecasting on UCR datasets", "Which papers published after 2019 focus on the reverse kNN problem in trajectory search?", "Papers using learned indexes for time series similarity search that evaluate on both synthetic and real-world datasets").
    - Comparative queries (6 queries): Require relative reasoning about performance, baselines, datasets, or results. These should highlight comparisons (e.g., "Which papers outperform DTW on UCR datasets in runtime?", "Papers comparing against DeepMatcher and achieving higher F1 on Wiki dataset", "Which papers report better recall than BERT-based methods for schema matching?").  
5. Attribute coverage: Collectively, the 20 queries must cover every attribute at least once across metadata and aspects.  
6. No trivial queries: Avoid overly broad queries ("papers on machine learning") or impossible queries with no matching papers.  
7. Realism: Queries must reflect how an actual researcher would search for papers, especially focusing on problem formulations, proposed methods, datasets, baselines, and results. Time- or venue-based filters can be added for realism but should not dominate.  
8. Diversity of wording: Vary query style and phrasing (e.g., "Which papers...", "Show me papers that...", "Find papers that..."), and avoid repetitive templates.
                                         
## Output Format
Return results strictly as a JSON list of objects with the following schema:
[
  {
    "query_id": "1",
    "query": "<natural language query>",
    "ground_truth": ["<paper id>", "<paper id>", ...],
    "attributes": ["<attribute 1>", "<attribute 2>", ...],
    "query_type": "<single-attribute factual | multi-attribute factual  | in-topic comparative>"
  },
  {
    "query_id": "2",
    "query": "<natural language query>",
    "ground_truth": ["<paper id>", "<paper id>", ...],
    "attributes": ["<attribute 1>", "<attribute 2>", ...],
    "query_type": "<single-attribute factual | multi-attribute factual  | in-topic comparative>"
  }
]
\end{evbox}
\vspace{-5mm}
\caption{Prompt template for generating factual and in-topic analytical queries.}
\label{fig:search_prompt1}
\end{figure*}

\section{Supplements for Evaluation of Retrieval}

\subsection{Supplements for Query Generation}
\label{app:retrieval_setup}

\noindent\textbf{Query Types}. 
We consider two categories of queries for evaluation:
\begin{itemize}[leftmargin=*]
\item \emph{Factual queries} target explicit facts concerning query attributes, such as paper metadata or specific research aspects, including: 
(i)~\emph{single-attribute} queries (e.g., ``which papers use the UCR data\-set?'') and 
(ii)~\emph{multi-attribute} queries (e.g., ``which SIGMOD papers propose transformer-based models for time series forecasting after 2020?'').
\item \emph{Analytical queries} involve comparative or reasoning-based understanding including:
(i)~\emph{in-topic comparative} queries (e.g., ``which papers achieve higher accuracy than DTW on time series forecasting?'') and 
(ii)~\emph{cross-topic comparative} queries spanning multiple topics (e.g., ``which papers evaluate transformer-based methods on tasks beyond natural language processing?'').
\end{itemize}
 
\smallskip
\noindent\textbf{Query and Ground Truth Generation Prompts}. 
We design two structured prompt templates (Figures~\ref{fig:search_prompt1} and~\ref{fig:search_prompt2}) to generate natural language queries and identify corresponding ground truth sets from the paper corpus. Each prompt is conditioned on summaries of paper attributes, including metadata (e.g., title, authors, publication year, etc.) and research aspects (e.g., research topic, proposed method, experimental datasets, etc.). The first template constructs factual queries and analytical queries confined to a single research topic, while the second focuses on analytical queries that generalize across topics. Both templates instruct the LLM to output (query, ground truth) pairs, which are subsequently verified for correctness and completeness.

\begin{figure*}[t]
\centering
\begin{evbox}
You are an expert assistant for generating evaluation data for a paper search system. You are given a set of papers with structured attributes. Your task is to generate 20 realistic natural language queries that a human researcher might ask to find these papers.
## Attributes Available for Query Construction:                         
Metadata
- "title": Full title of the paper.
- "authors": List of author names.
- "affiliations": Institutions or organizations corresponding to the authors.
- "publication_year": Year of publication.
- "venue": The conference, workshop, or journal where the paper appeared.
Aspects
- "research_topic": The broad area or domain of study (e.g., "time series similarity search", "graph neural networks").
- "problem_formulation": The precise problem formulation addressed by the paper, including how it is defined or formalized.
- "proposed_method": The main algorithm, model, or framework introduced by the paper, including its name (if available) and a brief description of its core idea or mechanism.
- "experimental_datasets": Datasets used in experiments.
- "experimental_baselines": Methods, models, or algorithms that the paper explicitly compares against in experiments (excluding ablations or internal variants).
- "experimental_results": A concise but specific summary of the reported outcomes, including metrics and improvements.
## Query Requirements:
1. Cross-topic exploratory intent: All 20 queries must explore how methods, datasets, or baselines transfer or compare across different research topics. Queries should sound like a researcher trying to understand connections between broad areas, not like rigid database filters. Examples:
    - "Which papers have Transformer-based methods that have been applied to both time-series and graph data?"
    - "Show me papers where contrastive learning is used in both recommendation and retrieval tasks."
    - "Find papers that evaluate on ImageNet and also apply similar architectures to text or tabular data."
    - "Which papers apply graph neural networks across different domains?"
2. Aspect focus: Queries should primarily involve research aspects - especially proposed_method, experimental_datasets, experimental_baselines, and experimental_results. Metadata (authors, venue, year) may appear but should only act as secondary constraints.  
3. Attribute combinations:
    - Most queries (~80\%) should combine two broad attributes, such as:
        - proposed_method + research_topic
        - experimental_datasets + research_topic
        - experimental_baselines + problem_formulation
    - The remaining (~20\%) can include three attributes for richer comparisons, e.g.:
        - proposed_method + experimental_datasets + experimental_results
    - Avoid using more than three core attributes per query.
4. Ground truth size: Each query must correspond to at least one paper, with an ideal ground truth set of 2-8 papers. Avoid queries that would return either no papers or nearly all papers.  
5. Attribute coverage: Collectively, the 20 queries should involve all aspect attributes (research_topic, problem_formulation, proposed_method, experimental_datasets, experimental_baselines, experimental_results) at least once.  
6. Level of specificity: Keep each query general enough to sound natural and return multiple relevant results.  
    - Avoid over-constraining with unnecessary filters like year, venue, or exact numeric metrics.  
    - Focus on semantic curiosity rather than structured filtering.
    - Avoid: "Find 2021 VLDB papers that outperform BERT by over 10\% F1 on the WikiTables dataset."
    - Prefer: "Which methods outperform BERT across different text-related tasks?"
7. Realism: Queries must sound like authentic research questions a scholar would ask, not like artificial database filters. They should capture meaningful comparisons that help researchers find papers that identify methods, results, or datasets across different domains.
8. Diversity of wording: Use varied query phrasing styles (e.g., "Which papers...", "Show me papers that...", "Find papers that...") to avoid repetitive templates.  
## Output Format
Return results strictly as a JSON list of objects with the following schema:
[
  {
    "query_id": "1",
    "query": "<natural language query>",
    "ground_truth": ["<paper id>", "<paper id>", ...],
    "attributes": ["<attribute 1>", "<attribute 2>", ...],
    "query_type": "cross-topic comparative"
  },
  {
    "query_id": "2",
    "query": "<natural language query>",
    "ground_truth": ["<paper id>", "<paper id>", ...],
    "attributes": ["<attribute 1>", "<attribute 2>", ...],
    "query_type": "cross-topic comparative"
  }
]
\end{evbox}
\vspace{-5mm}
\caption{Prompt template for generating cross-topic analytical queries.}
\label{fig:search_prompt2}
\end{figure*}

\subsection{Evaluation Results}\label{app:search_eval_results}

Table~\ref{tab:search_prec_per_topic} reports per-topic R-Precision and MAP across all retrieval models. {\sysname} consistently outperforms both lexical and dense baselines across all domains. The improvement is consistent even in analytically demanding queries (i.e., \emph{Cross-Topic Comparative}), demonstrating that {\sysname} generalizes effectively across heterogeneous research topics.

\setlength{\textfloatsep}{0pt}
\begin{table*}[t]
    \setlength{\tabcolsep}{3pt}
    \setlength{\abovecaptionskip}{0cm}
    \setlength{\belowcaptionskip}{0cm}
    \caption{Per-topic effectiveness across models.}\label{tab:search_prec_per_topic}
    \centering
    \small
    \begin{tabular}{ccccccccccccccccc}
        \toprule
        \multirow{2}{*}{Topic} & \multicolumn{2}{c}{BM25} & \multicolumn{2}{c}{GTR} & \multicolumn{2}{c}{E5} & \multicolumn{2}{c}{Instructor} & \multicolumn{2}{c}{Emb.Gm} & \multicolumn{2}{c}{ColBERT} & \multicolumn{2}{c}{Qwen3} & \multicolumn{2}{c}{\textbf{Ours}} \\ 
        \cmidrule(lr){2-3} \cmidrule(lr){4-5} \cmidrule(lr){6-7} \cmidrule(lr){8-9} \cmidrule(lr){10-11} \cmidrule(lr){12-13} \cmidrule(lr){14-15} \cmidrule(lr){16-17}
        & R-P & MAP & R-P & MAP & R-P & MAP & R-P & MAP & R-P & MAP & R-P & MAP & R-P & MAP & R-P & MAP \\ 
        \midrule
        Cross-Modal Ret.   & 0.50 & 0.56 & 0.51 & 0.54 & 0.47 & 0.55 & 0.51 & 0.51 & 0.45 & 0.47 & 0.60 & 0.61 & \underline{0.67} & \underline{0.69} & \textbf{0.80} & \textbf{0.81} \\ 
        Dataset Discovery       & 0.54 & 0.59 & 0.43 & 0.43 & 0.49 & 0.49 & 0.47 & 0.48 & 0.52 & 0.55 & 0.55 & 0.57 & \underline{0.56} & \underline{0.63} & \textbf{0.86} & \textbf{0.88} \\ 
        Entity Resolution       & 0.45 & 0.47 & 0.36 & 0.35 & 0.39 & 0.40 & 0.42 & 0.46 & 0.40 & 0.41 & 0.44 & 0.45 & \underline{0.48} & \underline{0.49} & \textbf{0.60} & \textbf{0.62} \\ 
        Health IR               & 0.68 & 0.71 & 0.45 & 0.50 & 0.62 & 0.67 & 0.50 & 0.54 & 0.60 & 0.63 & 0.69 & 0.71 & \underline{0.74} & \underline{0.77} & \textbf{0.85} & \textbf{0.90} \\
        Learned Index           & 0.50 & 0.48 & 0.28 & 0.31 & 0.54 & 0.52 & 0.50 & 0.44 & 0.42 & 0.41 & \underline{0.62} & 0.59 & 0.58 & \underline{0.60} & \textbf{0.71} & \textbf{0.72} \\ 
        Recommendation          & 0.35 & 0.43 & 0.45 & 0.48 & 0.57 & 0.57 & 0.51 & 0.52 & 0.46 & 0.47 & 0.56 & 0.56 & \underline{0.62} & \underline{0.64} & \textbf{0.66} & \textbf{0.70} \\ 
        TS Forecasting & 0.45 & 0.43 & 0.48 & 0.53 & 0.49 & 0.51 & 0.52 & 0.54 & 0.45 & 0.44 & 0.57 & \underline{0.58} & \underline{0.58} & \underline{0.58} & \textbf{0.74} & \textbf{0.78} \\ 
        TS Sim. Search & 0.41 & 0.44 & 0.23 & 0.27 & 0.30 & 0.38 & 0.35 & 0.40 & 0.39 & 0.41 & 0.43 & 0.45 & \underline{0.48} & \underline{0.51} & \textbf{0.55} & \textbf{0.60} \\ 
        Trajectory Search       & 0.37 & 0.52 & 0.35 & 0.44 & 0.42 & 0.53 & 0.57 & 0.62 & 0.50 & 0.53 & 0.61 & 0.65 & \underline{0.61} & \underline{0.69} & \textbf{0.79} & \textbf{0.77} \\ 
        Vector Search           & 0.37 & 0.39 & 0.30 & 0.34 & 0.39 & 0.40 & 0.30 & 0.33 & 0.36 & 0.35 & 0.49 & 0.49 & \underline{0.50} & \underline{0.51} & \textbf{0.68} & \textbf{0.64} \\ 
        \emph{Cross-Topic Comp.}& 0.46 & 0.35 & 0.45 & 0.39 & \underline{0.52} & \underline{0.46} & \underline{0.52} & 0.43 & 0.41 & 0.38 & 0.38 & 0.35 & 0.44 & 0.44 & \textbf{0.56} & \textbf{0.51} \\ 
        \hline
        \textbf{Mean}           & 0.46 & 0.50 & 0.39 & 0.42 & 0.47 & 0.50 & 0.47 & 0.48 & 0.45 & 0.46 & 0.53 & 0.53 & \underline{0.57} & \underline{0.59} & \textbf{0.71} & \textbf{0.72} \\ 
        \bottomrule
    \end{tabular}
\end{table*}

\begin{figure*}[t]
\setlength{\abovecaptionskip}{0cm}
\setlength{\belowcaptionskip}{0cm}
\centering
\subfigure[Input Tokens (Time Series Forecasting)]{\label{subfig:TF_input} \includegraphics[height=23mm, width=28mm]{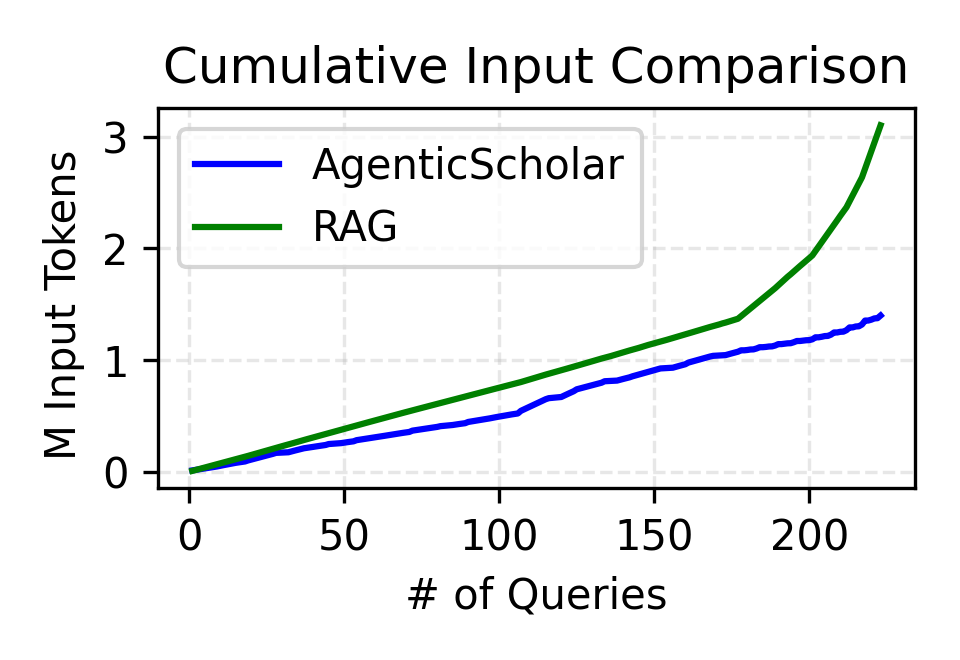}}
\subfigure[Output Tokens (Time Series Forecasting)]{\label{subfig:TF_output} \includegraphics[height=23mm, width=28mm]{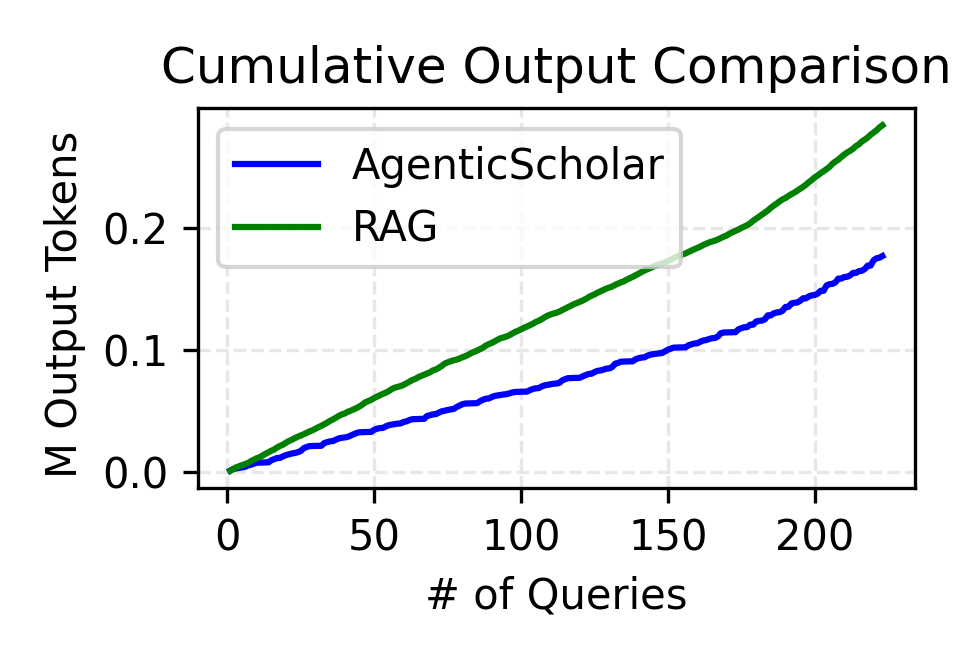}}
\subfigure[Response Time (Time Series Forecasting)]{\label{subfig:TF_time} \includegraphics[height=23mm, width=28mm]{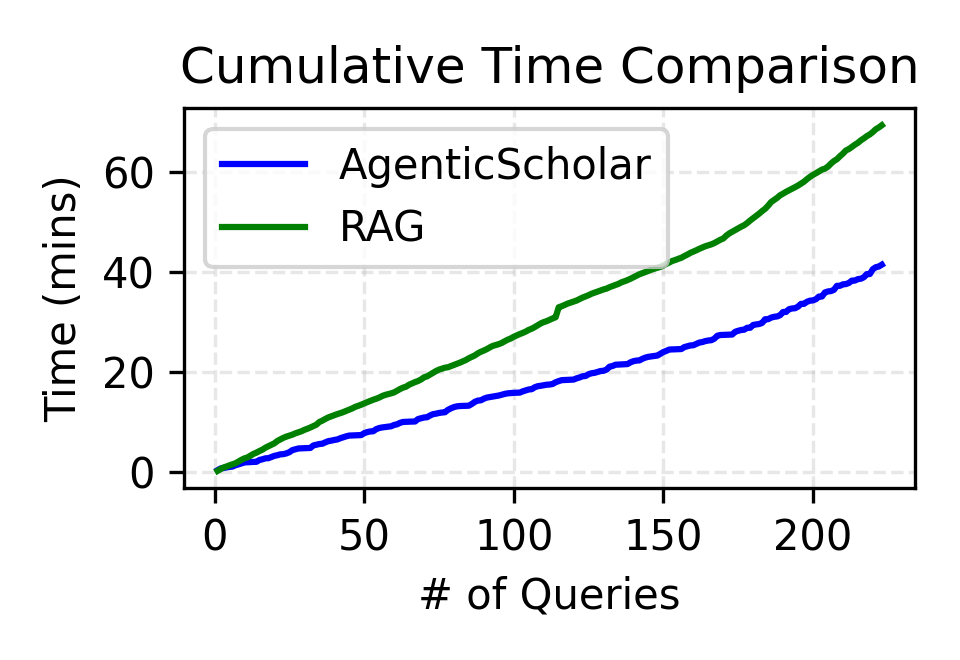}} 
\subfigure[Input Tokens (Vector Search)]{\label{subfig:VS_input} \includegraphics[height=23mm, width=28mm]{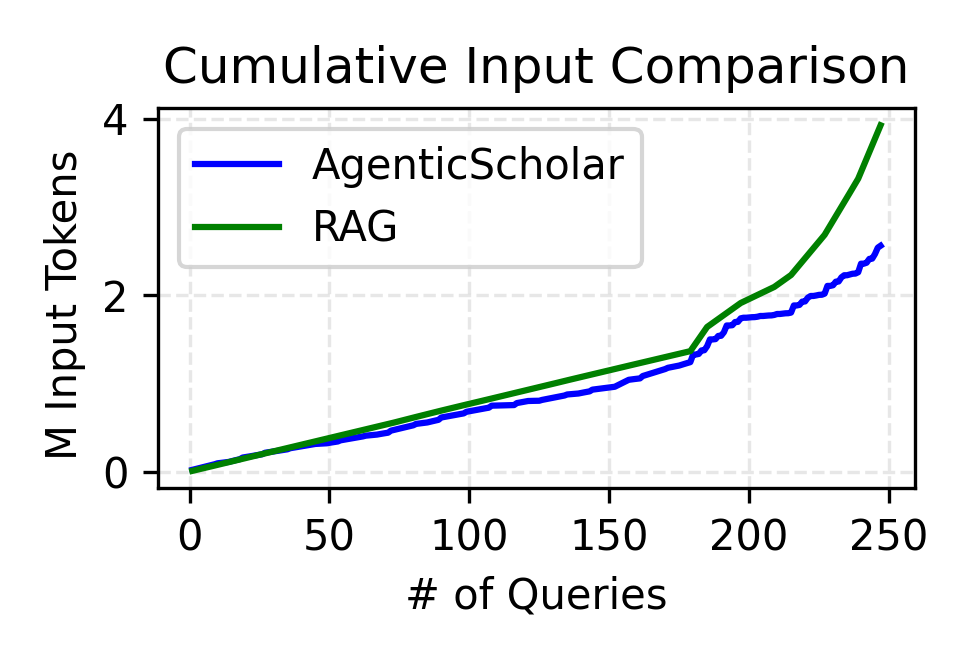}}
\subfigure[Output Tokens (Vector Search)]{\label{subfig:VS_output} \includegraphics[height=23mm, width=28mm]{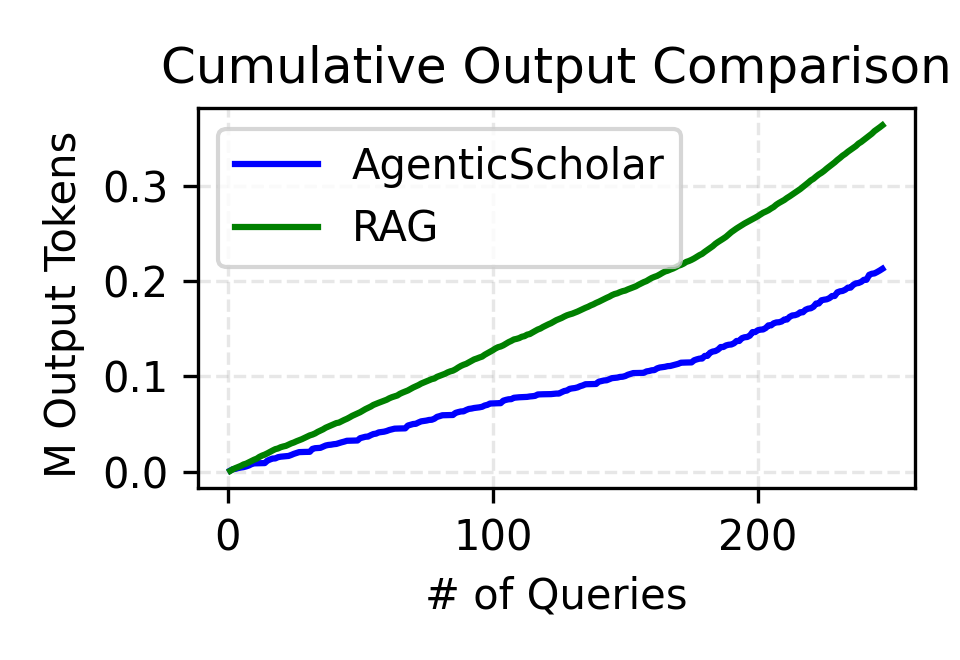}}
\subfigure[Response Time (Vector Search)]{\label{subfig:VS_time} \includegraphics[height=23mm, width=28mm]{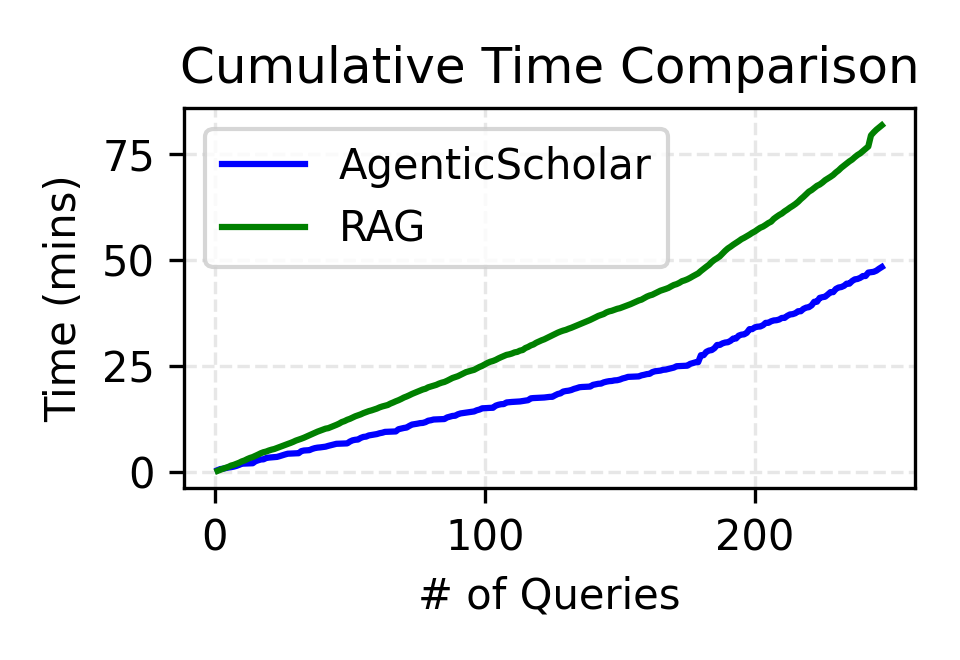}}
\caption{\label{fig:qa1}{Additional results of Tier-2 queries.}
}
\end{figure*}

\section{Supplements for Evaluation of Information Extraction and
Synthesis}\label{sec:app_qa}

Figure~\ref{fig:qa1} shows additional results of Tier-2 queries.

\section{Supplements for Evaluation of Knowledge Discovery and
Generation} \label{app:gen_query_kb}
\noindent\textbf{Query Generation.}  
For each topic, we design three concise prompt types to form the query, covering distinct Tier-3 objectives: research trend analysis, research idea exploration, and milestone paper selection, as presented in Figures~\ref{fig:problem_oriented_research_trend_analysis}, \ref{fig:method_oriented_research_trend_analysis}, \ref{fig:research_idea_generation_prompt}, and \ref{fig:milestone_paper_selection_prompt}.

\begin{itemize}[leftmargin=*,noitemsep]
    \item \textbf{Milestone Paper Selection:} Identifies landmark works since a given year, emphasizing problem or method novelty and long-term impact through influence or citations.
    \item \textbf{Research Trend Analysis:} Analyzes the most recent 2--3 years of literature to reveal emerging \textit{problems} and \textit{methods} within a topic, explaining their motivation, significance, and driving factors such as new models, applications, or environments.
    \item \textbf{Research Idea Exploration:} Encourages creative brainstorming of new research questions, summarizing each idea’s core concept, novelty, and key technical challenge.
\end{itemize}

\begin{figure*}[t]
\begin{evbox}
I’m exploring current research in computer science. Please analyze the most recent 2–3 years of literature in top conferences and identify the top 10 emerging research trends in the problems being addressed within the topic of {Input Topic}. For each trend, briefly describe what specific problem or subproblem it focuses on, why it matters (e.g., practical bottlenecks, new application needs, scalability or reliability gaps), and what is driving current progress (e.g., new environments, workloads, constraints, or user demands). Make sure the trends are distinct, specific, and evidence-based—avoid generic statements. Focus on problem formulations, tasks, and objectives rather than solution techniques.
\end{evbox}
\caption{The prompt template for problem-oriented research trend analysis query}\label{fig:problem_oriented_research_trend_analysis}
\end{figure*}

\begin{figure*}[t]
\begin{evbox}
I’m exploring current research in computer science. Please analyze the most recent 2–3 years of literature in top conferences and identify the top 10 emerging research trends in the techniques employed within the topic of {Input Topic}. For each trend, briefly describe what approach, framework, or algorithmic technique it focuses on, why it matters (e.g., efficiency, interpretability, adaptability, scalability), and what is driving current progress (e.g., new models, optimization paradigms, hardware accelerators, or integration with foundation models). Make sure the trends are distinct, specific, and evidence-based—avoid generic statements. Focus on methods, architectures, and enabling techniques rather than problem formulations.
\end{evbox}
\caption{The prompt template for method-oriented research trend analysis query}\label{fig:method_oriented_research_trend_analysis}
\end{figure*}

\begin{figure*}[t]
\begin{evbox}
Please act as a creative research advisor and brainstorm 10 promising research directions stemming from the following research topic: query optimization. 
For each direction, please frame it as a research question and briefly describe: 
The Core Idea: The main technical or conceptual approach. The Novelty: What makes this angle new or different from existing work? 
The Key Challenge: The main difficulty that would need to be overcome.
\end{evbox}
\caption{The prompt template for research idea exploration query.}\label{fig:research_idea_generation_prompt}
\end{figure*}
\begin{figure*}[t]
\begin{evbox}
Please act as a creative research advisor and brainstorm 10 promising research directions stemming from the following research topic: query optimization. 
For each direction, please frame it as a research question and briefly describe: 
The Core Idea: The main technical or conceptual approach. The Novelty: What makes this angle new or different from existing work? 
The Key Challenge: The main difficulty that would need to be overcome.
\end{evbox}
\caption{The prompt template for milestone paper selection query.}\label{fig:milestone_paper_selection_prompt}
\end{figure*}

\begin{table*}[t]
\centering
\small
\caption{Survey papers used for benchmark construction}\label{tab:survey_paper}
\begin{tabular}{llcl}
\hline
\textbf{Paper Title} & \textbf{First Author et al.} & \textbf{Year} & \textbf{Domain} \\ \hline

Instruction Tuning for Large Language Models: A Survey
& Zhang et al.  & 2025 & \multirow{3}{*}{ML} \\
A Primer on Temporal Graph Learning
& Rahman et al. & 2025 & \\
Multi-Step Reasoning with Large Language Models, a Survey
& Plaat et al.  & 2025 & \\ \hline

Indexing Techniques for Graph Reachability Queries
& Zhang et al. & 2025 & \multirow{2}{*}{DB} \\
A Survey of Learned Indexes for the Multi-dimensional Space
& Al-Mamun et al. & 2025 & \\ \hline

Review of Explainable Graph-Based Recommender Systems
& Markchom et al. & 2025 & IR \\ \hline

Machine Learning Systems: A Survey from a Data-Oriented Perspective
& Cabrera et al. & 2025 & SE \\ \hline

\end{tabular}
\end{table*}

\setlength{\textfloatsep}{0pt}
\begin{table}[t]
\setlength{\tabcolsep}{3pt}
\setlength{\abovecaptionskip}{0cm}
\setlength{\belowcaptionskip}{0cm}
\centering
\caption{Evaluation on taxonomies (Ass.: Assignment Accuracy; Hie.: Hierarchical Consistency; Sib.: Sibling Coherence).}
\label{tab:taxo-metrics}
\small
\begin{threeparttable}
\begin{tabular}{cccccccc}
\toprule
\multirow{2}{*}{\centering Level} & \multirow{2}{*}{\centering Methods} &
\multicolumn{3}{c}{Problem taxonomy} & \multicolumn{3}{c}{Method taxonomy} \\
\cmidrule(lr){3-5}\cmidrule(lr){6-8}
& & Ass. & Hie. & Sib. & Ass. & Hie. & Sib. \\
\midrule
\multirow{4}{*}{1}
  & TaxoAdapt        & 0.76 & 0.41 & 0.25 & 0.87 & 0.61 & 0.25 \\
  & \sysnameR{}      & 0.76 & 0.96 & \textbf{1.00} & 0.74 & 0.76 & 0.00 \\
  & \sysnameC{}      & \textbf{0.90} & 0.77 & 0.75 & \textbf{0.93} & 0.75 & 0.50 \\
  & \textbf{\sysname{}}           & 0.84 & \textbf{0.97} & \textbf{1.00} & 0.90 & \textbf{0.87} & \textbf{1.00} \\
\midrule
\multirow{4}{*}{2}
  & TaxoAdapt        & \textbf{0.85} & 0.72 & 0.29 & \textbf{0.95} & 0.80 & 0.79 \\
  & \sysnameR{}      & 0.66 & \textbf{0.98} & \textbf{1.00} & 0.79 & 0.76 & 0.76 \\
  & \sysnameC{}      & 0.84 & 0.94 & \textbf{1.00} & 0.92 & 0.76 & 0.96 \\
  & \textbf{\sysname{}}           & 0.81 & 0.97 & \textbf{1.00} & 0.87 & \textbf{0.82} & \textbf{0.97} \\
\midrule
\multirow{4}{*}{3}
  & TaxoAdapt        & \textbf{0.95} & 0.67 & 0.67 & 0.86 & 0.63 & 0.60 \\
  & \sysnameR{}      & 0.47 & 0.85 & 0.90 & 0.78 & \textbf{0.80} & 0.91 \\
  & \sysnameC{}\tnote{$\dagger$} & --   & --   & --   & --   & --   & --   \\
  & \textbf{\sysname{}}           & 0.80 & \textbf{0.92} & \textbf{1.00} & \textbf{0.94} & 0.77 & \textbf{0.92} \\
\bottomrule
\end{tabular}
\begin{tablenotes}[flushleft]
\footnotesize
\item[$\dagger$] Since \sysnameC{} typically yields shallow structures and no level-3 nodes, its level-3 results are omitted.
\end{tablenotes}
\end{threeparttable}
\end{table}

\section{Supplements for Validating Architectural Design}

\subsection{Quality Evaluation of Taxonomies}

\subsubsection{Benchmark Construction For Quantitative Analysis}
\label{ap:kb_benchmark}
Following~\cite{zhu2025context}, we construct the ground-truth taxonomy by manually extracting expert-curated hierarchies from authoritative survey papers for each topic and validating the resulting nodes and parent--child relations. We select high-quality surveys from top venues (e.g., CSUR) spanning \emph{Databases (DB)}, \emph{Machine Learning (ML)}, \emph{Software Engineering (SE)}, and \emph{Information Retrieval (IR)}. We instantiate the taxonomy over the paper corpus covered by these surveys. The survey sources are listed in Table~\ref{tab:survey_paper}.

\subsubsection{LLM-judged Quality Evaluation}
\label{app:ab_metric_defs}

As discussed in Section~\ref{sec:storage}, the knowledge representation layer organizes the papers into hierarchical taxonomies.

\noindent \textbf{Baselines.} 
We evaluate their quality against (1) \textit{TaxoAdapt}~\cite{kargupta2025taxoadapt}, which is a state-of-the-art method that refines LLM-generated taxonomies for corpus alignment. (2) two ablated variants of our reference-enhanced taxonomy construction method,
\sysnameR{} and \sysnameC{}, which differ in taxonomy
construction: \sysnameR{} directly adopts the LLM-generated
reference taxonomies, whereas \sysnameC{} derives
them entirely from the corpus — i.e., solely from the input
papers without any reference guidance.

\noindent \textbf{Evaluation Metrics.} For each taxonomy, we use GPT-5 as the LLM evaluator (prompts in Figures~\ref{fig:eval-ab-prompts-problem} and~\ref{fig:eval-ab-prompts-method}) and assess its quality across the four metrics following \cite{kargupta2025taxoadapt}:  

\begin{itemize}[leftmargin=*,noitemsep]
\item \textbf{Assignment Accuracy.}
This metric measures whether each node’s assigned papers are predominantly relevant.
For each node $n_i$ with an assigned paper set $S_i$, let $y_{ij}\in\{0,1\}$ denote the evaluator’s binary judgment that paper $j\in S_i$ is semantically relevant to $n_i$.
A node is counted as correct if a strict majority (i.e., $>50\%$) of its assigned papers are labeled relevant.
The metric is calculated as
\(
\text{Acc}_{\text{assign}}
= \frac{1}{N}\sum_{i=1}^{N}
\mathbf{I}\!\left[\frac{1}{|S_i|}\sum_{j\in S_i} y_{ij} > 0.5\right]
\),
where $\mathbf{I}[\cdot]$ is the indicator function.

\item \textbf{Hierarchical Consistency.} 
This metric evaluates whether each child node is semantically subsumed by its parent.
Let $\mathcal{PC}$ denote the set of directed parent--child pairs $(n_p,n_c)$, the metric is calculated as \(
\text{Cons}_{\text{hier}} = \frac{1}{|\mathcal{PC}|}\sum_{(n_p,n_c)\in\mathcal{PC}} \text{LLMScore}(n_p,n_c)
\),
where $\text{LLMScore}(n_p, n_c)$ is the evaluator's binary judgment that the concept of $n_p$ meaningfully generalizes that of $n_c$.

\item \textbf{Sibling Coherence.} 
This metric evaluates the conceptual coherence of sibling groups (children under the same parent).
Let $P$ be the set of parent nodes, and for each $p\in P$, let $C_p=\{n_1,\ldots,n_{k_p}\}$ denote its sibling group.
This metric is calculated as
\(
\text{Coh}_{\text{sib}} \;=\; \frac{1}{|P|}\sum_{p\in P}\text{LLMScore}_{\text{sib}}(C_p, p),
\)
, where $\text{LLMScore}_{\text{sib}}(C_p)$ is the evaluator's binary judgement that the nodes in $C_p$ are semantically comparable and instantiate the same organizing aspect relative to $p$.

\item \textbf{Description Accuracy.} 
This metric measures how accurately each node’s textual description captures its intended concept. 
For a set of taxonomy tree nodes $\{n_i\}_{i=1}^{N}$, the metric is defined as
\(
\text{Acc}_{\text{desc}} = \frac{1}{N} \sum_{i=1}^{N} \text{LLMScore}(d_i,\, c_i)
\),
where $\text{LLMScore}(d_i, c_i) \in \{0, 1\}$ indicates whether the evaluator judges the extracted attributes $d_i$ of node $n_i$ to be semantically correct with respect to the canonical concept $c_i$.
\end{itemize}

\smallskip
\noindent \textbf{Results.} Given the hierarchical nature of the taxonomy, we report quality metrics by depth and average them across constructed trees. 
Detailed Description Accuracy is omitted since all methods are near-perfect ($\approx 1$).
The results are shown in Table~\ref{tab:taxo-metrics}.
Observe that: (1) \sysname{} consistently surpasses TaxoAdapt on Hierarchical Consistency and Sibling Coherence, indicating stronger structure from cross-paper standardization and reference alignment. Note that \our{} shows slightly lower Assignment Accuracy in a few cases (e.g., problem–level 3). This stems from TaxoAdapt’s fragmentation (e.g., many nodes with ~2 papers), which can inflate the accuracy, whereas \our{} consolidates papers into fewer nodes for better structural quality.
(2) \sysname{} maintains uniformly strong results compared to its two variants, confirming that fusing corpus-grounded evidence with an LLM scaffold yields higher-quality taxonomies.

\noindent \textbf{Case Study.} To further qualitatively assess taxonomy quality, we conduct a case study using the problem taxonomy of the topic ``Vector Search''.  
As illustrated in Figure \ref{fig:appendix:ab_study}, \our{} produces a more coherent, semantically organized hierarchy with more consistent sibling groupings, and closer alignment to source papers than other approaches.

\subsection{Case Study for Self-correction}
\label{app:case_study_self_correction}

\begin{figure*}[!ht]
\centering
\includegraphics[width=0.9\linewidth, keepaspectratio]{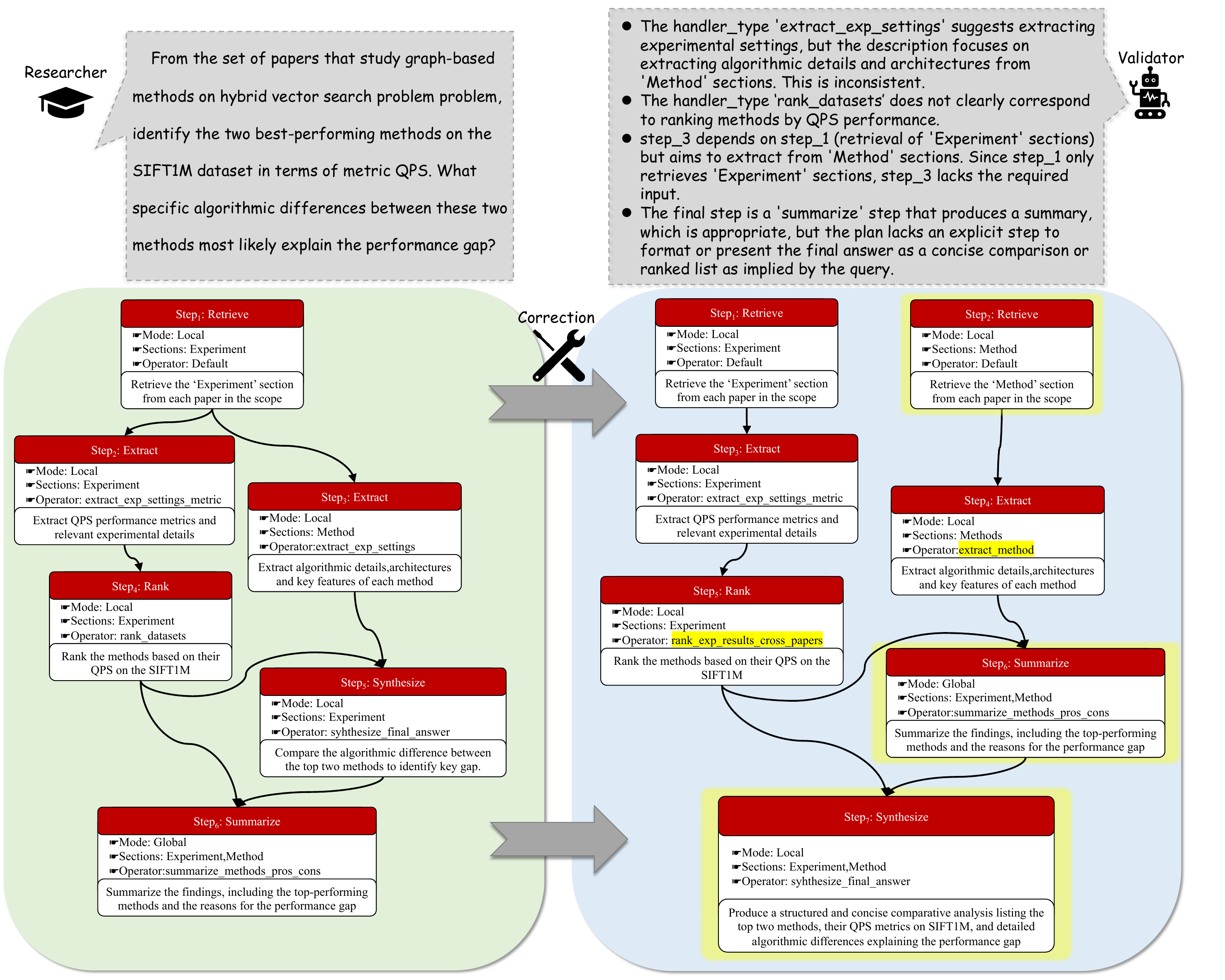} 
\caption{Case study for self correction.}
\label{fig:case_study_for_self_correction}
\end{figure*}

Figure \ref{fig:case_study_for_self_correction} presents a real case from our testing process. For the researcher's query: \textit{``From the set of papers that study graph-based methods on the hybrid vector search problem, identify the two best-performing methods on the SIFT1M dataset in terms of the QPS metric. What specific algorithmic differences between these two methods most likely explain the performance gap?''},
the system initially generated a plan tree as shown in the lower-left corner. 
\textit{Plan Validator} then inspected the plan and identified the following major issues:
\begin{itemize}[leftmargin=*,noitemsep]
    \item \textbf{Step-internal issues:} issues in internal consistency within each step.
    \begin{itemize}
        \item \textit{``The handler\_type `extract\_exp\_settings' suggests extracting experimental settings, but the description focuses on extracting algorithmic details and architectures from `Method' sections. This is inconsistent.''}
        \item \textit{``The handler\_type `rank\_datasets' does not clearly correspond to ranking methods by QPS performance.''}
    \end{itemize}

    \item \textbf{Inter-step issues:} issues in logical coherence between steps.
    \begin{itemize}
        \item \textit{``step\_3 depends on step\_1 (retrieval of `Experiment' sections) but aims to extract from `Method' sections. Since step\_1 only retrieves `Experiment' sections, step\_3 lacks the required input.''}
    \end{itemize}

    \item \textbf{Overall plan issues:} issues in the completeness or correctness of the entire plan.
    \begin{itemize}
        \item \textit{``The final step is a `summarize' step that produces a summary, which is appropriate, but the plan lacks an explicit step to format or present the final answer as a concise comparison or ranked list as implied by the query.''}
    \end{itemize}
\end{itemize}
The detected issues were then passed to the \textit{self-correction} loop for repairing, 
ultimately producing a corrected plan tree as shown in the lower-right corner (the modified parts are highlighted in yellow for clarity).

\begin{figure*}[t]
  \centering

  % Row 1
  \subfigure[TaxoAdapt\label{fig:vs-taxoadapt}]{
    \includegraphics[width=0.48\textwidth,clip,trim=0 0 0 2]{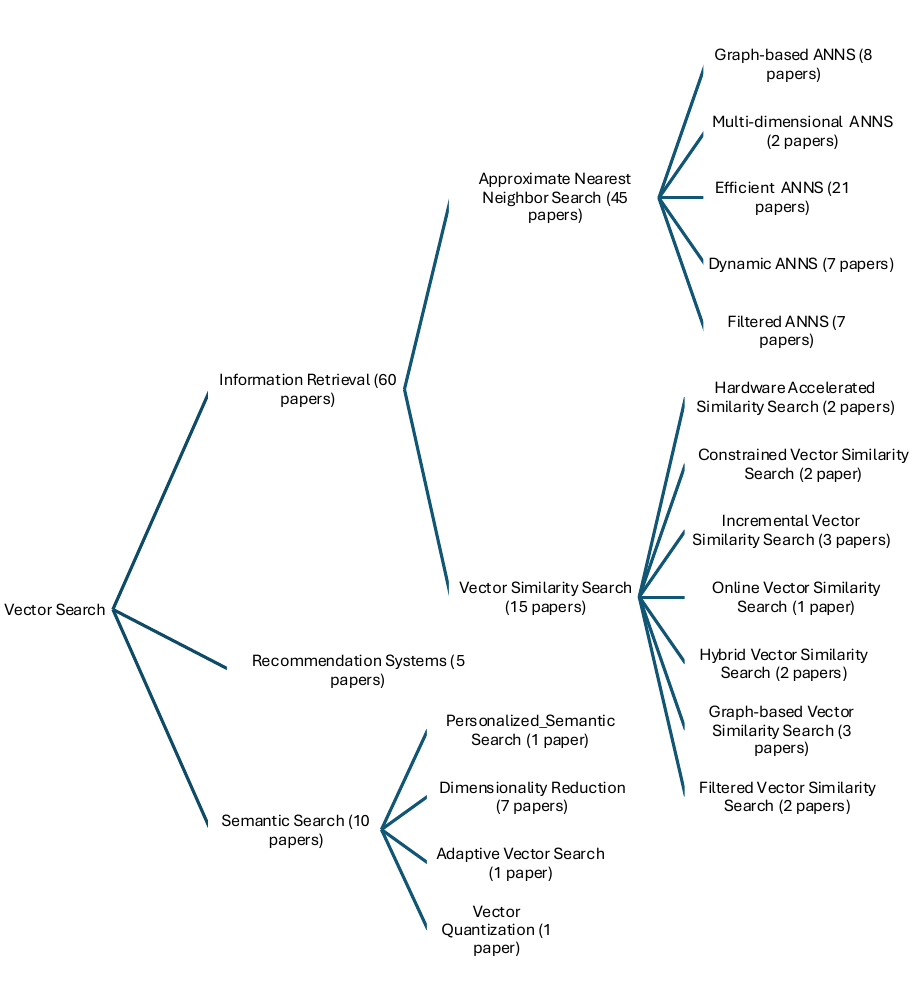}
  }\hfill
  \subfigure[\sysnameR{}\label{fig:vs-llm}]{
    \includegraphics[width=0.48\textwidth,clip,trim=0 0 0 2]{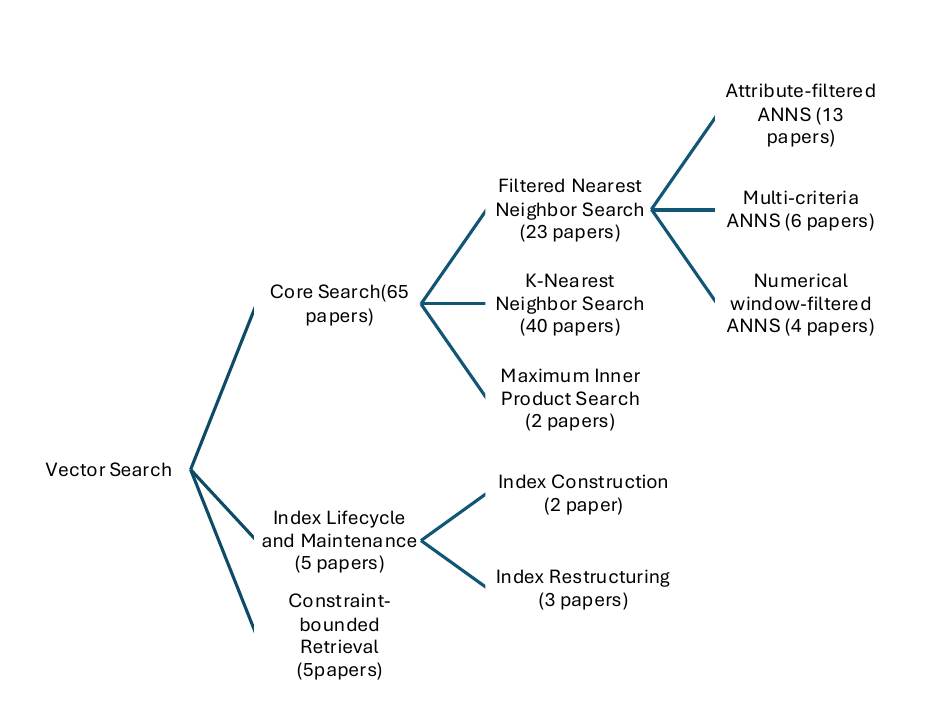}
  }

  \vspace{2mm}

  % Row 2
  \subfigure[\sysnameC{}\label{fig:vs-corpus}]{
    \includegraphics[width=0.48\textwidth,clip,trim=0 0 0 2]{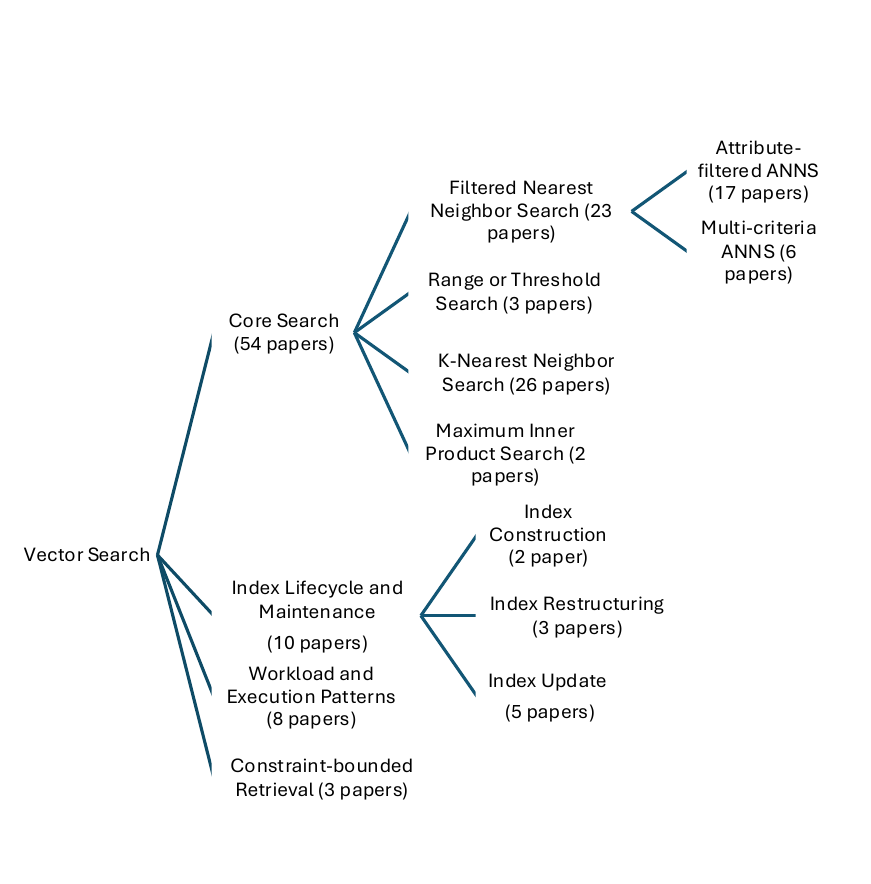}
  }\hfill
  \subfigure[\our{}\label{fig:vs-our}]{
    \includegraphics[width=0.48\textwidth,clip,trim=0 0 0 2]{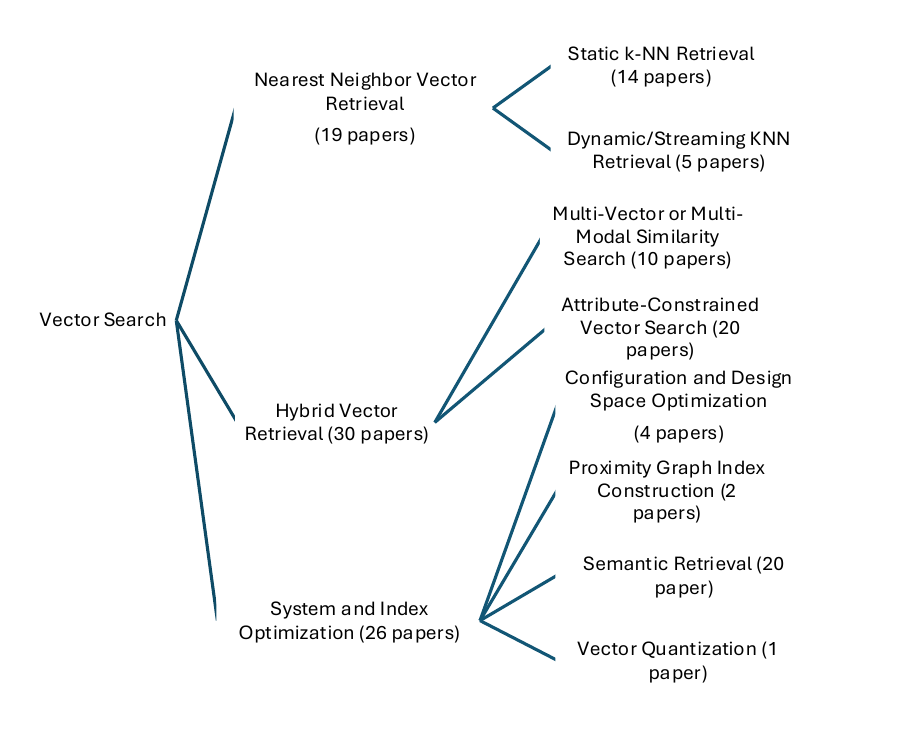}
  }

  \caption{Problem taxonomy comparison on “Vector Search”.}
  \label{fig:appendix:ab_study}
\end{figure*}

\begin{figure*}[t]
\centering
\begin{evbox}
Assignment Accuracy — You are checking whether one paper matches a problem-definition node.
Rules: use only fields present among name, description; treat "", " ", "n/a", "none", "null", "-" as missing; mark consistent only if title/abstract clearly targets the node's scope (use both name and description); if both title and abstract are effectively missing, mark inconsistent.
Scoring: "score" = 1 only if paper and node are consistent; else 0.
Input: Problem node { name: {{name}}, description: {{description}} }; Paper: {{paper}}.
Output (strict JSON; no prose): { "score": 0|1 }

Hierarchical Consistency — Decide if Node B BELONGS_TO Node A (problem definitions).
Rule: return { "score": 0 } if B is NOT a strict specialization/subset of A, or if this cannot be established; otherwise { "score": 1 }.
Notes: use only provided fields; treat "", " ", "n/a", "none", "null", "-" as missing; do not infer beyond broadly accepted topic knowledge.
Input: Node A: {{node_a}} ; Node B: {{node_b}}.
Output (strict JSON): { "score": 0 | 1 }

Sibling Coherence — Evaluate the sibling set under a parent problem-definition node.
Definition: children are siblings iff each fits under the parent and no child strictly subsumes/duplicates/contradicts another; granularity should be comparable.
Return 0 if ANY: parent misfit; subsumption (strict subset/superset); near-duplicate; cross-child conflict; insufficient info. Otherwise 1.
Input: Parent: {{parent_node}} ; Candidate siblings: {{sibling_nodes}}.
Output (strict JSON): { "score": 0 | 1 }

Description Accuracy — Evaluate a problem-definition node.
Objective: assess quality/consistency among present fields {description, input, output}; ignore missing ( "", " ", "n/a", "none", "null", "-" ).
Checks: (1) clarity/specificity; (2) alignment in scope; (3) feasibility (output achievable from input when both present); (4) no contradictions; (5) no method leakage/tautology.
Scoring: if no field present, return 1 (root/placeholder); if more than 1 present and all pass and are mutually consistent, return 1; else 0.
Input: description: {{description}} ; input: {{input}} ; output: {{output}}.
Output (strict JSON): { "score": 0|1 }
\end{evbox}
\caption{LLM evaluation prompts used to compute Assignment Accuracy, Hierarchical Consistency, Sibling Coherence, and Description Accuracy over problem taxonomy.}
\label{fig:eval-ab-prompts-problem}
\end{figure*}

\begin{figure*}[t]
\centering
\begin{evbox}
Assignment Accuracy — Check whether one paper matches a method node.
Rules: use only fields present among name, description; treat "", " ", "n/a", "none", "null", "-" as missing; mark consistent only if title/abstract clearly describes the same method/approach (or a clear variant/implementation) within the node’s stated method; if both title and abstract are effectively missing, mark inconsistent.
Scoring: "score" = 1 only if the paper is consistent with the method node; else 0.
Input: Method node { name: {{name}}, description: {{description}} }; Paper: {{paper}}.
Output (strict JSON; no prose): { "score": 0 | 1 }

Hierarchical Consistency — Decide if Node B BELONGS_TO Node A.
Rule: return { "score": 0 } if B is NOT a specific subtopic/variant/implementation of A; otherwise { "score": 1 }.
Notes: use only provided fields; treat "", " ", "n/a", "none", "null", "-" as missing; do not infer beyond broadly accepted topic knowledge.
Input: Node A: {{node_a}} ; Node B: {{node_b}}.
Output (strict JSON): { "score": 0 | 1 }

Sibling Coherence — Evaluate the sibling set under a parent method node.
Definition: children are siblings iff each fits the parent’s methodological family, no child strictly subsumes/duplicates/contradicts another, and all are at comparable granularity (peer variants).
Return 0 if ANY: parent misfit; subsumption (strict subset/superset); near-duplicate; core conflict with defining mechanisms/assumptions; pros/cons directly negate defining claims; insufficient information. Otherwise 1.
Input: Parent method: {{parent_node}} ; Candidate sibling methods (JSON array): {{sibling_nodes}}.
Output (strict JSON): { "score": 0 | 1 }

Description Accuracy — Evaluate a method node.
Objective: assess quality/consistency among present fields {description, pros, cons}; ignore missing ("", " ", "n/a", "none", "null", "-").
Checks: (1) clarity & specificity; (2) alignment of pros/cons with the method; (3) no contradictions (e.g., “low latency” vs “very slow”); (4) no near-duplicate bullets and consistent granularity; (5) plausibility of claims for the method class.
Scoring: if \textbf{more than one} field is present and all present fields pass and are mutually consistent, return 1; otherwise 0.
Input: description: {{description}} ; pros: {{pros}} ; cons: {{cons}}.
Output (strict JSON): { "score": 0 | 1 }
\end{evbox}
\caption{LLM evaluation prompts used to compute Assignment Accuracy, Hierarchical Consistency, Sibling Coherence, and Description Accuracy over the method taxonomy.}
\label{fig:eval-ab-prompts-method}
\end{figure*}

\subsection{Impact of Backbone LLMs on End-to-End Performance}
\label{ap:backbone}

Table~\ref{tab:backbone_llms1} reports results for research trend analysis, which exhibit trends similar to those in Table~\ref{tab:backbone_llms}.

\setlength{\textfloatsep}{0pt}
\begin{table*}[t]
\setlength{\tabcolsep}{3pt}
\setlength{\abovecaptionskip}{0cm}
\setlength{\belowcaptionskip}{0cm}
\footnotesize
\centering
\caption{The impact of backbone LLMs for research trend analysis. The colors express relative values (\belowavg{worse than average}, \avg{average}, and \aboveavg{better than average}).}
\label{tab:backbone_llms1}
\begin{tabular}{lcccccccccc}
\toprule
\textbf{Metric} &
\makecell{\textbf{Claude}\\\textbf{4.5 Sonnet}} &
\makecell{\textbf{Claude}\\\textbf{4.5 Opus}} &
\makecell{\textbf{GPT-5}\\\textbf{Mini}} &
\textbf{GPT-5} &
\makecell{\textbf{Gemini-3.0}\\\textbf{Flash}} &
\makecell{\textbf{Gemini-3.0}\\\textbf{Pro}} &
\makecell{\textbf{Grok-4.1}\\\textbf{Fast (NR)}} &
\makecell{\textbf{Grok-4.1}\\\textbf{Fast (R)}} & \textbf{Avg.} & \textbf{Std. Dev} \\
\midrule

\multicolumn{11}{l}{\textit{Problem-oriented}} \\
Token cost & \avg{11,270} & \belowavg{11,645} & \avg{10,328} & \belowavg{11,345} & \aboveavg{9,915} & \avg{11,203} & \aboveavg{9,908} & \belowavg{13,650} & 11,158 & $\pm$10.9\% \\
Monetary cost (\$) & \belowavg{0.164} & \belowavg{0.282} & \avg{0.021} & \avg{0.110} & \avg{0.030} & \belowavg{0.130} & \aboveavg{0.005} & \aboveavg{0.007} & 0.094 & $\pm$104.3\% \\
Efficiency (s) & \belowavg{402.6} & \belowavg{531.2} & \avg{255.8} & \avg{387.9} & \avg{243.9} & \belowavg{416.7} & \aboveavg{107.6} & \aboveavg{190.3} & 317.0 & $\pm$44.2\% \\
Correctness & \avg{7.74} & \avg{7.9} & \belowavg{7.48} & \aboveavg{8.08} & \avg{7.56} & \aboveavg{7.98} & \belowavg{7.18} & \belowavg{7.28} & 7.65 & $\pm$4.3\% \\
Relevance & \avg{8.7} & \avg{8.84} & \belowavg{8.32} & \aboveavg{9.08} & \avg{8.42} & \aboveavg{8.92} & \belowavg{7.96} & \belowavg{8.06} & 8.54 & $\pm$4.8\% \\
Diversity & \avg{7.62} & \avg{7.78} & \belowavg{7.42} & \aboveavg{7.92} & \avg{7.54} & \aboveavg{8.02} & \belowavg{7.12} & \belowavg{7.22} & 7.58 & $\pm$4.2\% \\
Specificity & \avg{8.22} & \avg{8.36} & \belowavg{7.88} & \aboveavg{8.62} & \avg{7.98} & \aboveavg{8.48} & \belowavg{7.52} & \belowavg{7.62} & 8.09 & $\pm$5.0\% \\
\midrule

\multicolumn{11}{l}{\textit{Method-oriented}} \\
Token cost& \avg{17,397} & \belowavg{18,150} & \avg{16,552} & \avg{17,463} & \aboveavg{15,760} & \belowavg{17,652} & \aboveavg{15,134} & \belowavg{21,605} & 17,464 & $\pm$11.2\% \\
Monetary cost (\$) & \belowavg{0.052} & \belowavg{0.090} & \avg{0.007} & \avg{0.035} & \avg{0.009} & \belowavg{0.042} & \aboveavg{0.001} & \aboveavg{0.002} & 0.030 & $\pm$105.1\% \\
Efficiency (s) & \avg{357.4} & \belowavg{518.5} & \avg{231.3} & \belowavg{370.9} & \avg{221.4} & \belowavg{391.8} & \aboveavg{133.6} & \aboveavg{216.9} & 305.2 & $\pm$41.0\% \\
Correctness & \avg{7.66} & \avg{7.82} & \belowavg{7.38} & \aboveavg{7.96} & \avg{7.44} & \aboveavg{7.88} & \belowavg{7.08} & \belowavg{7.18} & 7.55 & $\pm$4.4\% \\
Relevance & \avg{8.58} & \avg{8.76} & \belowavg{8.18} & \aboveavg{8.98} & \avg{8.26} & \aboveavg{8.84} & \belowavg{7.84} & \belowavg{7.92} & 8.42 & $\pm$5.1\% \\
Diversity & \avg{7.58} & \avg{7.7} & \belowavg{7.28} & \aboveavg{7.86} & \avg{7.36} & \aboveavg{7.74} & \belowavg{6.98} & \belowavg{7.06} & 7.45 & $\pm$4.4\% \\
Specificity & \avg{8.14} & \avg{8.32} & \belowavg{7.78} & \aboveavg{8.54} & \avg{7.88} & \aboveavg{8.58} & \belowavg{7.4} & \belowavg{7.48} & 8.02 & $\pm$5.7\% \\
\bottomrule
\end{tabular}
\end{table*}

\end{document}